\pdfoutput=1

\documentclass[11pt,twoside,a4paper,cmspaper,final,collab]{cms-tdr}
\def\svnVersion{294872}\def\cmsCernNo{2013-037}\def\cmsCernDate{\today}\def\cmsMessage{Published in the Journal of High Energy Physics as \href{http://dx.doi.org/10.1007/JHEP06(2015)080}{\doi{10.1007/JHEP06(2015)080.}}}

\begin{document}\cmsNoteHeader{B2G-14-002}

\hyphenation{had-ron-i-za-tion}
\hyphenation{cal-or-i-me-ter}
\hyphenation{de-vices}
\RCS$Revision: 292388 $
\RCS$HeadURL: svn+ssh://svn.cern.ch/reps/tdr2/papers/B2G-14-002/trunk/B2G-14-002.tex $
\RCS$Id: B2G-14-002.tex 292388 2015-06-15 10:22:53Z alschmid $
\newcommand{\SFb}{\ensuremath{SF_\cPqb}\xspace}
\newcommand{\SFc}{\ensuremath{SF_\cPqc}\xspace}
\newcommand{\SFlight}{\ensuremath{SF_\text{light}}\xspace}
\newcommand{\T}{\ensuremath{\mathrm{T}}\xspace}
\providecommand{\NA}{---}
\providecommand{\HTTagger}{\textsc{HEPTopTagger}\xspace}
\hyphenation{ATLAS}
\cmsNoteHeader{B2G-14-002}
\title{Search for vector-like T quarks decaying to top quarks and Higgs bosons in the all-hadronic channel
using jet substructure}

\date{\today}

\abstract{
A search is performed for a vector-like heavy T quark that is produced in pairs and that decays to a top quark and
a Higgs boson.  The data analysed correspond to an integrated luminosity
of 19.7\fbinv collected with the CMS detector in proton-proton collisions at $\sqrt{s} = 8$\TeV.
 For T quarks with large mass values the top quarks and Higgs bosons can have  significant Lorentz boosts, so that their individual decay products often overlap and merge. Methods  are applied to resolve the substructure of such merged  jets. Upper limits on the production cross section of a T quark with mass between 500 and
1000\GeVcc are derived. If the T quark decays exclusively to tH, the observed (expected)
lower  limit on the mass of the T quark is 745 (773)\GeVcc at
95\% confidence level.  For the first time an algorithm is used for tagging boosted Higgs bosons that is based on a combination of jet
substructure information and b tagging.
}

\hypersetup{%
pdfauthor={CMS Collaboration},%
pdftitle={Search for vector-like T quarks decaying to top quarks and Higgs bosons in the all-hadronic channel
using jet substructure},%
pdfsubject={CMS},%
pdfkeywords={CMS, physics, BSM, top quark}}

\maketitle
\section{Introduction}
The discovery of a Higgs boson with a mass of 125\GeVcc~\cite{Chatrchyan:1471016,Aad:2012tfa} motivates the
search for exotic states involving the newly discovered particle.  The
mechanism that stabilizes the mass of the Higgs particle is not entirely clear and could be explained by little Higgs models
\cite{ArkaniHamed:2001nc,Schmaltz:2005ky}, models with extra dimensions \cite{Antoniadis:2001cv,Hosotani:2004wv}, and
composite Higgs models \cite{Antoniadis:2001cv,Hosotani:2004wv,Agashe:2004rs}. These theories predict the existence of heavy
vector-like quarks that may decay into top
quarks and Higgs bosons. This article presents  a search for exotic resonances decaying into Higgs
bosons and top quarks. A model of
vector-like T quarks with charge 2/3\,$e$, which are produced in pairs by the strong
interaction, is used as a benchmark for this analysis.

The left-handed and right-handed components of vector-like quarks transform in the same way under the standard model (SM) symmetry
group $SU(3)_c \times SU(2)_L \times U(1)_Y$.  This allows direct
mass terms in the Lagrangian of the form $m \overline{\psi}\psi$ that
do not violate gauge invariance. As a consequence, vector-like quarks do not
acquire their mass via Yukawa couplings, in contrast to the other
quark families. A fourth generation of chiral fermions, replicating one of the three generations
of the SM with identical quantum numbers, is disfavoured by electroweak fits within the framework of the SM~\cite{PhysRevD.86.013011}. This is because of the large modifications  to the Higgs production cross sections and branching fractions, if a single SM-like Higgs doublet is assumed. Vector-like
heavy quarks are not similarly constrained by the measurements of the
Higgs boson properties~\cite{PhysRevD.88.094010}.

Vector-like T quarks can decay into three different final states: tH,
tZ, and bW~\cite{PhysRevD.88.094010}. The assumption of decays with 100\%
branching fraction ($\mathcal{B}$) has been used in various searches by the ATLAS and CMS collaborations
\cite{Chatrchyan2012307,Chatrchyan2012103,Aad20131284,PhysRevLett.107.271802}. Other
searches that do not make specific assumptions on the branching
fractions have also been performed \cite{tagkey2014149}.  In the
present analysis  the event selection is optimized to be sensitive to exclusive T quark
decays to tH. In addition,   the results are quoted
as a function of the branching fractions to the three decay modes: tH, tZ, and bW.

While searches for T quarks have been performed in leptonic final
states~\cite{Chatrchyan2012307,Chatrchyan2012103,Aad20131284,PhysRevLett.107.271802,tagkey2014149},
this article presents the first
analysis that exploits the all-hadronic final state in the search for
vector-like quarks.
In the SM the Higgs boson decays  predominantly into
b quark pairs with a branching fraction of 58\%  for a mass of
125\GeVcc, while the top quark
decays almost exclusively into a bottom quark and a W boson, which in turn decays
hadronically  67.6\% of the time. The main final state is therefore
the all-hadronic final state $\T\to \PQt\PH \to (\PQb jj)  (\bbbar) $,
where $j$ denotes the light-flavour jets of the W boson
decay and b denotes the b-flavour jets from the top quark or Higgs
boson decays. For sufficiently large T quark mass values, the decay
products  can be highly Lorentz-boosted, leading to final states with overlapping and
merged jets. In the extreme case, all top quark decay products are
merged into a single jet. A similar topology may arise for the Higgs
boson decaying into b quarks. A related analysis concept has been proposed in Ref.~\cite{Kribs:2010ii}. In recent years, the methodology of jet substructure analysis has  proved to be very powerful in resolving such  boosted
topologies~\cite{Butterworth:2008i,catop_theory,catop_cms,Plehn:2011tg}. For example, the analysis of high-mass Z' resonances
decaying into top quark pairs became feasible in the all-hadronic final
state as a result of the application of jet substructure methods
\cite{ZprimeSpringer,PhysRevLett.111.211804,ATLASZprime}.  A
similar strategy is followed in this analysis by applying algorithms for the
identification of boosted top quarks (t tagging) and
boosted Higgs bosons (H tagging)  in combination with
algorithms for the identification of b quark jets (b tagging).  In particular, the application of
b tagging in subjets has enhanced the identification of boosted
$\bbbar$  final states, for instance $\PH\to \bbbar$ decays. This is the first analysis to
apply an algorithm for tagging boosted Higgs bosons that is based on a combination of jet
substructure information and b tagging.

\section{The CMS detector}
The central feature of the CMS apparatus is a superconducting solenoid of 6\unit{m} internal diameter. Within the superconducting solenoid volume are a silicon pixel and strip tracker, a lead tungstate crystal electromagnetic calorimeter (ECAL), and a brass and scintillator hadron calorimeter (HCAL), each composed of a barrel and two endcap sections. Muons are measured in gas-ionization detectors embedded in the steel flux-return yoke outside the solenoid. Extensive forward calorimetry complements the coverage provided by the barrel and endcap detectors.

The energy resolution for photons with $\ET {\approx} 60$\GeV varies between 1.1 and 2.6\% over the solid angle of the ECAL barrel, and from 2.2 to 5\% in the endcaps. The HCAL, when combined with the ECAL, measures jets with a resolution $\Delta E/E \approx 100\% / \sqrt{E\,[\GeVns{}]} \oplus 5\%$~\cite{Chatrchyan:2013dga}.

In the region $\abs{ \eta }< 1.74$, the HCAL cells have widths of 0.087 in $\eta$ and 0.087 in azimuth ($\phi$). In the $\eta$-$\phi$ plane, and for $\abs{\eta}< 1.48$, the HCAL cells map on to $5 \times 5$ ECAL crystal arrays to form calorimeter towers projecting radially outwards from close to the nominal interaction point. At larger values of $\abs{ \eta }$, the size of the towers increases and the matching ECAL arrays contain fewer crystals. Within each tower, the energy deposits in ECAL and HCAL cells are summed to define the calorimeter tower energies, subsequently used to provide the energies and directions of hadronic jets.

The silicon tracker measures charged particles within the pseudorapidity range $\abs{\eta}< 2.5$. It consists of 1440 silicon pixel and 15\,148 silicon strip detector modules and is located in the 3.8\unit{T} field of the superconducting solenoid. For nonisolated particles of $1 < \pt < 10\GeVc$ and $\abs{\eta} < 1.4$, the track resolutions are typically 1.5\% in \pt and 25--90 (45--150)\mum in the transverse (longitudinal) impact parameter \cite{TRK-11-001}.

A more detailed description of the CMS detector, together with a definition of the coordinate system used and the relevant kinematic variables, can be found in Ref.~\cite{Chatrchyan:2008zzk}.

\section{Event samples \label{sec:samples}}
The data used for this analysis were collected by the CMS experiment
using pp collisions provided by the CERN LHC with a centre-of-mass energy of 8 TeV, and correspond to an integrated
luminosity of 19.7\fbinv.  Events are selected online by a trigger
algorithm that requires \HT, the scalar
sum of the transverse momenta of reconstructed jets in the detector,  to be
greater than 750\GeVc. The online \HT is calculated from calorimeter jets with $\pt > 40$\GeVc. Calorimeter jets are reconstructed from the energy deposits in the calorimeter towers, clustered by the anti-\kt algorithm~\cite{Cacciari:2008gp, fastjet} with a size parameter of 0.5.

Simulated samples are used to determine signal selection efficiencies
as well as the background contribution from $\ttbar$ plus
jets,  $\ttbar H$, and  hadronically decaying W/Z plus b jet
production. The background from QCD multijet production is
derived from data.

Events from  T quark decays are generated
for mass hypotheses between 500 and 1000\GeVcc in steps of 100\GeVcc. The inclusive cross
sections for the signal samples and $\ttbar$  samples are calculated at
next-to-next-to-leading order (NNLO) for the reaction $\Pg\Pg \to \ttbar +
X$. The fixed order calculations are supplemented with soft-gluon
resummation with next-to-next-to-leading logarithmic
accuracy~\cite{Czakon:2013goa}.  The  $\ttbar$ cross sections are computed based on the
\textsc{Top++}~v2.0 implementation using the MSTW2008nnlo68cl
parton distribution
functions (PDF) and the 5.9.0 version of
LHAPDF~\cite{Czakon:2013goa,Czakon:2011xx}. The evaluated $\ttbar$ cross section is 252.9\unit{pb}, assuming a top quark mass of 172.5\GeVcc. The
theoretical pair-production cross sections for the signal samples are
listed in Table \ref{tab:signalEfficiency}.

The
mass of the Higgs boson in the signal samples is set to 120\GeVcc, as the samples were produced before the discovery of the Higgs boson.  The
branching fractions of the Higgs boson decays  are corrected to the expected
values for a Higgs boson with a mass of 125\GeVcc  using the
recommendations from Ref.~\cite{Heinemeyer:2013tqa}.
The difference between the
 actual mass of the Higgs boson (125\GeVcc) and the simulated mass (120\GeVcc) has no impact
on the analysis results.

The $\ttbar$ background sample is generated with \POWHEG v1.0~\cite{powheg1,powheg2,powheg3} interfaced to
\PYTHIA~6.426 \cite{1126-6708-2006-05-026} to simulate the parton shower and
hadronisation.
All other
background samples and the signal samples are
simulated with  \MADGRAPH~5.1~\cite{MadgraphRef},
interfaced with \PYTHIA~6.426. The
CTEQ6L1~\cite{1126-6708-2002-07-012} PDF set is used with \MADGRAPH, while the \POWHEG
samples have been produced with CTEQ6M. For \PYTHIA, the Z2* tune is
used to simulate the underlying event~\cite{Field:2010bc}.

Simulated QCD multijet samples are used to validate the estimation of
this background from data. These samples are simulated with
\MADGRAPH in the same way as the other background samples described above.

\section{Event reconstruction \label{sec:reco}}

Tracks are reconstructed using an iterative tracking procedure~\cite{TRK-11-001}. The primary vertices are reconstructed with a deterministic annealing method~\cite{IEEE_DetAnnealing} from all  tracks in the event that are compatible with the location of the proton-proton interaction region.
The vertex with the highest $\sum (\pt^\text{track})^2$ is defined as the
primary interaction vertex,
whose position is determined from an adaptive vertex
fit~\cite{AVFitter}.

The particle-flow event algorithm \cite{CMS-PAS-PFT-09-001,CMS-PAS-PFT-10-001}  reconstructs and identifies each individual particle with an optimized combination of information from the various elements of the CMS detector. The energy of photons is directly obtained from the ECAL measurement, corrected for zero-suppression effects. The energy of electrons is determined from a combination of the electron momentum at the primary interaction vertex as determined by the tracker, the energy of the corresponding ECAL cluster, and the energy sum of all bremsstrahlung photons spatially compatible with originating from the electron track. The energy of muons is obtained from the curvature of the corresponding track. The energy of charged hadrons is determined from a combination of their momentum measured in the tracker and the matching ECAL and HCAL energy deposits, corrected for zero-suppression effects and for the response function of the calorimeters to hadronic showers. Finally, the energy of neutral hadrons is obtained from the corresponding corrected ECAL and HCAL energy.

For each event, hadronic jets are clustered from these reconstructed particles with the infrared and collinear-safe anti-$k_\mathrm{t}$ algorithm  or with the  Cambridge--Aachen algorithm (CA jets) \cite{CACluster1}. The jet momentum is defined to be the vector sum of all particle momenta in this jet, and is found in the simulation to be within 5\% to 10\% of the true momentum over the whole \pt spectrum and detector acceptance. Jet energy corrections are derived from the simulation, and are confirmed with in situ measurements using the energy balance of dijet and photon+jet events~\cite{JEC2012}. The jet energy resolution amounts typically to 15\% at 10\GeV, 8\% at 100\GeV, and 4\% at 1\TeV, to be compared to about 40\%, 12\%, and 5\% obtained when the calorimeters  are used alone for jet clustering.

The jets contain neutral particles from additional collisions within the same beam crossing (pileup). The contribution from these additional
particles is subtracted based on the average expectation of the energy deposited from pileup
in the jet area, using the methods described in Ref. \cite{Cacciari:2008gn}.

For the identification of b jets, the combined secondary vertex (CSV)
algorithm is used and the medium
operating point (CSVM) is applied \cite{1748-0221-8-04-P04013}. With this operating point  the b tagging efficiency is 70\% and the light flavour jet  misidentification rate is 1\%  in \ttbar events. This algorithm uses information from
 reconstructed tracks and secondary vertices that are displaced from the primary interaction vertex. The information is combined into a single discriminating variable. The same b tagging algorithm is used in boosted topologies and the corresponding efficiencies and misidentification rates are tested in the relevant samples. More details on b tagging in boosted topologies are given in Section~\ref{sec:substructure}.

\section{Analysis strategy \label{sec:strategy}}
Event selection criteria that make use of novel jet substructure methods are applied to reduce the large background contributions from QCD multijet and \ttbar events in the analysis. The jet substructure methods are described in detail in Section \ref{sec:substructure} and the event selection criteria are summarized in Section \ref{sec:selection}.

Two variables are used to distinguish signal from background events after the event selection. These variables are \HT and the invariant mass $m_{\bbbar}$ of two b-tagged subjets in Higgs boson candidate jets. High \HT values characterize events with large hadronic activity as in the case of signal events.

The shape and normalization of the \HT and $m_{\bbbar}$ distributions of QCD multijet events in this analysis are derived using data in signal-depleted sideband regions. The sideband regions are defined by inverting the jet substructure criteria. Closure tests are performed with simulated QCD events to verify that the method predicts the  rates and shapes of \HT and $m_{\bbbar}$ accurately.  The background determination is discussed in detail in Section \ref{sec:background}.

The  \HT and $m_{\bbbar}$ variables are combined into a single discriminator that enhances the sensitivity of the analysis. This combination is performed using a likelihood ratio method, which is described in Section \ref{sec:results}.

Two event categories are used in the statistical interpretation of the results: a category with a single Higgs boson candidate and a category with at least two Higgs boson candidates. These are denoted as single and multiple H tag categories. They are chosen as such to be statistically independent and are combined in setting the final limit. For the multiple H tag category, the Higgs boson candidate with the highest transverse momentum is used in the likelihood definition. The procedure of the limit setting is discussed in detail in Section~\ref{sec:results}.

\section{Jet substructure methods \label{sec:substructure}}
\begin{figure}[hb]
                \includegraphics[width=.49\textwidth]{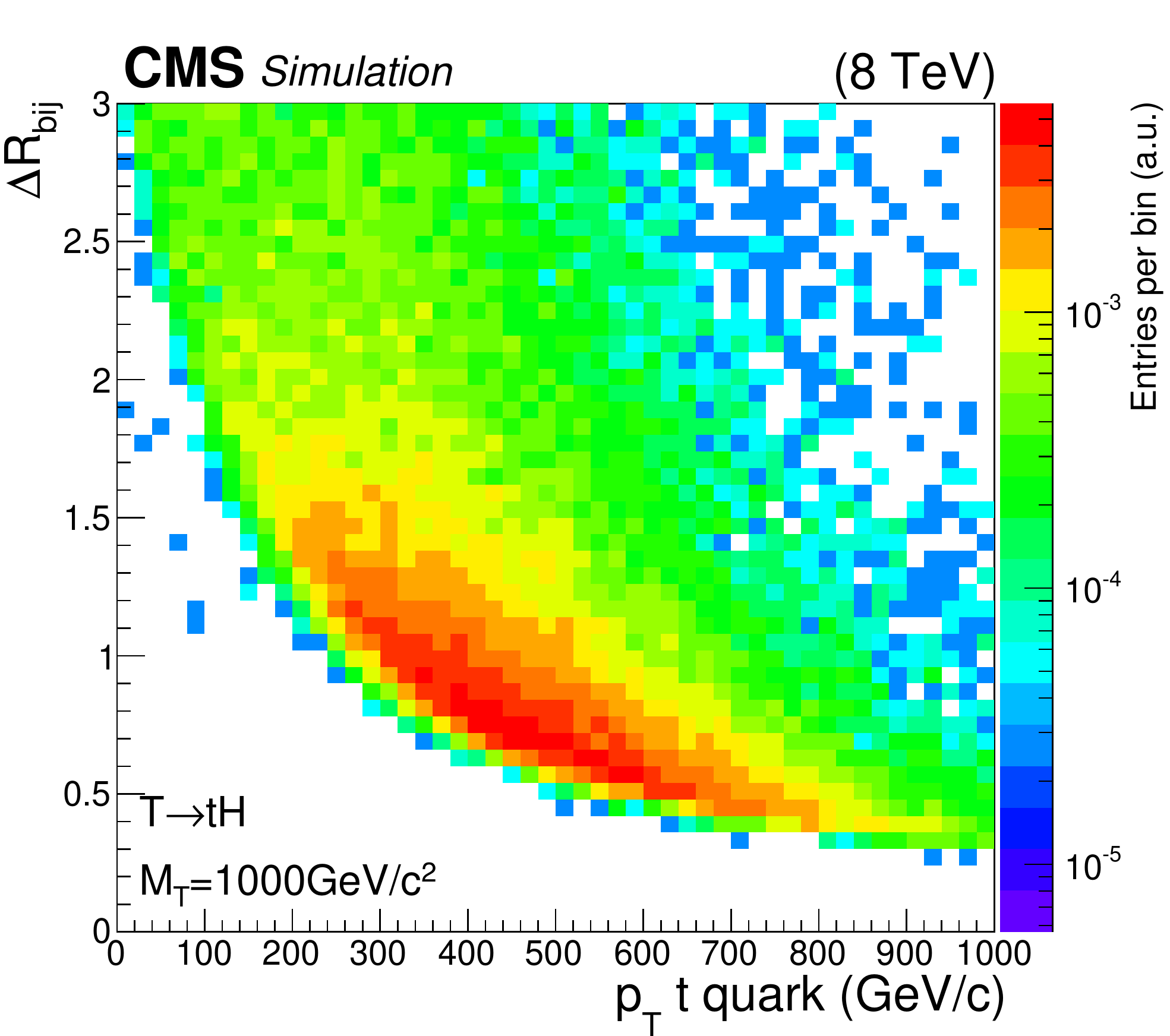}
                \includegraphics[width=.49\textwidth]{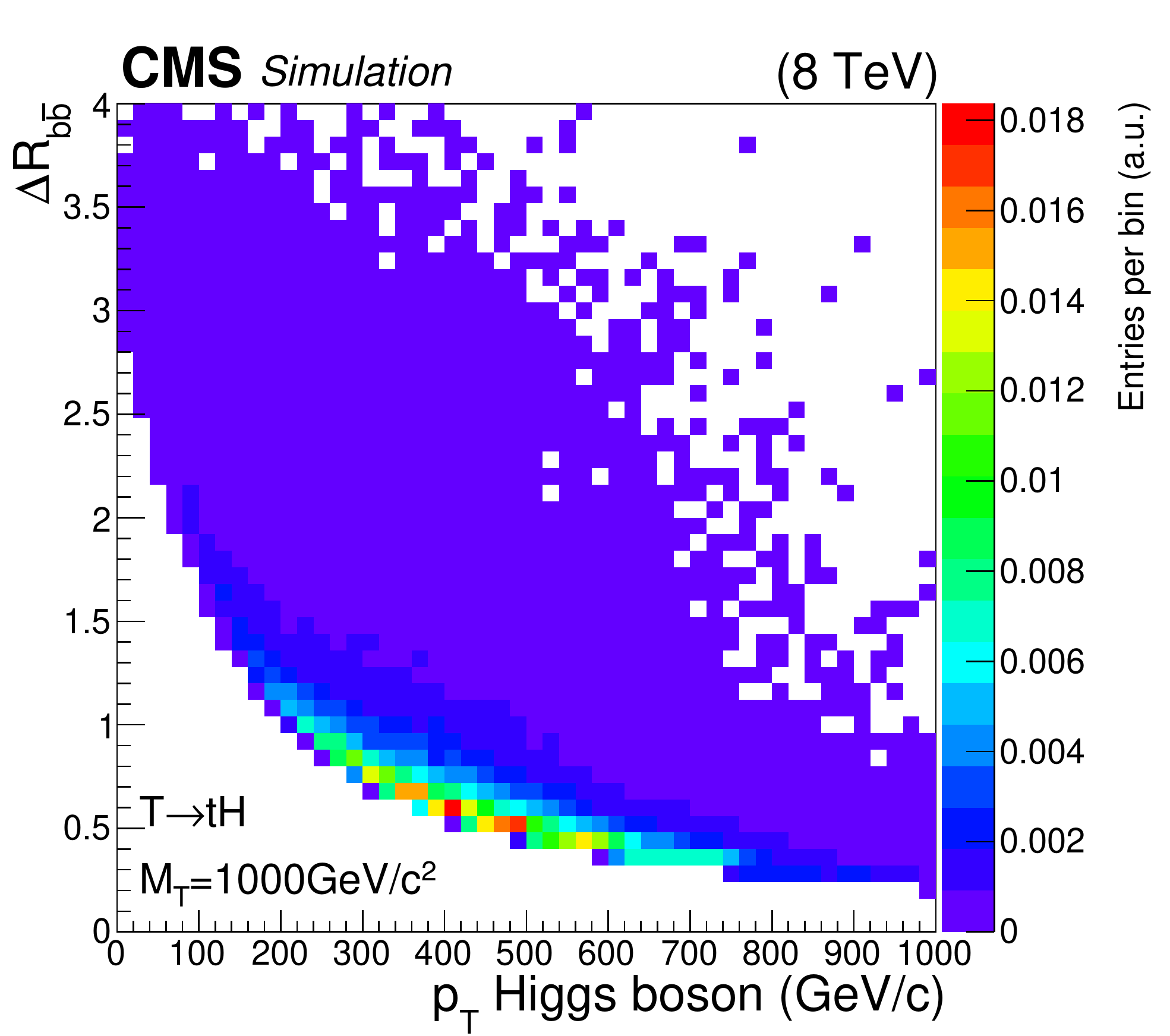}
        \caption{The distribution of the angular distance $\Delta
          R_{\cPqb ij}$ between the three top quark decay products
          as a function of the top quark \pt  for simulated
          T quark events with a T quark mass of 1000\GeVcc
          (left). Distribution of the angular distance  $\Delta R_{\bbbar}$ of the two generated b
          quarks from  Higgs boson decays versus the Higgs boson
          \pt, for the same event sample (right).}
    \label{fig:bananaPlots}
\end{figure}

Because of the large mass of the T quarks, the top quarks and Higgs bosons from T quark decays
would have  significant Lorentz boosts. Daughter particles of these
 top quarks are therefore not well separated. In many cases all of the
top quark decay products are clustered into a single, large jet  by the event reconstruction algorithms. The approximate spread of a hadronic top quark decay can be determined on simulated events from the $\Delta R$ distances between the quarks produced during its decay. The four-momenta of the two quarks with the smallest $\Delta R$ distance,
$\Delta R (\cPq_{1}, \cPq_{2})$, are vectorially summed and the
$\Delta R$ distance between the vector sum  and
the third quark, $\Delta R (\cPq_{1+2}, \cPq_{3})$, is evaluated. The maximum distance between
$\Delta R (\cPq_{1}, \cPq_{2})$ and $\Delta R (\cPq_{1+2}, \cPq_{3})$ indicates the approximate size
$\Delta R_{\cPqb ij}$ needed to cluster the entire top quark decay within one single CA jet. For the boosted decays of a Higgs boson in
$\PH\to \bbbar$ events, the corresponding quantity can be
defined as the angular distance $\Delta R_{\bbbar}= \sqrt{\smash[b]{(\Delta \eta)^2 + (\Delta \phi)^2}}$ between the two generated b
quarks. Figure~\ref{fig:bananaPlots} shows the distributions of these quantities plotted as a function of the transverse
momentum of the top quark and of the Higgs boson, generated from the decay of a T quark
with a mass of 1000\GeVcc. This shows that, for large transverse
momenta, and hence for large T quark mass values, the decay products from Higgs bosons and top quarks are generally collimated and are
difficult to separate using standard jet reconstruction algorithms.

The approach adopted by this analysis is to apply the CA
algorithm using a large size parameter $R=1.5$, in order to cluster the decay products from top quarks and Higgs bosons into single large CA jets,  using an implementation based on \textsc{FastJet}~3.0~\cite{fastjet}.
To identify these so called ``top jets'' and ``Higgs jets'', the
analysis uses dedicated jet substructure tools, in particular
a t tagging algorithm and a H tagging algorithm that relies on
b tagging of individual subjets. A more detailed
description of these algorithms is provided in the following sections.

\subsection{Subjet b tagging and H tagging\label{sec:subjetbtag}}

It is not possible to identify b jets in boosted top quark decays using the standard CMS b tagging algorithms, since these are based on
separated, non-overlapping jets. For dense environments where standard
jet reconstruction algorithms are not suitable, two dedicated
b tagging concepts have been investigated:
\begin{itemize}
\item tagging of CA jets, reconstructed using a distance parameter of
  0.8 (CA8 jets) or 1.5 (CA15 jets). The 0.8 and the 1.5 jet size
  parameters are used because they have been found to provide optimal
  performance for large and for intermediate boost
  ranges, respectively, as discussed in the following sections.
\item tagging of subjets that are reconstructed
  within CA jets.
\end{itemize}
The subjets of CA15 jets are reconstructed using the
``filtering algorithm''~\cite{Butterworth:2008i},  splitting jets into
subjets based on an angular distance of $R=0.3$. Only the three
highest \pt subjets are retained. This filtering algorithm has been found to provide the best mass resolution for CA15 jets compared to the jet pruning \cite{Ellis:2009su} and trimming \cite{Krohn:2009th}  algorithms. The pruning, trimming, and filtering algorithms are often referred to as jet grooming algorithms and their main purpose is to remove soft and wide-angle radiation as well as pileup contributions. Subjets of CA8 jets are reconstructed using the pruning algorithm, which is found to give the best performance for the reduced jet size.

For the application of b tagging to CA jets, tracks in a wide region around
the jet axis are considered. The association region corresponds to the
size of the CA jet. For the application of b tagging to subjets, tracks in a region of $\Delta
R<0.3$ around the subjet axis are used by the b tagging algorithm. This is the cone size employed by the standard CMS b tagging algorithms, and has also been found to give good performance for subjet b tagging.

The advantage of subjet b tagging is that it allows two subjets within a single CA jet to be identified as b jets. This
is the main component of the H tagging algorithm that
distinguishes between boosted Higgs bosons
decaying to $\bbbar$ and boosted top quarks.

\subsubsection{Algorithm performance\label{sec:subbtagperformance}}

\begin{figure}
\centering
\includegraphics[width=0.49\textwidth]{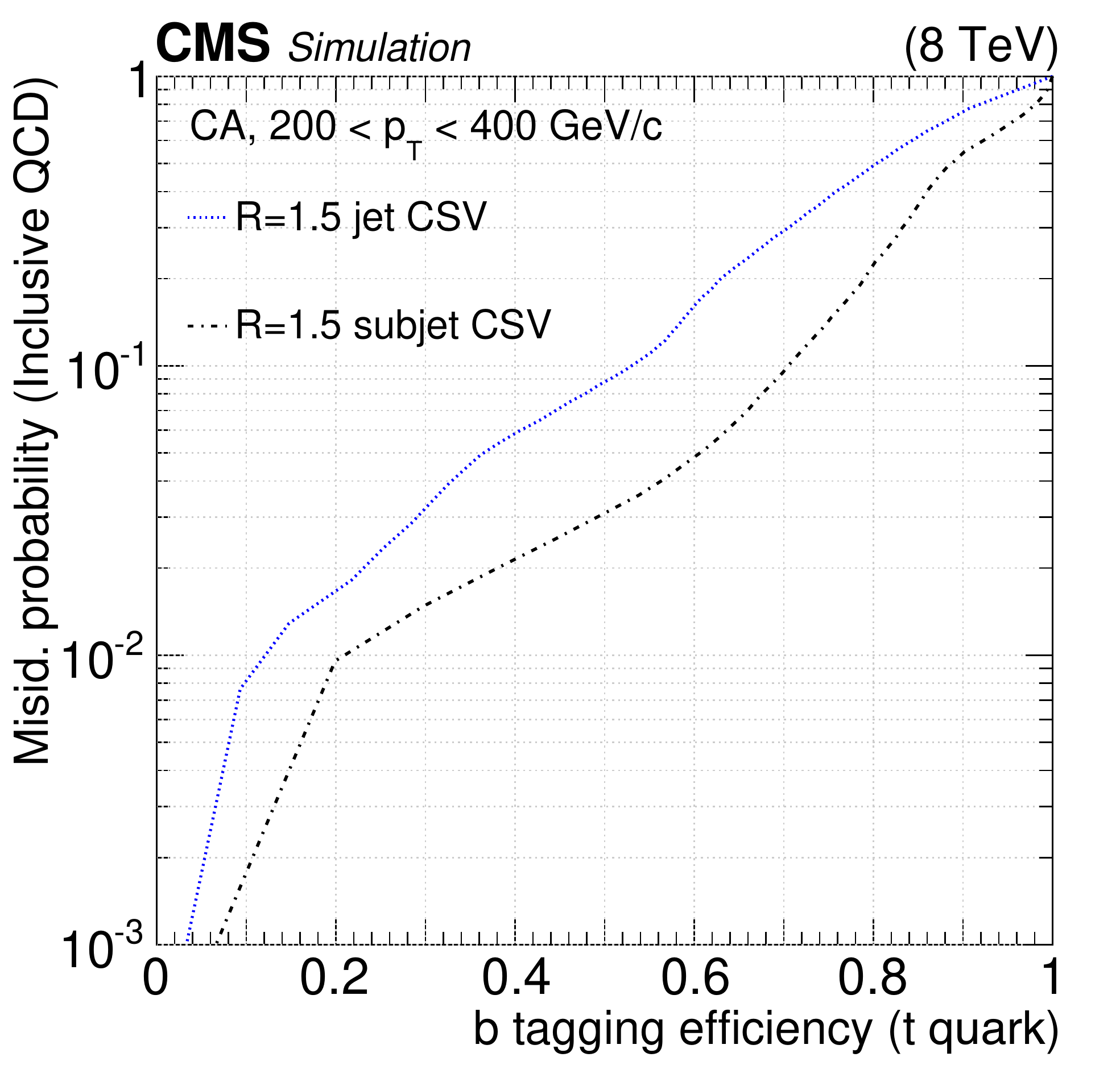}
 \includegraphics[width=0.49\textwidth]{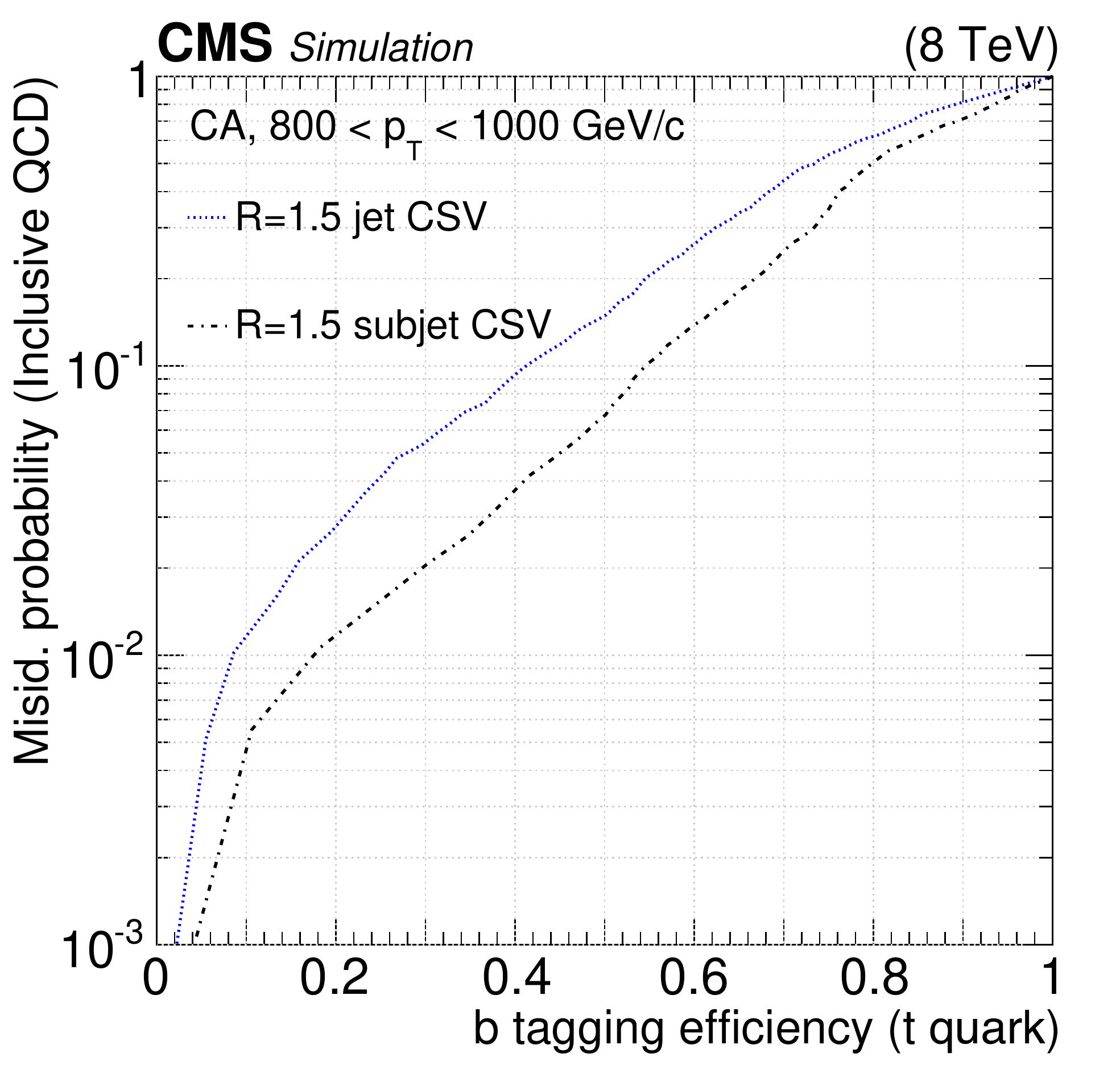}
\caption{Performance of the CSV b tagging algorithm in simulated
  events with CA15 jets and with subjets within the same CA15
  jet. The misidentification probability for inclusive QCD jets is
  shown versus the b tagging efficiency for boosted top quarks
  originating from T quark decays, for CA15 jet transverse momentum ranges of (left) $200 < \pt < 400$\GeVc and (right) $800 < \pt < 1000$\GeVc.}
\label{fig:boostedTopBTag}
\end{figure}

Figure~\ref{fig:boostedTopBTag} shows the performance of subjet b tagging compared to CA15 jet b tagging for events with boosted top quarks
that originate from T quark decays. The choice of the clustering algorithm and the cone size is driven by the t tagging algorithm, described in Section~\ref{sec:toptag}. The b tagging efficiency is plotted
versus the misidentification probability for inclusive QCD
jets. Two different regions of transverse jet
momentum are shown. It can be seen that   subjet b tagging outperforms the CA15 jet b tagging.

\begin{figure}
\centering
\includegraphics[width=0.49\textwidth]{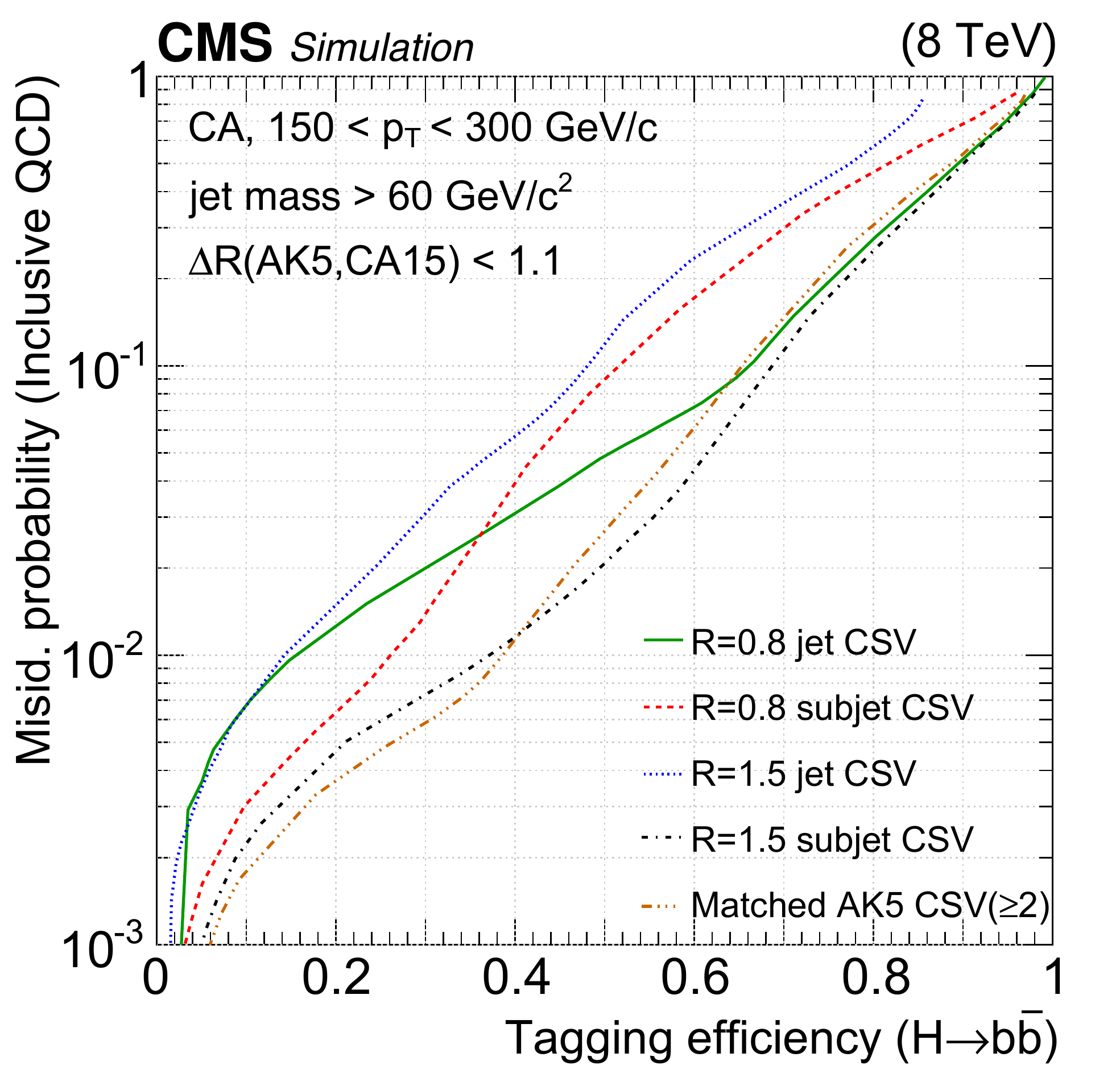}
 \includegraphics[width=0.49\textwidth]{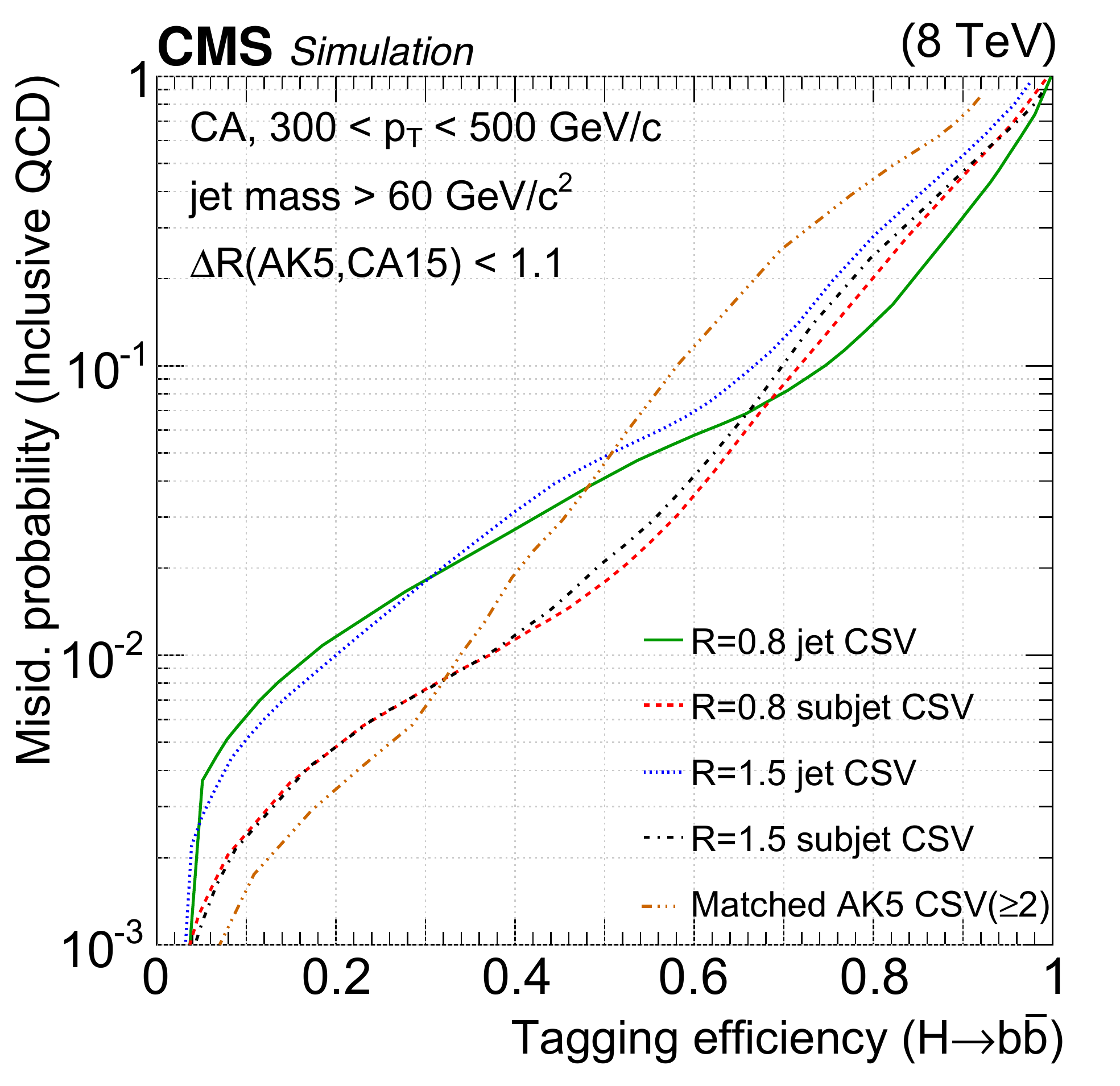}
\caption{ Performance of different H tagging algorithms in simulated
signal  events, with a signal mass hypothesis of 1000\GeVcc.  The misidentification probability for inclusive QCD jets is
  shown versus the  tagging efficiency for boosted Higgs boson decays, for jet transverse momentum ranges of (left)
$150<\pt<300$\GeVc and (right) $300<\pt<500$\GeVc. Different b tagging options are compared: standard b tagging of AK5 jets, subjet b tagging of CA15 and CA8 jets, and b tagging of CA15 jets and CA8 jets. For the case of
  subjet b tagging, two subjets are required to pass the b tagging
  criteria. Similarly, two AK5 jets are required to pass the b tagging criteria for standard b tagging.}
\label{fig:Higgstag}
\end{figure}

\begin{figure}
\centering
\includegraphics[width=0.49\textwidth]{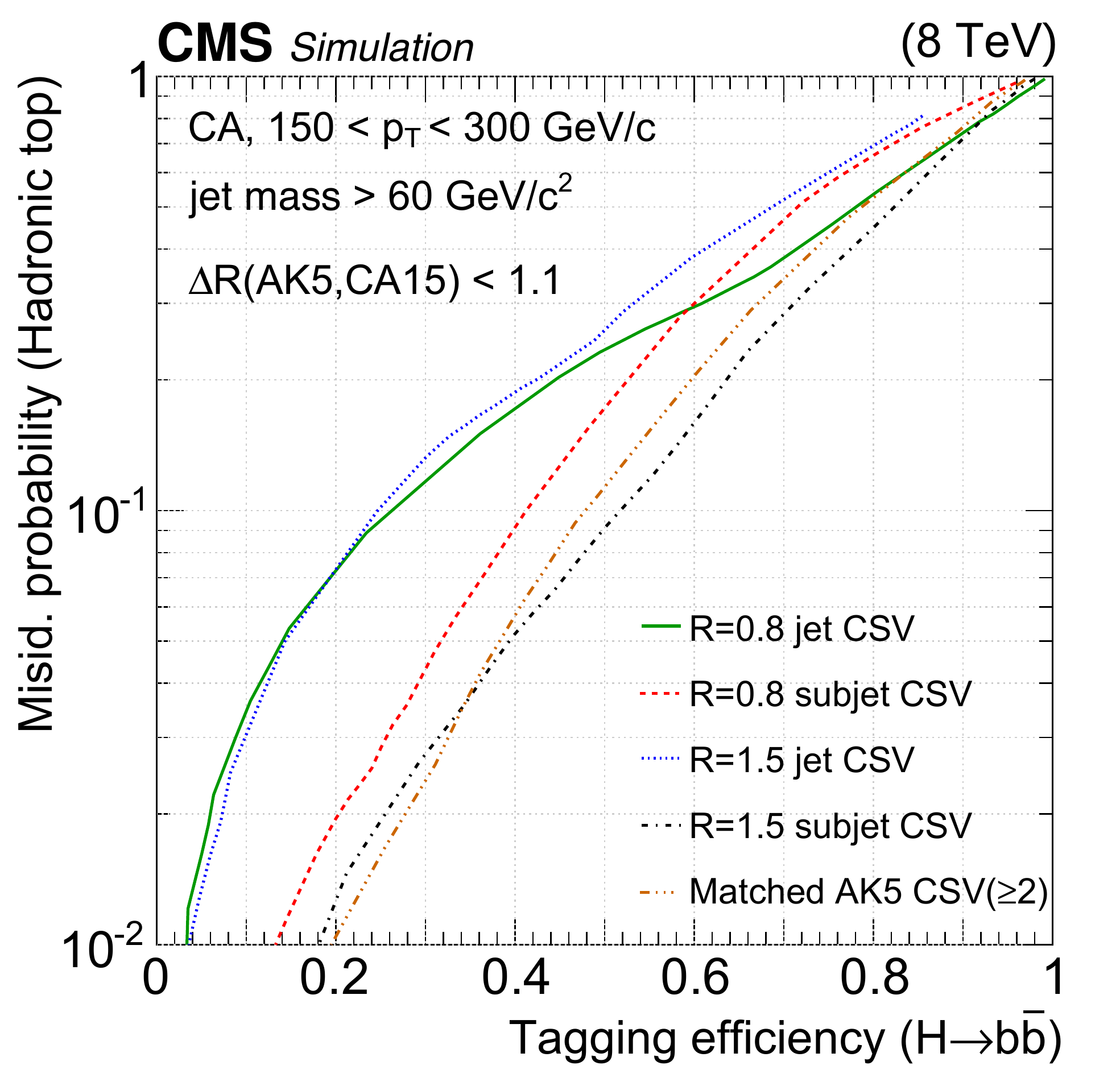}
 \includegraphics[width=0.49\textwidth]{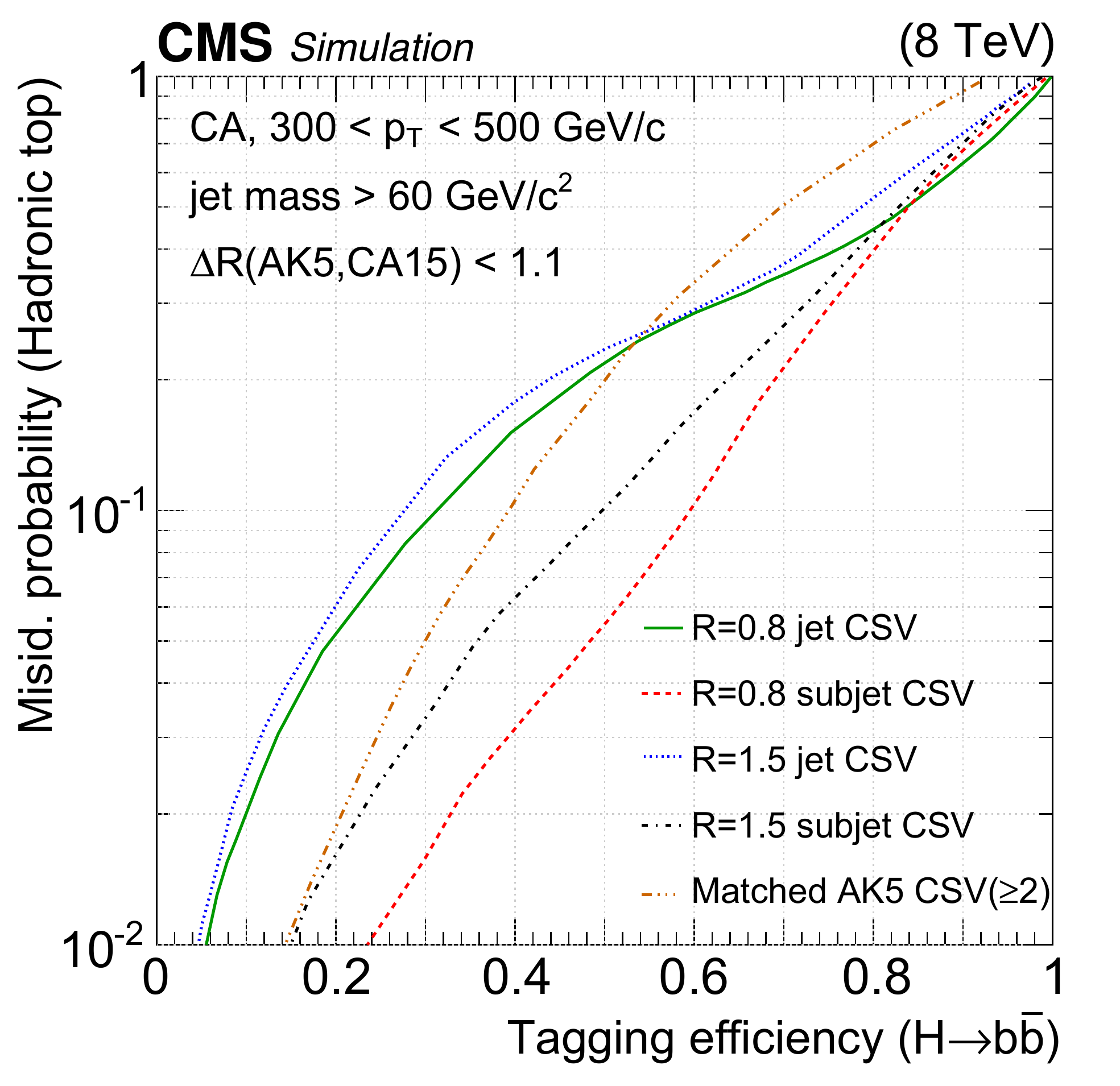}
\caption{ Performance of different H tagging algorithms in simulated
signal  events, with a signal mass hypothesis of 1000\GeVcc.  The misidentification probability for the \ttbar background is
  shown versus the   tagging efficiency for boosted Higgs boson decays, for jet transverse momentum ranges of (left)
$150<\pt<300$\GeVc and (right) $300<\pt<500$\GeVc. Different b tagging options are compared: standard b tagging of AK5 jets, subjet b tagging of CA15 and CA8 jets, and b tagging of CA15 jets and CA8 jets. For the case of
  subjet b tagging, two subjets are required to pass the b tagging
  criteria. Similarly, two AK5 jets are required to pass the b tagging criteria for standard b tagging.}
\label{fig:Higgstagtop}
\end{figure}

For the identification of boosted Higgs bosons, two subjets must be  b tagged and their invariant mass must be greater than 60\GeVcc.
Both CA8 jets and CA15 jets are considered. The performance of the H tagging algorithm is shown in Fig.~\ref{fig:Higgstag} for two different regions of transverse jet
momentum. The  tagging efficiency is shown
versus the misidentification probability for inclusive QCD
jets. Figure~\ref{fig:Higgstagtop} shows the performance obtained when
evaluating the misidentification probability from \ttbar events. The
performance of the standard b tagging algorithm based on AK5 jets is
also shown. A CA15 jet is considered as satisfying the H tagging requirement if two AK5 jets satisfy the b tagging requirement and have a $\Delta R$ distance $<$1.1 from the CA15 jet. Overall, subjet b tagging is found to provide better performance than b tagging based on AK5 jets.  The choice of the optimal CA jet size parameter $R$ depends on the $\pt$ region considered.  A size of $R=1.5$ is found to be optimal for most signal mass hypotheses and is chosen for the analysis.

\subsubsection{Scale factors\label{sec:subbtagSF}}
The subjet b tagging efficiency has been measured in data using a
sample of semileptonic \ttbar events.  Scale factors
have been derived to correct the efficiency predicted by simulation to that measured in data. The ``flavor-tag consistency'' (FTC)
method~\cite{1748-0221-8-04-P04013} has been used to measure
these scale factors. The FTC method requires
consistency between the number of b-tagged jets in data and
simulation for  boosted top quark events. A maximum likelihood fit is performed
in which the b tagging efficiency scale factor \SFb and the $\ttbar$ cross section are free parameters.
Usually the light flavour misidentification scale factor \SFlight is
fixed to a value obtained independently, but in this case the
simultaneous fit of \SFlight, \SFb, and the $\ttbar$ cross section has
been performed for the first time.  This method relies on
simulation for the flavour of the subjets. A systematic
uncertainty of 2\% in the subjet flavour composition is taken into account.

The FTC method is
applied  to three different \pt regions
of the CA15 jet: $150\le\pt<350$\GeVc, $\pt\ge350$\GeVc, and
$\pt\ge450$\GeVc. No significant deviation of the scale factors for the three different samples is observed. Both the scale factors \SFb and \SFlight are found to be in agreement with the scale factors measured for standard b tagging of AK5 jets in the non-boosted regime.

The efficiency of the  invariant mass selection requirement for  the two b-tagged subjets
of the Higgs boson candidate is validated with a sample of semileptonic $\ttbar$
events. Since no  sample of Higgs bosons decaying into b quark pairs can
be obtained in data, the validation procedure is based on the
selection of a pure sample of W bosons.

\begin{figure}[!htb]
\begin{center}
   \includegraphics[width=0.48\textwidth]{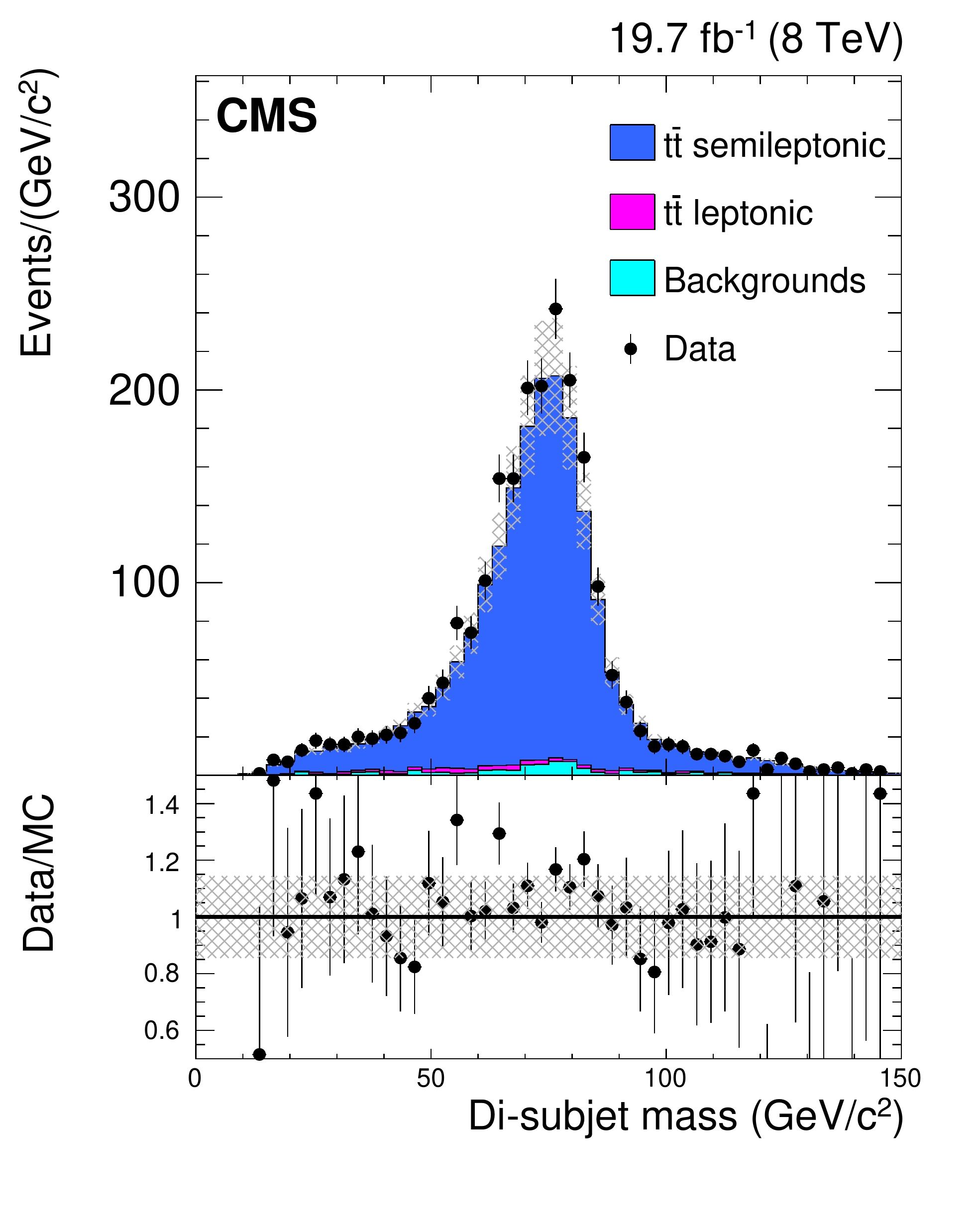}
 \caption{Distribution of the invariant id-subjet  mass of a
   hadronically decaying W boson obtained from a semi-leptonic
   $\ttbar$ sample. The lower panel shows the ratio of data and
   simulation. The hatched area indicates the uncertainty in the signal and background cross sections.}
          \label{fig:wmass}
\end{center}
\end{figure}

The selection of semileptonic $\ttbar$  events requires a
muon and a b-tagged  AK5 jet. In addition, one CA15 jet is required to be selected by the t tagging algorithm (see Section
\ref{sec:toptag}). The t-tagged jet must have exactly one b-tagged subjet. The two subjets that are not b-tagged are
used to calculate the invariant mass of a W boson candidate.  The
distribution of the W boson candidate mass is shown in Fig.~\ref{fig:wmass}.
 The shape of the W boson candidate mass distribution is the same in data and simulation and
no additional scale factors or systematic uncertainties are assigned.

\subsection{t tagging \label{sec:toptag}}

The \HTTagger algorithm, described in
Ref.~\cite{Plehn:2011tg},  is applied based on the implementation in \textsc{FastJet}~3.0~\cite{fastjet}. The algorithm uses CA15 jets as input.
This choice of jet size is suitable
for the region of phase space with intermediate boosts  (with a jet \pt slightly
above 200\GeVc). When the T quark mass is below 1\TeVcc, a
considerable fraction of the decay products populate the intermediate
boost range.  Such resolved events could in principle be reconstructed with
standard methods using AK5 jets.  The  \HTTagger provides a seamless transition
between the non-boosted and boosted domains.

\begin{figure}[htbp!]
\centering
\includegraphics[width=0.49\textwidth]{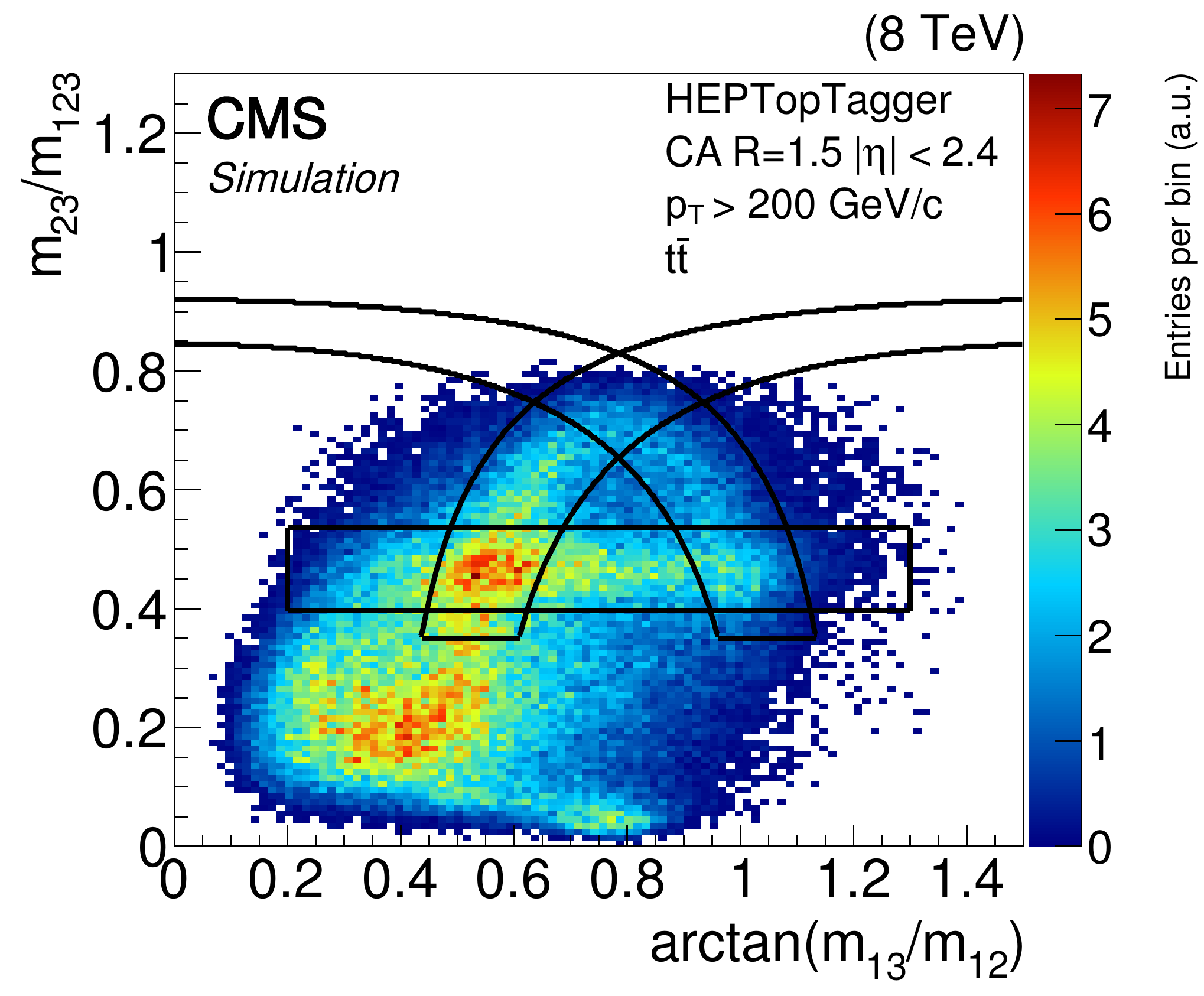}
\includegraphics[width=0.49\textwidth]{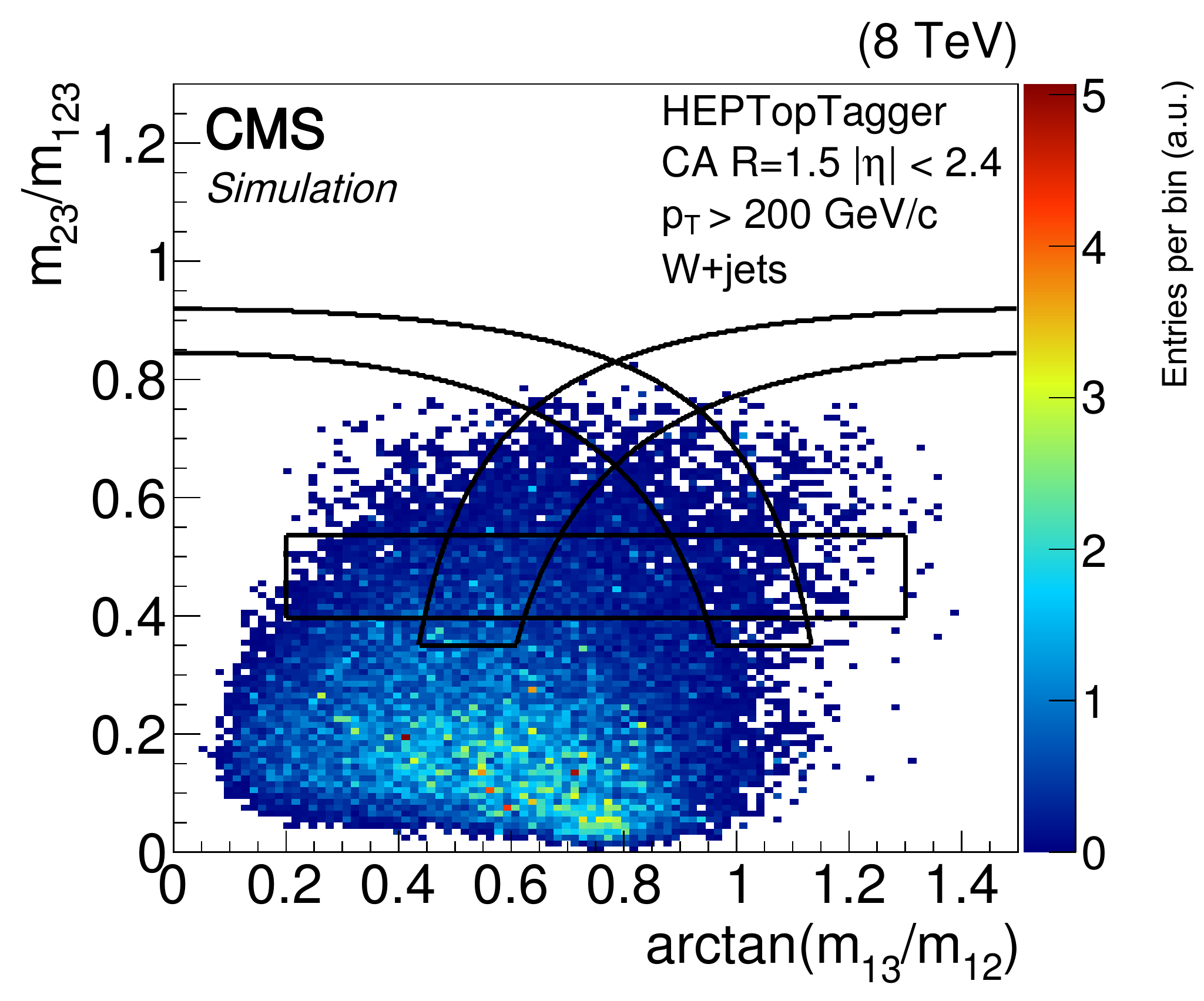}
 \caption{
  Two-dimensional distributions of
  $m_{23}$/$m_{123}$ versus $\mathrm{arctan}(m_{13}/m_{12})$ for
   \HTTagger jets in simulated  \ttbar events (left) and in
   simulated background events (right). The simulated background consists of  boson+jets, di-boson,
   single top quark, \ttbar all-hadronic, and \ttbar leptonic. The area enclosed by the thick solid lines denotes the region selected by the  \HTTagger.
}
\label{SemileptonicHEP:unmerged_p_t}
\end{figure}

  For each jet, the  \HTTagger   analyses the substructure by
stepping backward through the clustering history of the jet in an
iterative procedure until the conditions for splitting are no
longer fulfilled  and the subjets are not split any further. The filtering algorithm is applied to each combination of three
subjets that are found. The filtering
algorithm reclusters the constituents with a variable distance parameter
$R_\text{filt} = \min(0.3, \Delta R_{ij}/2)$,
where $i$ and $j$ are the closest subjets in $\Delta R$ in the subjet triplet. The five reclustered subjets with the largest \pt are
retained and the sum  yields the invariant mass of the top quark candidate. The configuration that has an invariant
mass closest to the top quark mass is chosen. The constituents of the five leading reclustered subjets are further reclustered
using the exclusive CA algorithm, which forces the
jet to have exactly three final subjets. The  \HTTagger uses
these three final subjets
and selects top quark jets based on the pairwise and three-way subjet masses.
Selections are applied in the two-\-dimensional plane defined by the ratio $m_{23}$/$m_{123}$ and the arctangent of $m_{13}$/$m_{12}$.
Here $m_{23}$ is the pairwise mass of the second and third leading subjets.
The variables $m_{12}$, $m_{13}$, and $m_{123}$ are defined in a similar fashion.
The distribution of events in this plane is shown for simulated \ttbar
events in Fig.~\ref{SemileptonicHEP:unmerged_p_t}  (left) and for a
mixture of background (boson+jets, di-boson,
   single top quark, \ttbar all-hadronic, and \ttbar leptonic) events in Fig.~\ref{SemileptonicHEP:unmerged_p_t}  (right).
A region with a well enhanced structure is only present for \ttbar
events. The region is highlighted by the thick black lines in
Fig.~\ref{SemileptonicHEP:unmerged_p_t}. This structure can  be used to  suppress  backgrounds that do not contain boosted top quarks by
rejecting events that lie outside of this region. Additionally, a selection on the top candidate mass,
$140<m_{\text{123}}<250\GeVcc$, is applied. Another populated region
shows up below and to the left of the selected region because of unmerged top decays. This contribution disappears for boosted top quarks above $\pt>300\GeVc$.

\subsubsection{Algorithm performance\label{sec:toptagperformance}}
\begin{figure}
\begin{center}
\includegraphics[width=0.49\linewidth]{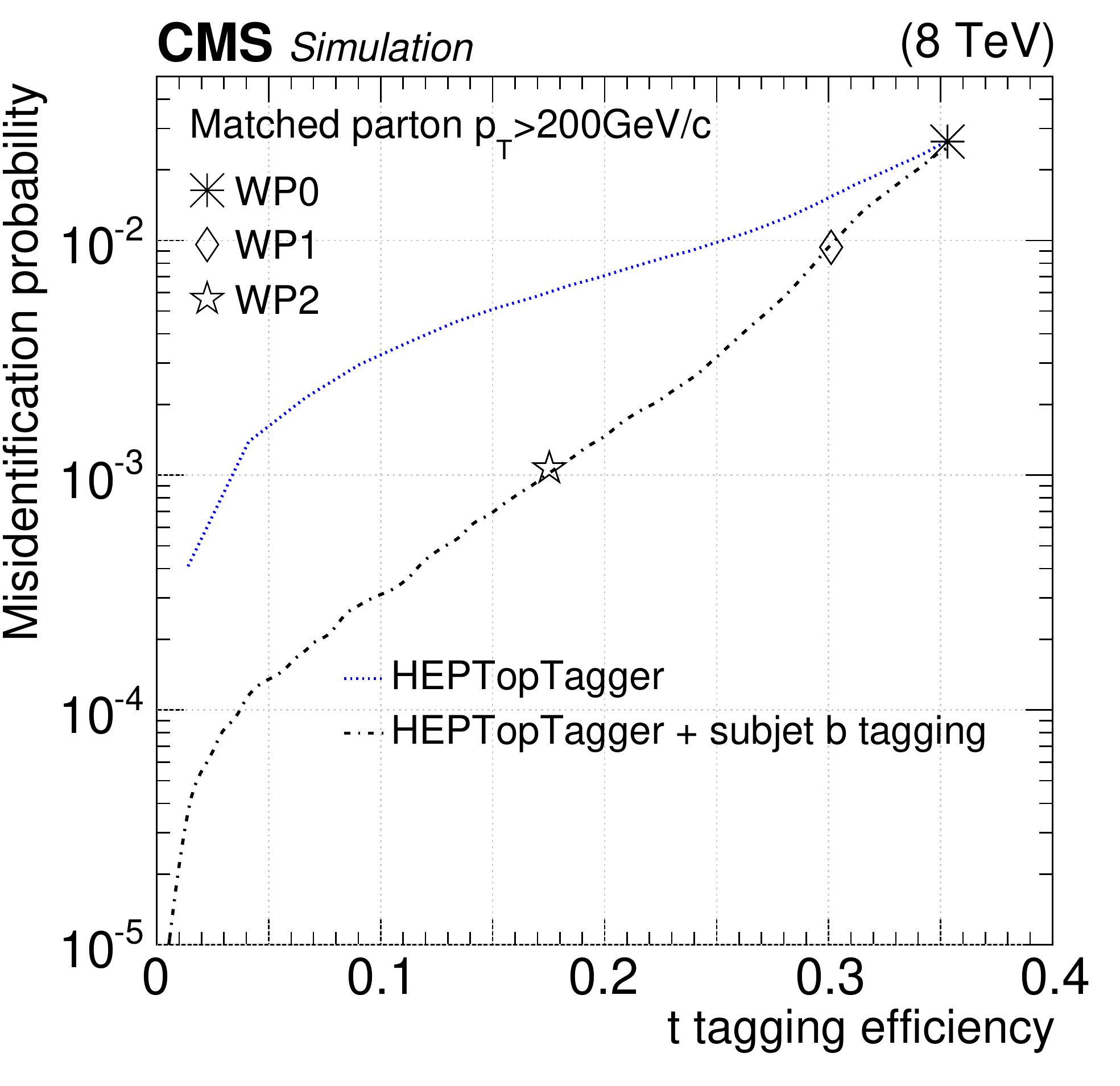}
\end{center}
\caption{Mistag rate versus t tagging efficiency for the  \HTTagger and the combination of the \HTTagger with subjet b tagging, for CA15 jets matched to generated partons with $\pt>200\GeVc$. The mistag rate is obtained from simulated QCD multijet events, while the efficiency is determined using simulated  $\ttbar$ events.}
\label{fig:roc_200}
\end{figure}

The selection criteria used in the algorithm are varied iteratively and the efficiency and mistag rate are calculated for each iteration.
The minimum mistag rate for a given signal efficiency is shown in Fig.~\ref{fig:roc_200}. The  \HTTagger curve is determined by fixing the $m_{\text{123}}$ selection ($140<m_{\text{123}}<250\GeVcc$)
and varying the width of the region selected by the algorithm. The other
curve is obtained by applying simultaneously the   \HTTagger  and
the subjet b tagging criteria and varying their requirements. Details of these selection criteria
are given in Ref.~\cite{CMS-PAS-JME-13-007}. Three
working points are defined as indicated by markers in the figure. The
working point used in this analysis is WP2, which is
defined by the standard  \HTTagger criteria in addition to a
b-tagged subjet identified with the CSVM b tagging algorithm. The
other working points (WP1 and WP0) use relaxed \HTTagger criteria and
relaxed b tagging, and are used to validate the scale factor
measurements which are described in the following section.

\subsubsection{Scale factors\label{sec:toptagSF}}
A  semileptonic \ttbar sample is used to study  boosted hadronic  top quark
decays  in  data.
This sample is then used to measure
data to simulation scale factors  for the t tagging efficiency using WP2.  This
procedure was introduced in Ref. \cite{ZprimeSpringer}.
The \ttbar sample is defined  by requiring one muon and at least one b-tagged AK5 jet.
Additionally, a top quark candidate CA15 jet is required, with high transverse momentum $\pt>200\GeVc$ and with at least one b-tagged subjet. This semileptonic selection is very pure and  background contributions
are negligible.  The efficiency of the \HTTagger is determined as the fraction of top quark candidate CA15 jets that pass all of the tagging requirements. These measurements yield scale factors ranging from 0.85 to 1.15 depending on the $\pt$ and the $\eta$ of the jet.

\section{Event selection \label{sec:selection}}

The \HT variable used in the analysis is calculated  from  the transverse momenta of all
subjets  within the reconstructed CA15 jets with $\pt>150$\GeVc. This definition is more accurate than
that used in the trigger
because particle-flow
reconstruction is exploited.  A threshold of $\HT >720$\GeVc is
applied in the offline analysis as the trigger is almost fully
efficient above this value. The simulation is corrected to match the data by weighting
events based on the ratio between the trigger efficiency calculated in data and in simulation. The
systematic uncertainty introduced by this procedure is discussed in Section \ref{sec:systematics}.

The full event selection requires the following criteria to be fulfilled:

\begin{itemize}
\item At least one   CA15 jet must be t-tagged by the
  \HTTagger algorithm and must contain at least one b-tagged subjet
  (identified by the   CSV b tagging algorithm at the medium operating
  point). The t-tagged jets must
  have $\pt > 200$\GeVc.
\item At least one   CA15 jet must have $\pt>150$\GeVc and must be H-tagged (at least two
  subjets identified by the CSVM b tagging algorithm).  The invariant mass
  of the two b-tagged subjets has to be  larger than 60\GeVcc.
This jet must not be identical to the top-quark candidate jet.
\end{itemize}

As mentioned in Section \ref{sec:strategy},  the event
selection is split further into two
categories: single and multiple H tags.

\begin{figure}[Htbp]
           \includegraphics[width=.45\textwidth]{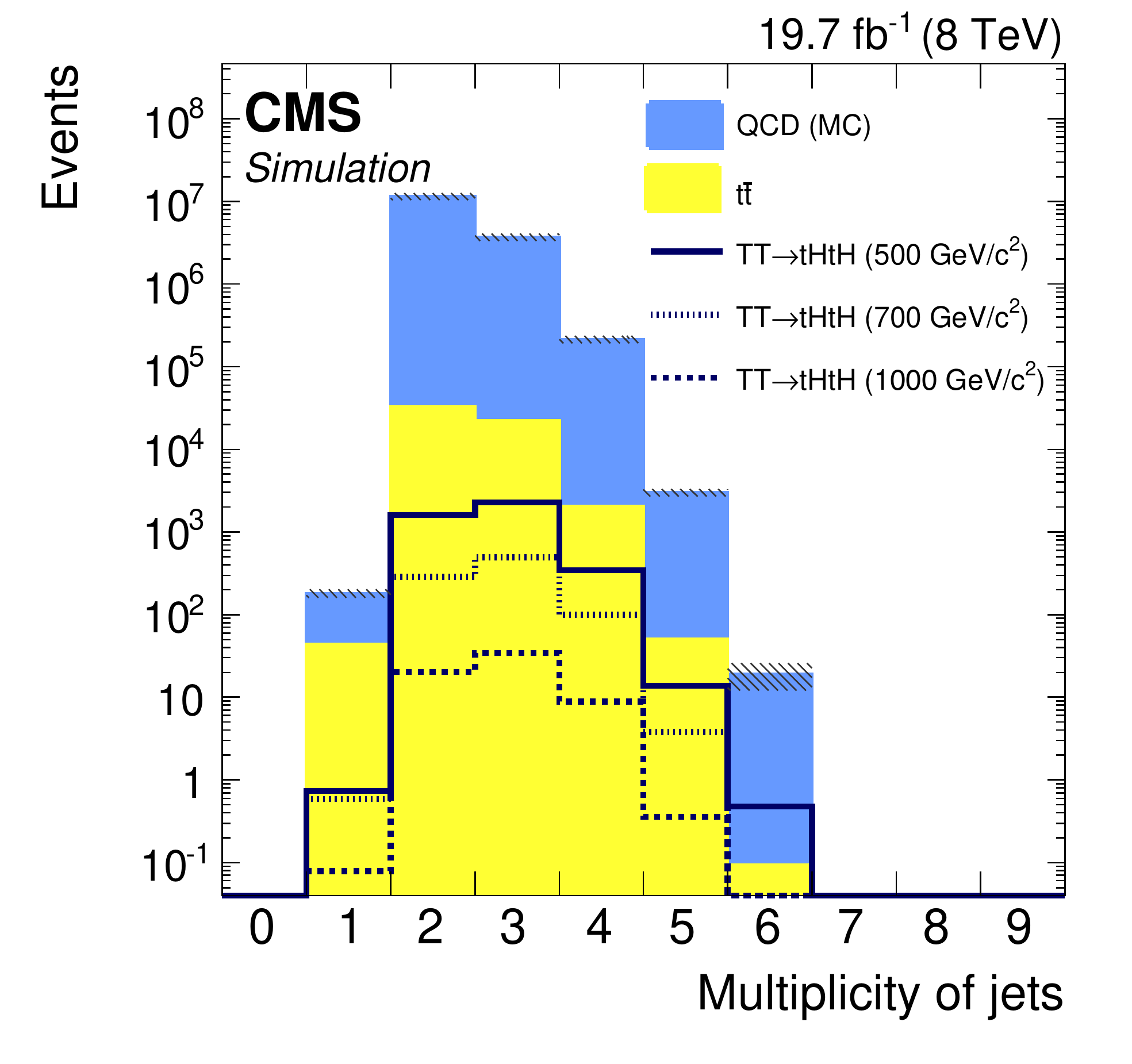}
\includegraphics[width=.45\textwidth]{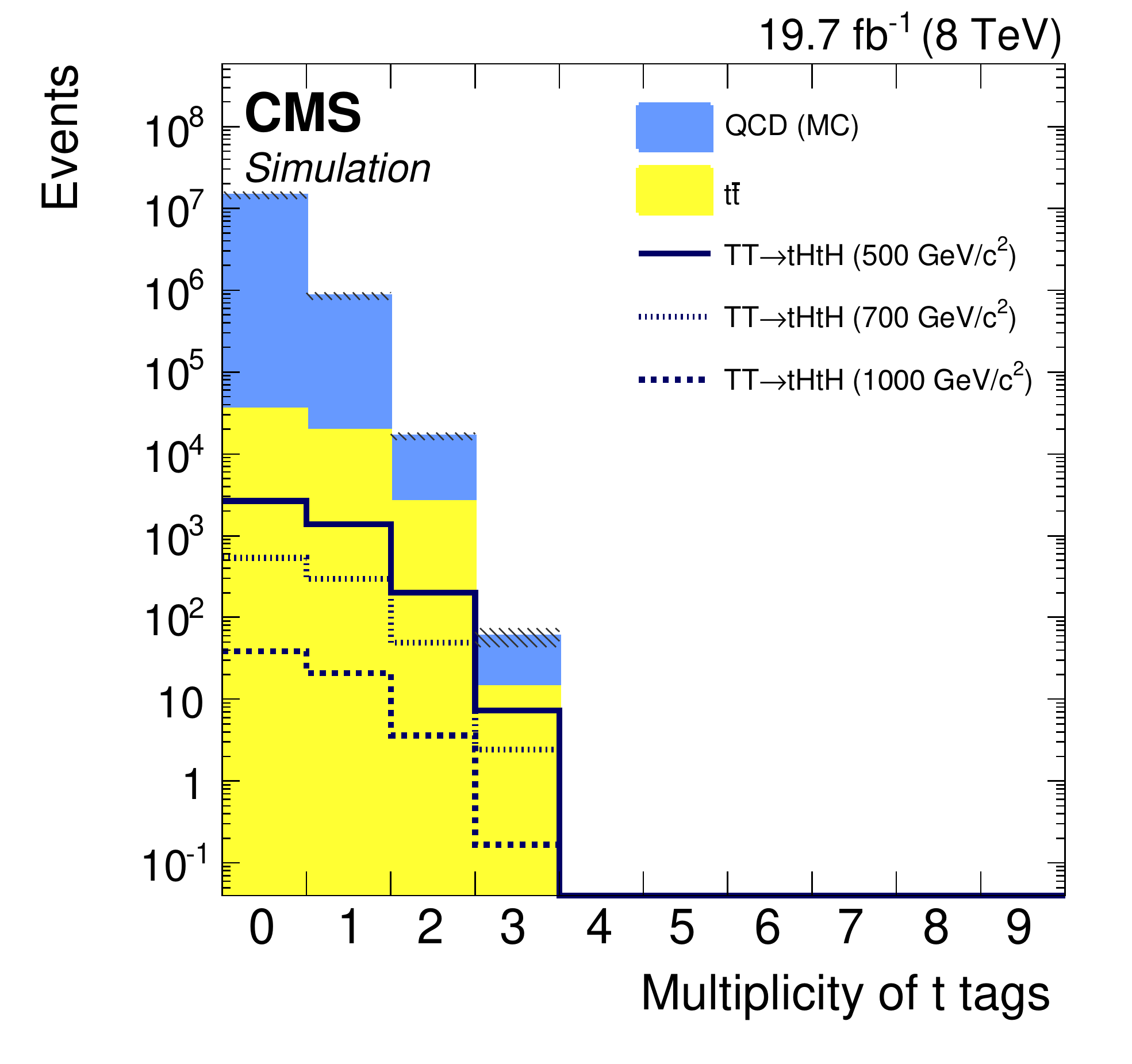}
         \caption{
           Left: multiplicity of CA15 jets with $\pt > 150$\GeVc. Events with at least two of these jets
           are selected. Right: multiplicity of CA15 jets with $\pt >
           200$\GeVc, that are selected by the  \HTTagger
           algorithm. The solid
           histograms represent the simulated background processes ($\ttbar$
            and QCD multijet). The hatched error
            bands show the statistical uncertainty of the simulated events.}
     \label{fig:SelectionStep1and2}
 \end{figure}

The number of reconstructed CA15 jets predicted by simulation with $\pt >150$\GeVc is shown
in the left plot of Fig. \ref{fig:SelectionStep1and2}, while the right plot
shows the number of jets passing the t tagging criteria.  In the following figures the hatched regions indicate
the statistical uncertainty in the simulated
background. The signal hypotheses are represented by the
solid and dashed  lines.

\begin{table}[htb]
 \centering
\topcaption{Cross section, expected numbers of selected events, and the selection
efficiencies for several signal samples with different values of the T
quark mass for an integrated luminosity of 19.7\fbinv. The
  signal samples assume $\mathcal{B}(\T \to \PQt\PH) = 100\%$. The efficiencies are
  calculated relative to an inclusive sample with no requirements on top quark or Higgs boson decay modes, and without any selection criteria applied. }
\begin{tabular}{llll}
\hline
T quark mass & production & expected  & selection  \\
    $(\GeVccns{})$  &    cross section (pb)   & events & efficiency \\
\hline
 500 &   0.59          &283.0    & 2.5\% \\
 600 &   0.174       &  152.0   & 4.4\% \\
 700 &   0.059     & 69.3     & 6.0\% \\
 800 &   0.021     &  30.3     & 7.2\% \\
 900 &   0.0083      &   12.1   & 7.3\% \\
 1000 &  0.0034   & 4.9       &  7.2\% \\
\hline
\end{tabular}
\label{tab:signalEfficiency}
\end{table}

\begin{figure}[Htbp]

\includegraphics[width=.45\textwidth]{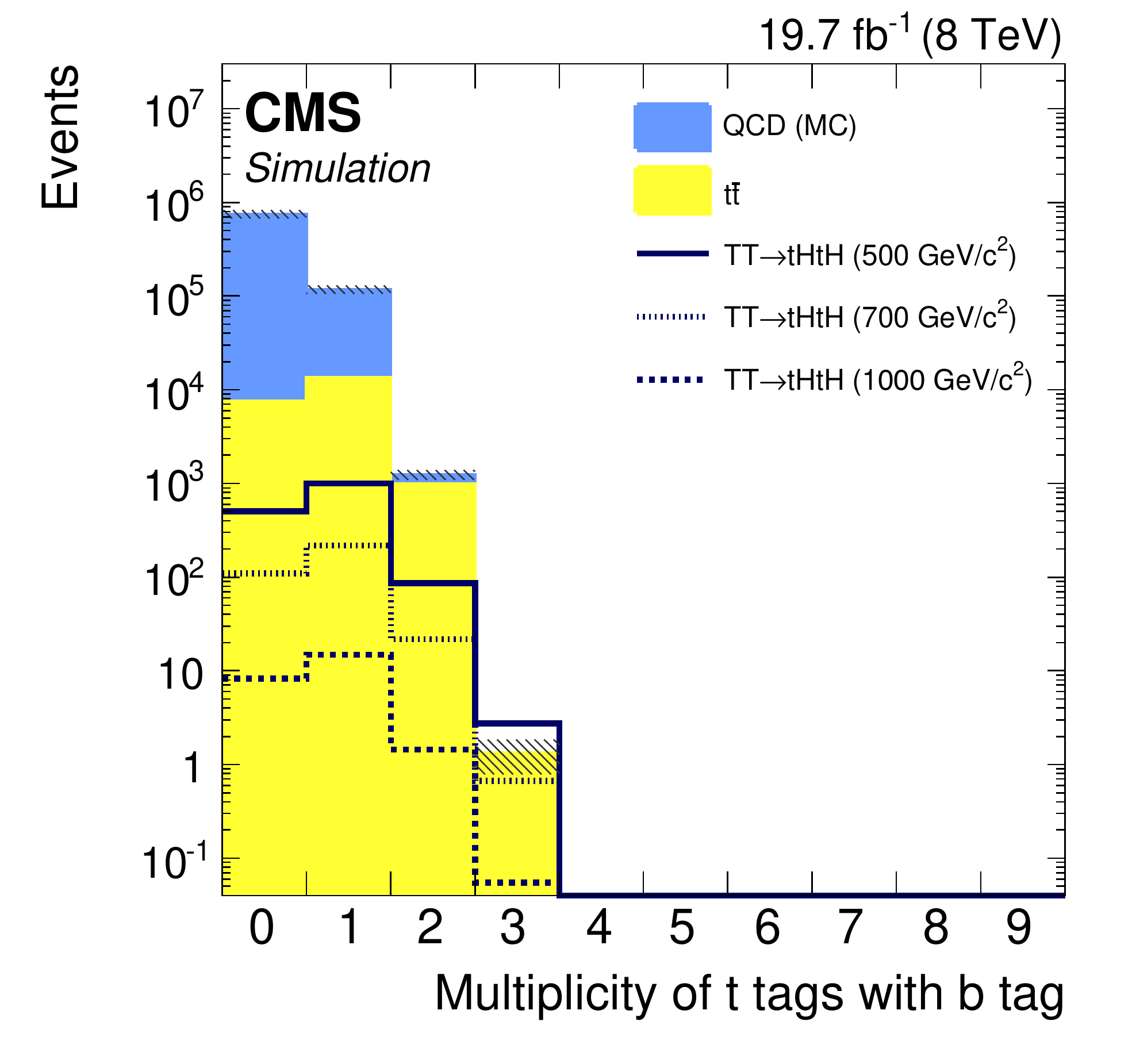}
  \includegraphics[width=.45\textwidth]{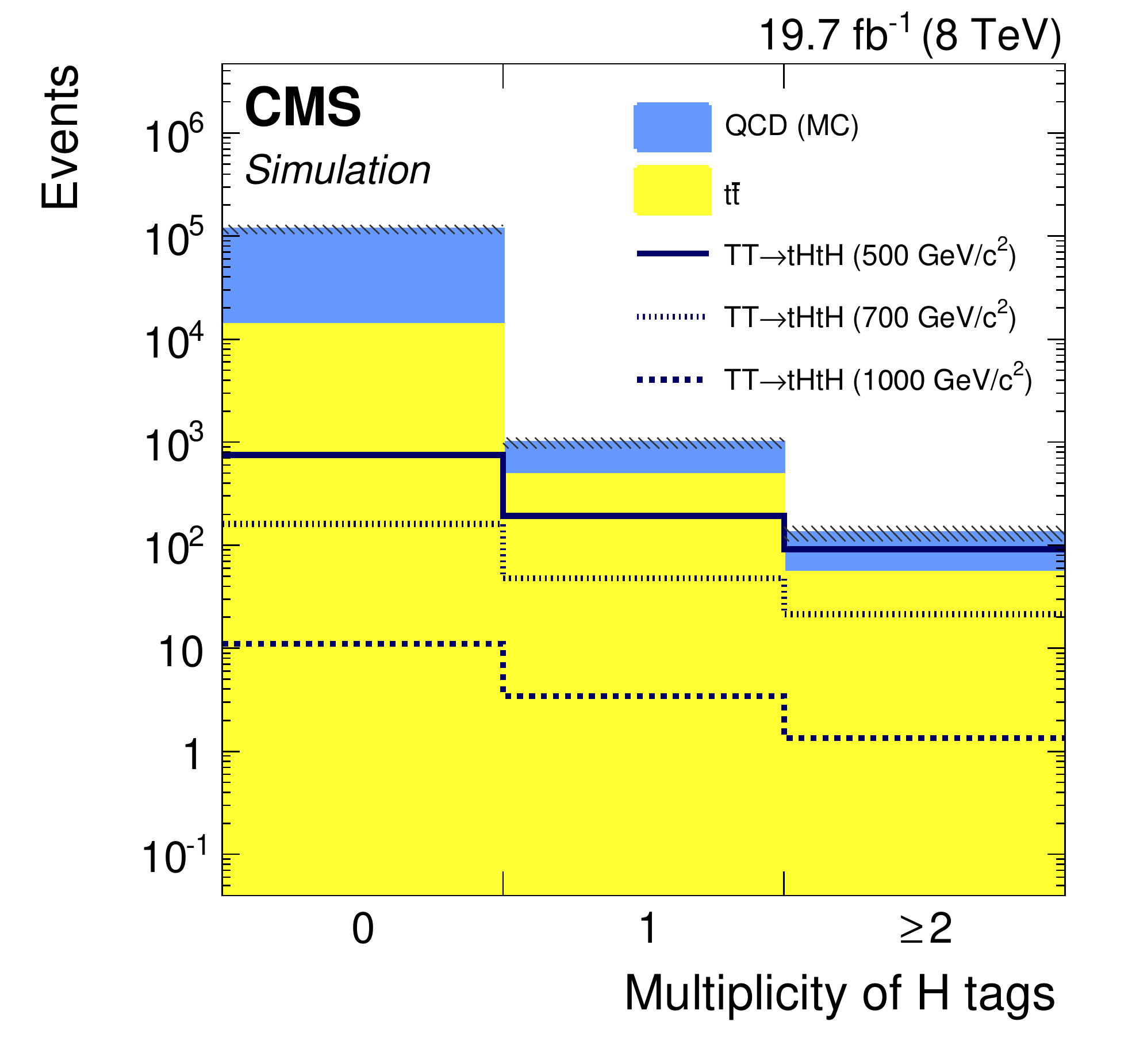}
        \caption{
Left: multiplicity of CA15 jets with  $\pt > 200$\GeVc
          that are tagged by the  \HTTagger and contain a b-tagged subjet, after requiring at least one jet per event to be selected by the  \HTTagger algorithm.
       Right:  multiplicity of CA15 jets with
          $\pt > 150$\GeVc satisfying the H tagging criteria.  Events with three or more H tags
          are included in the bin with two H tags. The solid
           histograms represent the simulated background processes ($\ttbar$
            and QCD multijet). The hatched error
            bands show the statistical uncertainty of the simulated events.}
    \label{fig:SelectionStep3and4}
\end{figure}

The impact of subjet b tagging is visible in Fig.~\ref{fig:SelectionStep3and4}. The left plot shows the number of t-tagged CA15 jets with a subjet b tag, while the right plot
shows the number of H-tagged jets for events that have at least one t-tagged CA15 jet with a subjet b tag.
These figures demonstrate the strong reduction of QCD multijet
background by the jet substructure criteria.

The number of selected events for each signal sample of the benchmark model and the selection
efficiencies, derived from simulated events, are given in Table
\ref{tab:signalEfficiency}.

\section{Background estimation \label{sec:background}}
The $\ttbar$ background is evaluated from
simulated events, corrected for differences between data and
simulation in b tagging and trigger efficiencies described above. The uncertainties in the normalization and shape of
$\ttbar$ events are discussed in Section
\ref{sec:systematics}. Background contributions from ttH and
hadronically decaying W/Z plus heavy flavour processes are found to
be below 1\% and are neglected.

The QCD multijet background is estimated in data using a
two-dimensional sideband extrapolation. In this method, two uncorrelated criteria in the event
selection are inverted to obtain sideband regions that are enriched
in QCD multijet events and depleted in signal events. Inverting each
criterion individually, as well as both  at the same time,
results in three exclusive sideband regions, denoted A, B and C:
\begin{itemize}
\item Sideband region B is obtained by inverting the selection criteria of the   \HTTagger
algorithm.  The top quark mass window as well as all requirements on the pairwise subjet
mass in the  \HTTagger are inverted.
Events outside of the selected region shown in Fig.~\ref{SemileptonicHEP:unmerged_p_t}  (Section~\ref{sec:substructure}) are used to define the inverted
 \HTTagger control region, while the events that are inside define the signal
region. Details of these selection criteria of the  \HTTagger are given in
Section \ref{sec:substructure} and \cite{CMS-PAS-JME-13-007}.

\item Sideband region C is obtained by inverting the H tagging
algorithm.  Only events with zero H tags  are
selected and the requirement on the pairwise subjet mass is removed.

\item Sideband region A is obtained by inverting both the
  H tagging and the t tagging algorithms as described above.

\item Events in the signal region D have all tagging requirements applied.
\end{itemize}

The \ttbar contamination in the sideband regions amounts to a
maximum of 8\% in  region C. This is accounted for by subtracting the
\ttbar contribution predicted by the simulation in each of the
sideband regions.  Backgrounds due to ttH and  hadronically
decaying W/Z plus heavy flavour processes are found  to have
 a negligible contribution in the sideband regions. A
signal injection test has been performed to evaluate the impact of a
hypothetical signal on the
background model. It has been found that the signal contamination in
the sideband regions leads to a small effect of less than 1.4\% for $m_{\mathrm{T}} = 700$
\GeVcc on the measured QCD multijet event rate, and  therefore the
possible signal contamination in the sideband regions is neglected in the analysis.

The QCD multijet yield in the signal region is calculated as
\begin{equation}
R_D = R_B \frac{R_C}{R_A},
\end{equation}
where $R_A$ denotes the rate of events in sideband A. The $\ttbar$ contamination in the
sideband regions is subtracted. The event rates in the three sideband regions and the signal region
are provided in Table \ref{yieldsFromABCD}. The resulting predictions of the QCD multijet backgrounds are  given in
Table \ref{tab:BackgroundsResults} for the two event categories.

\begin{table}[b]
 \centering
\topcaption{ Event rates in the signal and sideband regions obtained
  from the two-dimensional sideband extrapolation in data for the two
  H tag categories. The $\ttbar$ contamination
  is subtracted from the nominal yield in the sideband regions. The
  prediction of the QCD multijet event rate in the signal region D is
  given along with statistical uncertainties that arise from  the
  limited size of event samples in the sideband regions. The sideband regions A and C are common
to both H tag multiplicity categories.}
\begin{tabular}{l|c|l|c|l|c}

\cline{3-6}
\multicolumn{2}{c|}{} & \multicolumn{2}{c|}{single H tag category} & \multicolumn{2}{c}{multi H tag category} \\
\hline
\multicolumn{2}{c|}{region A} & \multicolumn{2}{c|}{region B} & \multicolumn{2}{c}{region B} \\
\hline
data                                        &  1152640  &  data & 8384 & data                                     &  1157 \\
data $-$ $\ttbar$  &  1146464  & data $-$ $\ttbar$ & 8089&    data $-$ $\ttbar$ & 1123\\
\hline
\multicolumn{2}{c|}{region C} & \multicolumn{2}{c|}{region D} & \multicolumn{2}{c}{region D}\\
\hline
data                                     &  140911 &  & &  &  \\
data $-$ $\ttbar$  &  129972 & prediction  &$917 \pm 11 $&  prediction &  $127 \pm 4$ \\
\hline
\end{tabular}

\label{yieldsFromABCD}
\end{table}

\begin{table}[b]
 \centering
\topcaption{Predicted background contributions in the signal region for
  the  two event categories with one and with multiple H tags. Statistical uncertainties in the background estimates are also shown.}
\begin{tabular}{lll}
\hline
 & single H tag category & multi H tag category \\
\hline
QCD (predicted from data) & $917 \pm 11$  &  $127\pm 4$ \\
$\ttbar$ (from simulation) & $486 \pm 8$ & $55\pm 3$\\
\hline
total background &$1403 \pm 14$ &  $182 \pm 5$  \\
\hline
data  & 1355  &  205 \\
\hline
\end{tabular}
\label{tab:BackgroundsResults}
\end{table}

The closure of this method is verified with simulated QCD multijet events. As the
method assumes the selection criteria defining the sideband regions to
be uncorrelated, the following condition must be fulfilled:
\begin{equation}
\frac{R_A}{R_B} = \frac{R_C}{R_D}.
\end{equation}
According to simulation, the ratios are $R_A / R_B = 185 \pm 5$ ($1417 \pm 97$) and
$ R_C / R_D = 185 \pm 17$ ($ 1203 \pm 250$) for the single (multi) H tag event category. The quoted uncertainties are statistical. It can be seen that the ratios agree within the statistical uncertainties. The largest uncertainties occur in the
$ R_C / R_D$ ratio and are about 10 (20)\% for the single (multi) H tag category.

In addition to the event yields, the shapes of the \HT and
$m_{\bbbar}$ distributions for the QCD multijet processes are also derived from the sideband regions. For both the \HT and
$m_{\bbbar}$ variables the sideband region B (inverted t tagger) is
used.  The expected contribution from $\ttbar$ events is
subtracted from the sideband.

\begin{figure}[h!tb]
  \centering{
                \includegraphics[width=.49\textwidth]{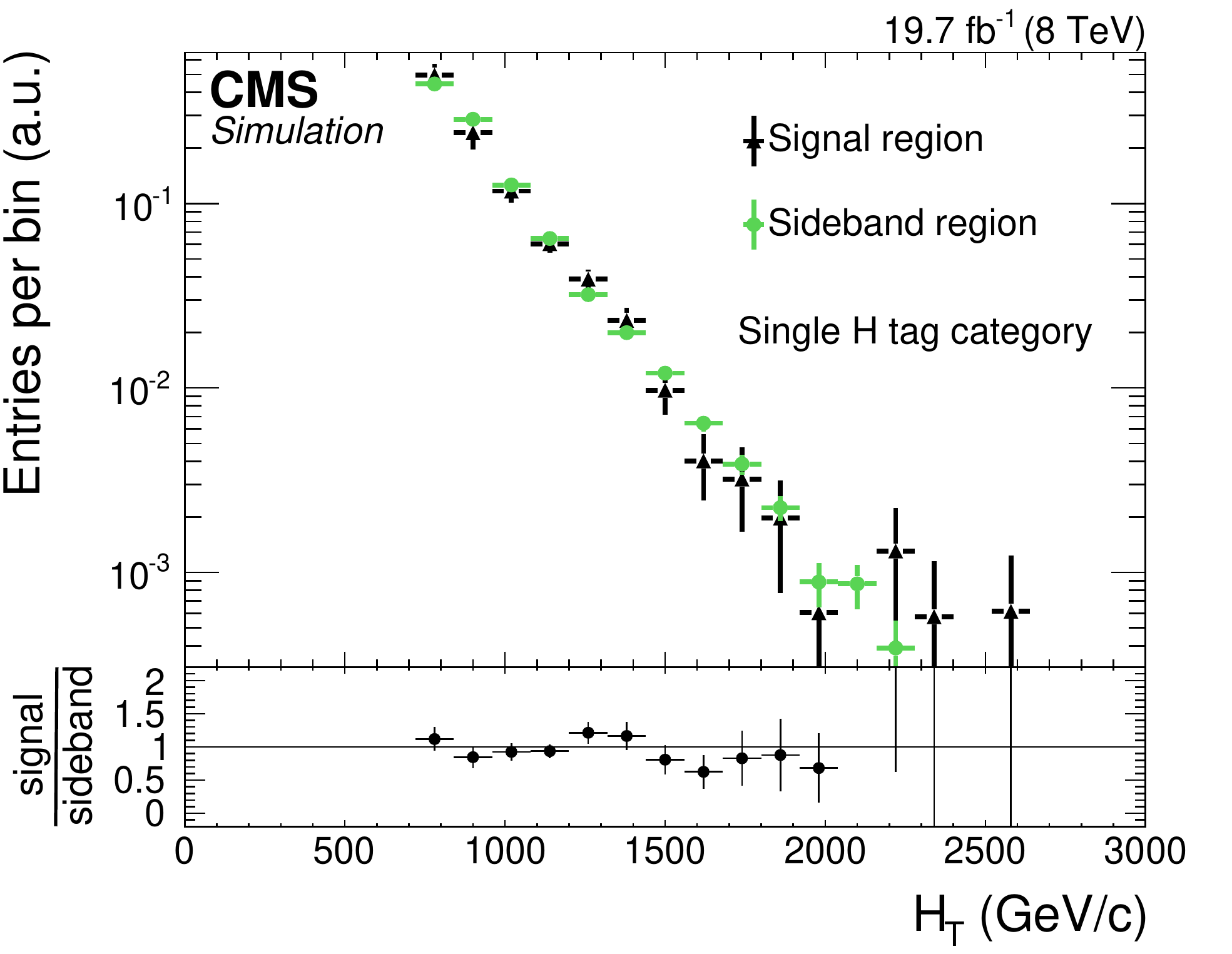}
                \includegraphics[width=.49\textwidth]{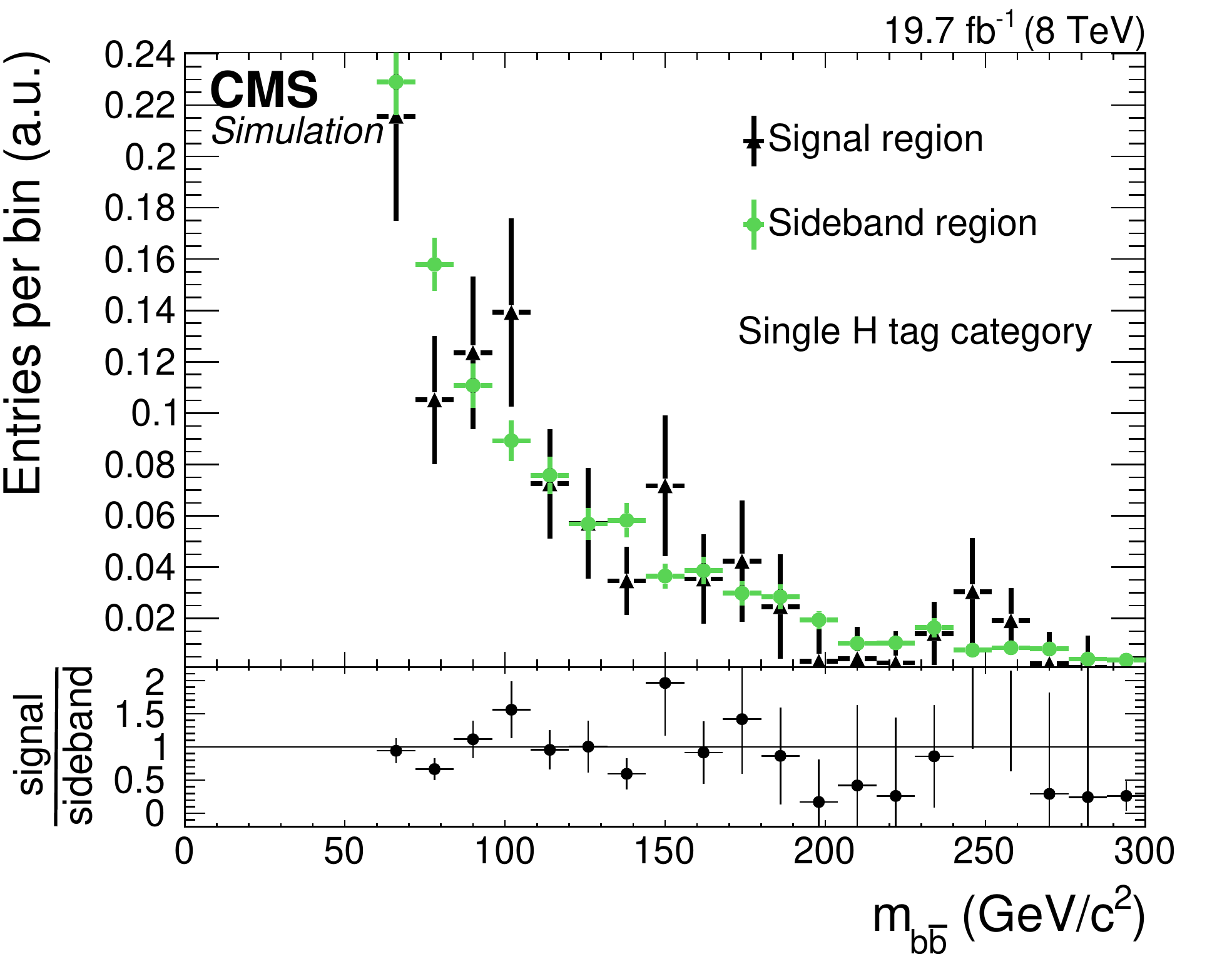}
}
\centering{
                \includegraphics[width=.49\textwidth]{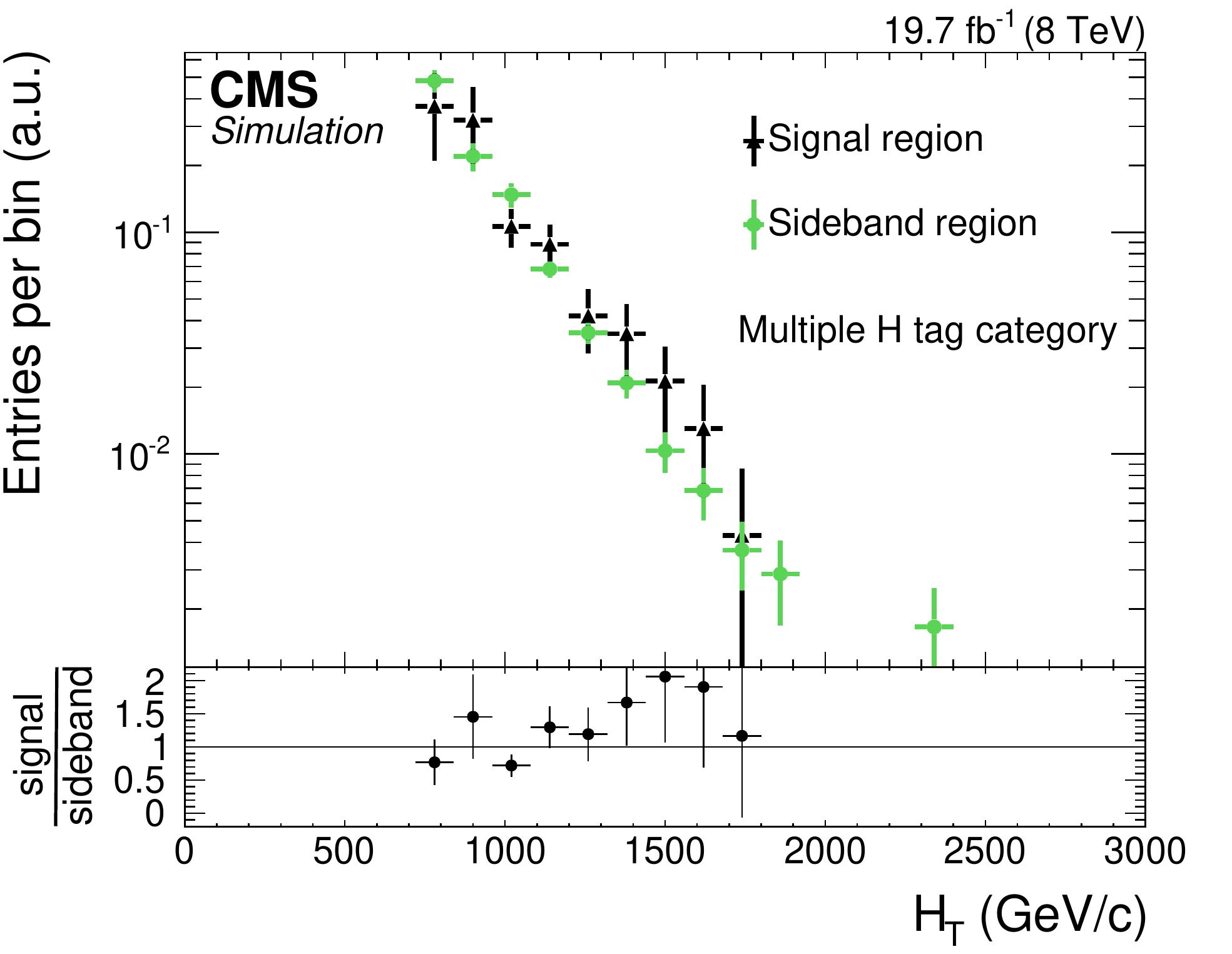}
                \includegraphics[width=.49\textwidth]{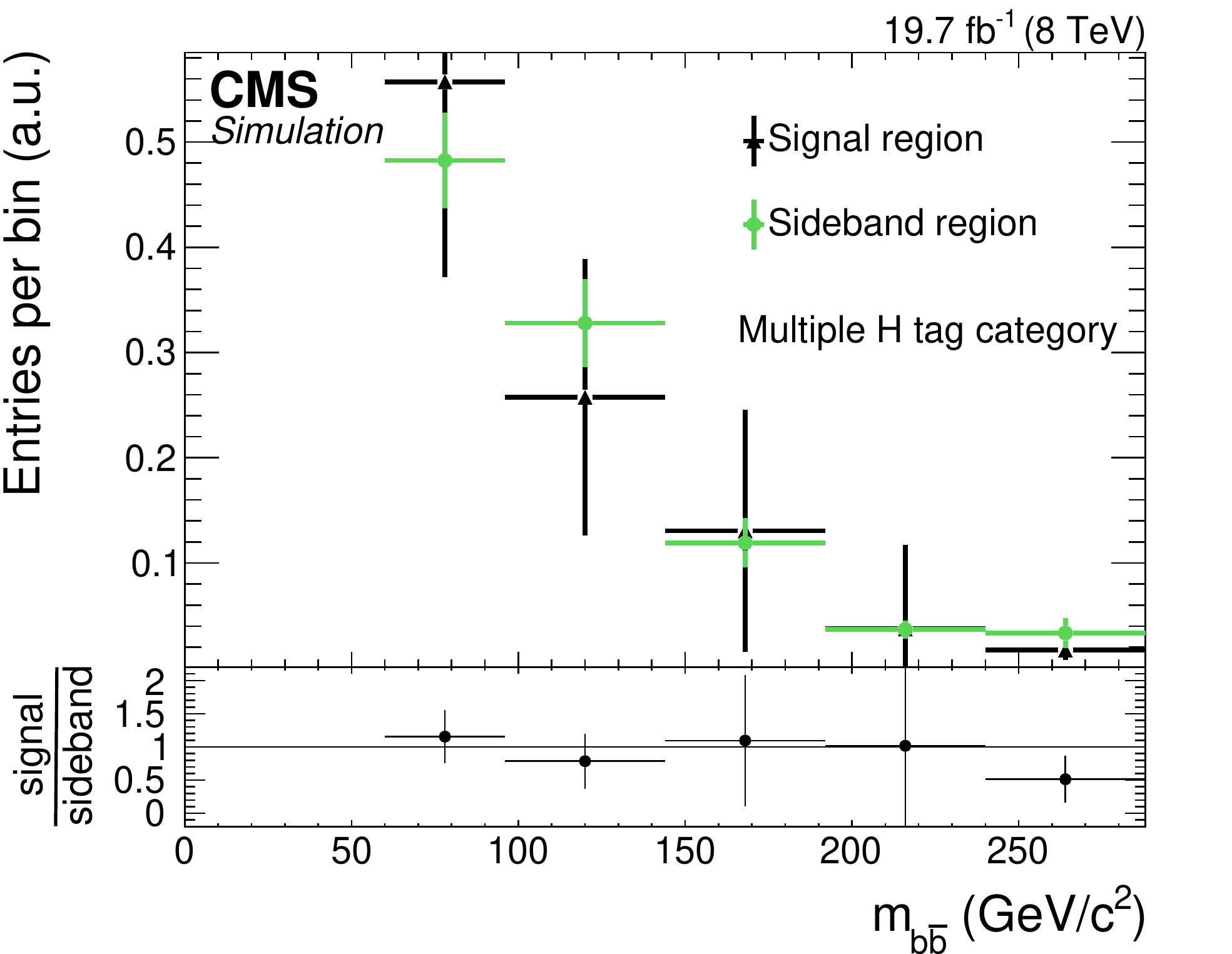}
}
        \caption{Comparison of the \HT (left) and $m_{\bbbar}$ (right) distributions in
        the sideband region B and signal region for the single (top) and multiple (bottom)
        H tag event categories for simulated QCD multijet
        events. All distributions are normalized to unity for shape
        comparison. The lower panels in the figures show the ratio of
        the signal and sideband regions. }
    \label{fig:closureM}
\end{figure}

Closure is also verified for the shape of \HT and $m_{\bbbar}$
distributions in the signal and sideband regions. Figure \ref{fig:closureM} shows
a comparison of the \HT and $m_{\bbbar}$ shapes in the sideband and signal
regions for the single and the multiple H tag event
categories. The distributions agree within statistical uncertainties.

The method has also been validated in data. The shapes of the simulated  \HT and $m_{\bbbar}$
distributions in the signal region agree well with the predicted
distributions in data.  The absolute rate of events shows a disagreement between simulation and
the data-derived rate of a factor of two. This disagreement is taken
into account when assigning systematic uncertainties in the
background, as explained in Section~\ref{sec:systematics}.

\section{Systematic uncertainties \label{sec:systematics}}

As the analysis relies on simulation
for the $\ttbar$ background prediction, a careful evaluation
of uncertainties affecting both the normalization and shape of the
\ttbar background events is needed. This is also required for the simulated signal events.

The QCD multijet background is obtained from data.
 The rate and shape of the  $\ttbar$ background have an
 effect on the measurement of the QCD multijet background because the $\ttbar$ contamination in the sideband
 region is subtracted from data.

The detailed list of systematic uncertainties is given
below.  Most of these uncertainties have an
impact on both the shapes and normalization of the sensitive
variables \HT and $m_{\bbbar}$, while the uncertainty in the integrated luminosity only
affects the normalization. The uncertainties are summarized in Table~\ref{summarysystematics}.

\begin{itemize}
\item  b tagging scale factor uncertainties: based on the measurements
  described in Section~\ref{sec:subjetbtag} and Ref.~\cite{CMS-PAS-BTV-13-001}, scale
factors with their corresponding uncertainties are applied to simulated
samples.  The scale factor uncertainties for the b tagging efficiency depend on \pt and
$\eta$. The typical size of these uncertainties is between 1 and
2\% while the mistag rate uncertainty is around 15\%.  The b tagging scale factor uncertainties affect both the
normalization and shape of the $\ttbar$ background and signal
events. Depending on the sample and signal mass point, the impact of
the b tagging scale factor uncertainty on the expected number of
selected signal and $\ttbar$ events is 5 to 8\% while
the impact of the mistag scale factor uncertainty is 0.3 to 4\%.

\item  \HTTagger scale factor uncertainty: the efficiency of the  \HTTagger has been measured and compared to
simulation to derive scale factors as described in Section~\ref{sec:toptag}. The
uncertainties in these scale factor measurements are between 3 and
6\%, and are parameterized as a function of \pt. These uncertainties affect both the normalization
and shape of the $\ttbar$ background and signal events. The
impact  on the expected number of signal and  $\ttbar$
events is 0.4 to 2.3\%.

\item Jet energy corrections: dedicated energy corrections for CA15
  jets are not available. Therefore, the energy corrections for jets
  reconstructed with the anti-\kt algorithm with size parameter
  $R=0.7$  (AK7) \cite{Cacciari:2008gp} have been used~\cite{JEC2012}. It has been verified that these corrections
  are valid by comparing the   reconstructed jets in simulation to the corresponding
  generator level jets where exactly the same clustering and
  grooming algorithms have been applied. The ratio between reconstructed and
  generated momentum for these jets is found to be consistent with unity, with variations that are less than 4\%. The impact of the uncertainty on the jet energy scale of filtered CA15 jets is evaluated by varying  the jet four-momentum
  up and down by the  jet energy scale uncertainties
  of AK7 jets, with  an additional 4\% systematic uncertainty. The
  uncertainty in the subjet energy scale is assumed to be similar to the
  energy scale uncertainty of AK5 jets. The impact on the expected
  number of selected  $\ttbar$ and  signal events is less than
  0.5\% for CA15 jets and less than  5\% for subjets.

\item  PDF uncertainties: simulated $\ttbar$ events are weighted
  according to the uncertainties parameterized by the CTEQ6
  eigenvectors \cite{1126-6708-2002-07-012}. The shifts produced by
  the individual eigenvectors are added
  in quadrature in each bin of the \HT and $m_{\bbbar}$ distributions. The
  resulting uncertainty in the number of expected
  $\ttbar$ events ranges from 2.4 to 8\%.

\item Scale uncertainties: the impact of the renormalization and
  factorization scale uncertainties on  the  $\ttbar$ simulation has been
  studied  using $\ttbar$ event samples generated  with two different values of these scales (moving them
  simultaneously up or down by a factor of two relative to the nominal
  value).  It has been verified that
  this uncertainty has no impact on the shapes of \HT and $m_{\bbbar}$
  distributions within the statistical uncertainties of
  the simulated samples. The resulting impact  on
  the selected number of $\ttbar$  events is 34\%.

\item QCD multijet background normalization: the normalization and
  shape of QCD multijet  events
 do not show any discrepancy between the predicted and observed shapes in
 the signal region  based on the closure test with simulated
 events,  as  discussed in Section~\ref{sec:background}. The
 comparison of the simulated sidebands with data shows a very good
 agreement of the shapes as well, but the normalization is not in
 agreement.  Therefore  a
 systematic uncertainty in  the normalization of QCD multijet events
 is taken into account. This uncertainty is derived from the
 statistical  precision of the closure test, which is limited by the
 finite size of simulated event samples. The uncertainty in the single
 H tag category is 10\% while the uncertainty is 20\% in the multi
 H tag category. The only systematic uncertainty in the shape
 of the QCD multijet background arises from the subtraction of $\ttbar$ events. The effect of the $\rm
 t\overline{t}$ scale uncertainty on the estimation of the QCD multijet
  background is less than 1\%. Uncertainties in the
  $\ttbar$ simulation and the corresponding propagated
  uncertainties in the QCD multijet  prediction are treated as
  correlated, but they have opposite effects.

\item Trigger reweighting: a scale factor $SF_\text{trig}$ is applied to
  correct for the different behaviour between data and simulation in
  the region in which the trigger is not fully efficient. A systematic uncertainty in the
  scale factor is obtained by varying $SF_\text{trig}$ by $\pm 0.5 (1 -SF_\text{trig})$. This
  uncertainty does not affect the plateau region of the trigger, where
  $SF_\text{trig}=1$. This uncertainty is taken into account both as a
  shape and as a rate uncertainty. It only affects the low-\HT range. The trigger  efficiency is measured in a \ttbar-enriched data sample. For $720<\HT\leq780\GeVc$ the efficiency is 75\%, with a $SF_\text{trig}$ of 80\%. For $780<\HT\le840\GeVc$ the trigger efficiency is 93\%, with a $SF_\text{trig}$ of 94\%. For $\HT>840\GeVc$  the trigger has an efficiency always greater than 99\% and a $SF_\text{trig}$ consistent with one. The overall impact of this uncertainty on the event yield is 3.5\%.

\item Luminosity: an uncertainty in the integrated luminosity of 2.6\% is taken into account~\cite{CMS-PAS-LUM-13-001}.

\item Cross section of the $\ttbar$ background: an uncertainty
  of 13\% is assigned to the $\ttbar$ cross section. This
  uncertainty is obtained with the technique used in the
  differential $\ttbar$ cross section   measurement~\cite{Khachatryan:2015oqa} for large invariant mass values of the $\ttbar$ system.
\end{itemize}

\begin{table}[Htb]
\centering
\topcaption{Systematic uncertainties and their effect on signal and
  background processes, expressed in percent. The uncertainties are
  described in detail in Section~\ref{sec:systematics}. This table
  shows  uncertainties in the normalization only.}
\begin{tabular}{cccccc}
\hline
uncertainty & $\ttbar$ & QCD multijet & signal & signal & signal \\
& & & 500\GeVcc & 700\GeVcc & 1000\GeVcc\\
\hline
b tagging: &  & &  &  & \\
heavy flavour & $+9.2$/$-7.5$ & \NA & $+6.0$/$-6.8$ & $+7.1$/$-6.5$ & $+7.8$/$-8.0$ \\
light flavour & $+4.2$/$-3.2$ & \NA & $+1.2$/$-0.7$ & $+0.9$/$-0.6$ & $+0.8$/$-1.0$ \\
\hline
\HTTagger & $+0.9$/$-0.4$ & \NA & $+1.6$/$-1.7$ & $+1.7$/$-1.8$ & $+1.8$/$-2.3$ \\
\hline
jet energy corrections & $+5.0$/$-4.1$ & \NA & $+3.7$/$-2.8$ & $+0.7$/$-0.7$ & $+0.1$/$-0.4$ \\
\hline
scale uncertainties & $\pm$34 & \NA & \NA & \NA & \NA \\
\hline
PDF & $+8.0$/$-4.4$ & \NA & \NA & \NA & \NA \\
\hline
trigger scale factors & $+3.6$/$-4.0$ & \NA & $+2.3$/$-2.3$ & $+0.7$/$-0.7$ & $+0.06$/$-0.08$ \\
\hline
luminosity & $\pm$2.6 & \NA & $\pm$2.6 & $\pm$2.6 & $\pm$2.6\\
\hline
$\ttbar$ cross section & $\pm$13 & \NA& \NA & \NA & \NA \\
\hline
background estimate: &  & &  &  & \\
single H tag & \NA & $\pm10$ & \NA & \NA & \NA \\
multi H tag & \NA & $\pm20$ & \NA & \NA & \NA \\
\hline
\end{tabular}
\label{summarysystematics}
\end{table}
\section{Results \label{sec:results}}

\begin{figure}[hHtb]
\centering{
                \includegraphics[width=.49\textwidth]{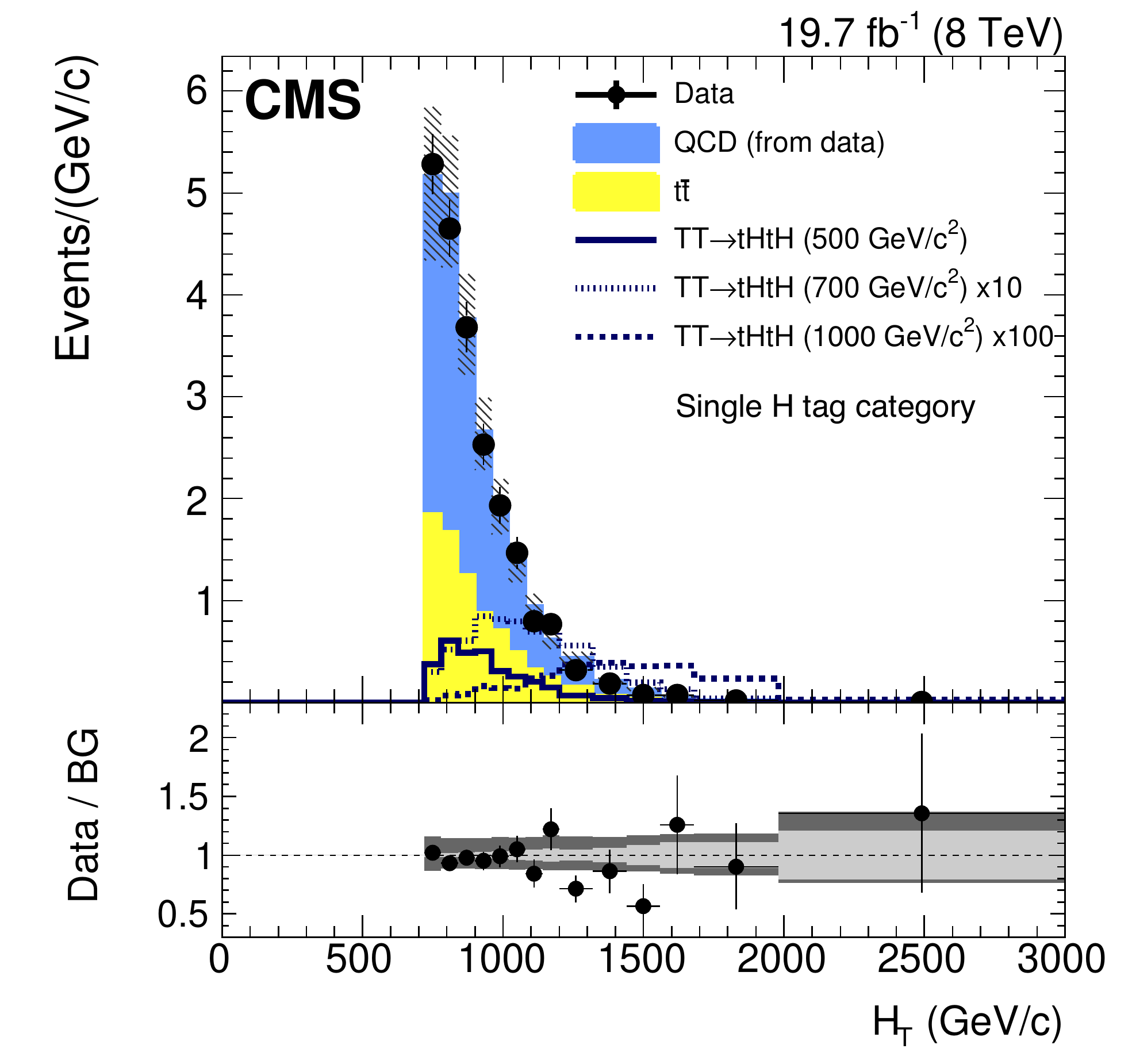}
                \includegraphics[width=.49\textwidth]{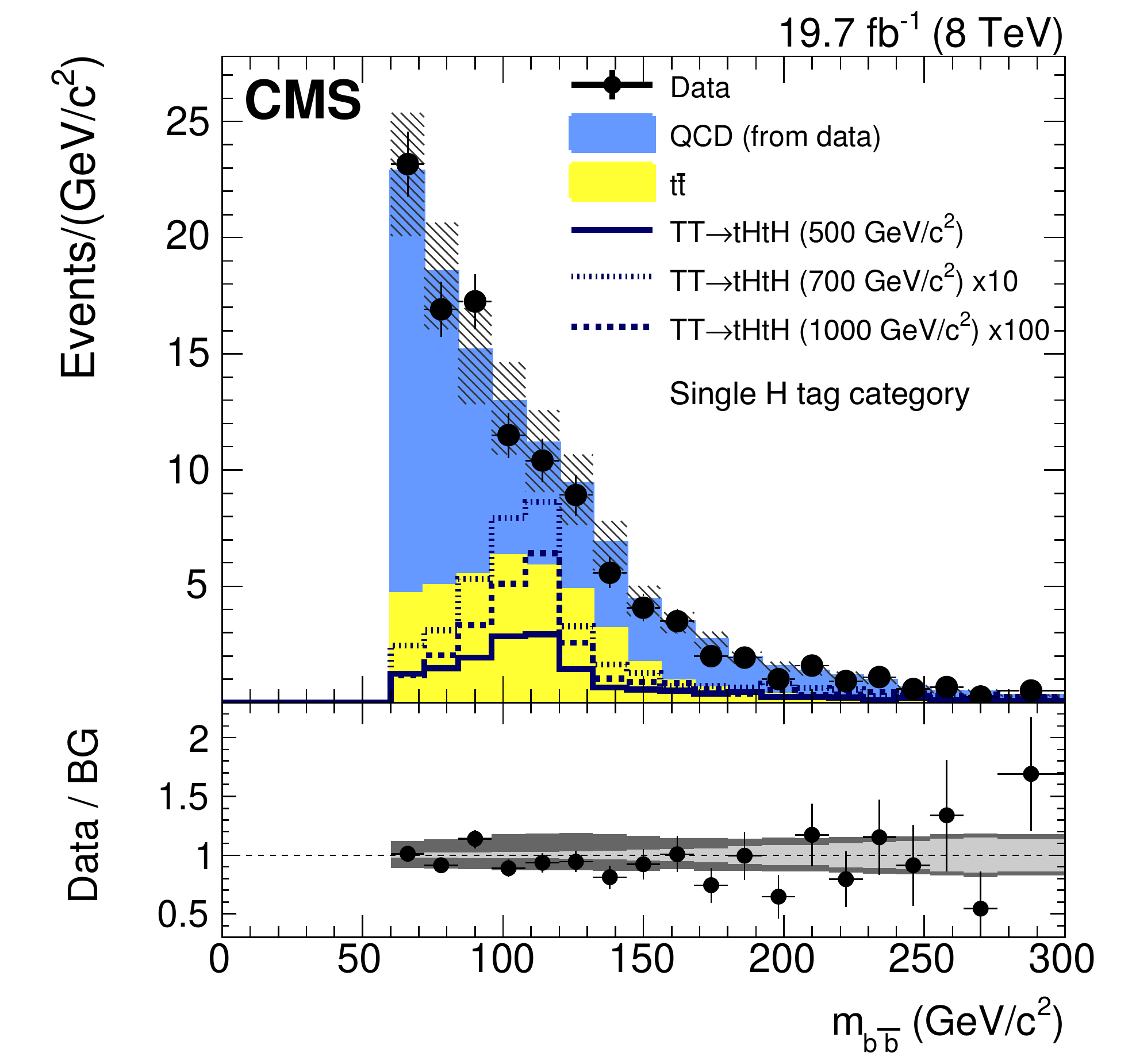}
}
\centering{
                \includegraphics[width=.49\textwidth]{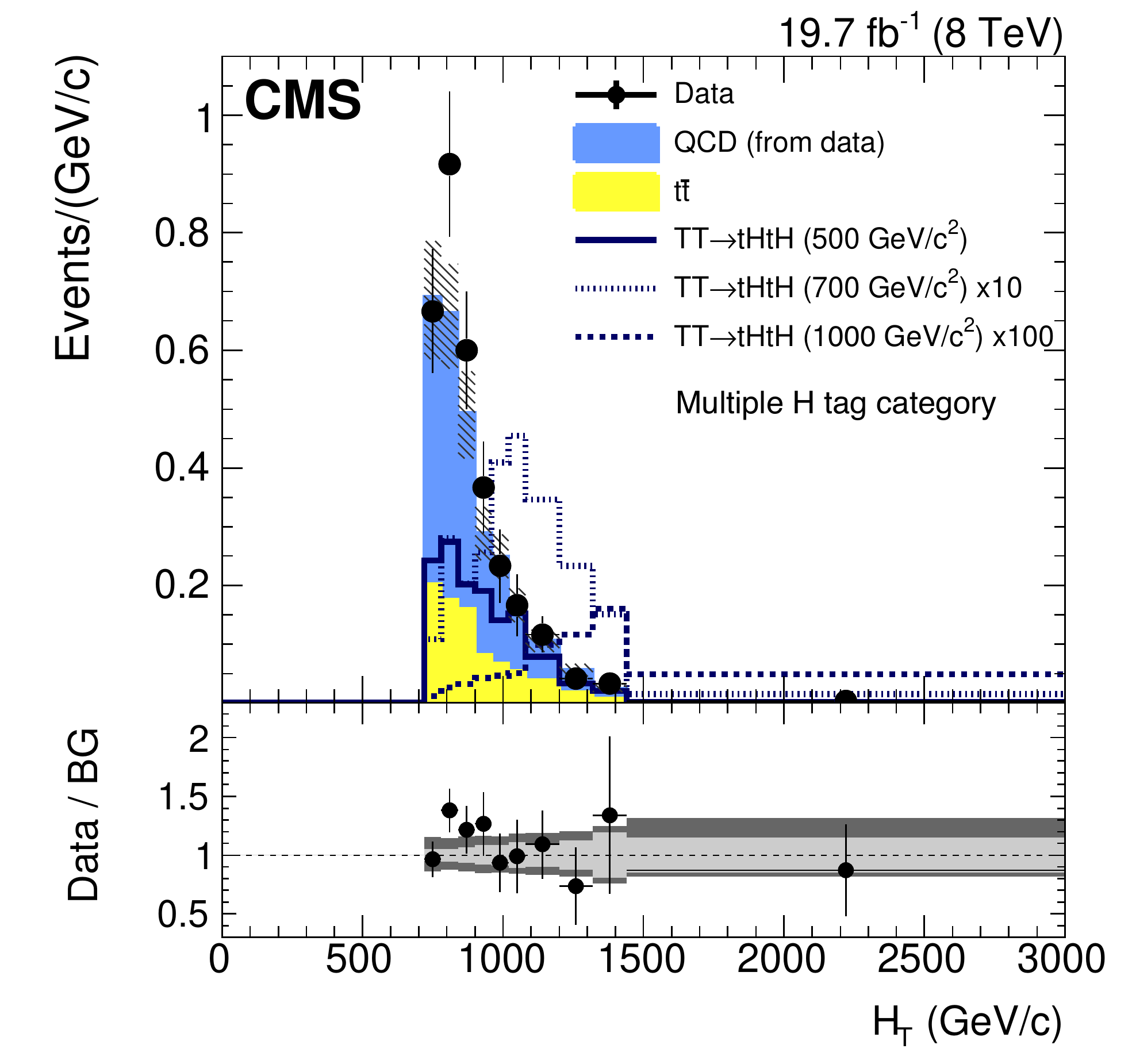}
                \includegraphics[width=.49\textwidth]{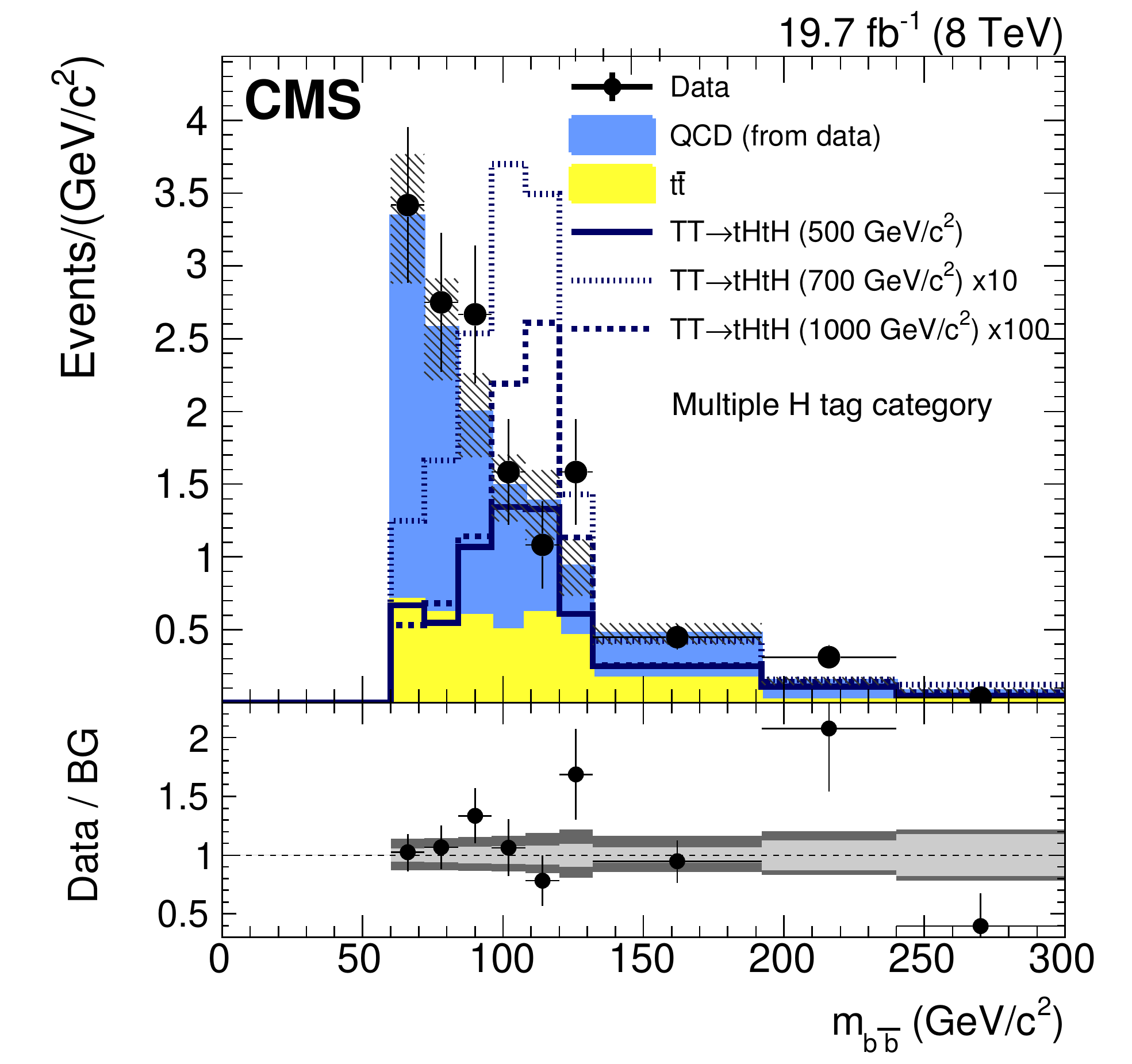}
}
        \caption{The \HT (left) and Higgs boson candidate mass (right) distributions
          for the single H tag category (top) and the multiple H tag category (bottom). The QCD multijet background  is derived
          from data. The $\ttbar$ background is taken
          from simulation.  The hypothetical signal is shown for three
          different mass points: 500, 700, and 1000\GeVcc. The hatched error bands
show the quadratic sum of all systematic and statistical uncertainties
in the background. In the ratio plot, the statistical uncertainty in
the background is depicted by the inner central band, while the
outer  band shows the quadratic sum of all systematic and
statistical uncertainties.}
    \label{fig:distributionsSingleMulti}
\end{figure}

Figure~\ref{fig:distributionsSingleMulti} shows the comparison between
data and the expected  background contributions for the single and
multiple H tag event categories after all event selection
criteria are applied. In the multiple H tag category only the Higgs
boson candidate with the highest transverse momentum is used.   The QCD multijet background has been derived from data as discussed
in Section \ref{sec:background}. Signal
samples at three different mass points are also shown. In these
plots only signal samples in which all T quarks decay into a top quark
and a Higgs boson are shown.

\begin{figure}[Htb]
\centering{
 \includegraphics[width=0.32\textwidth]{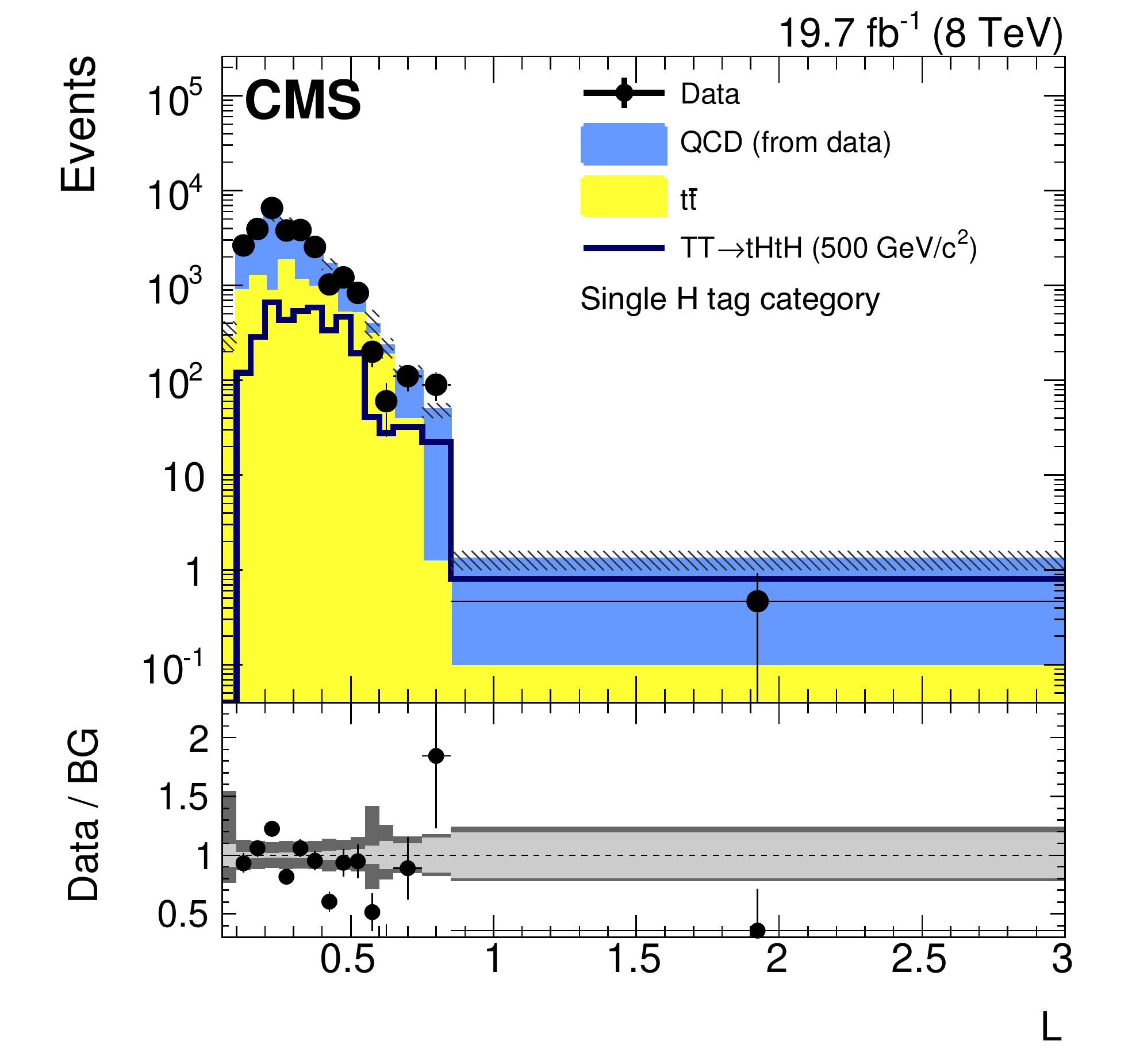}
 \includegraphics[width=0.32\textwidth]{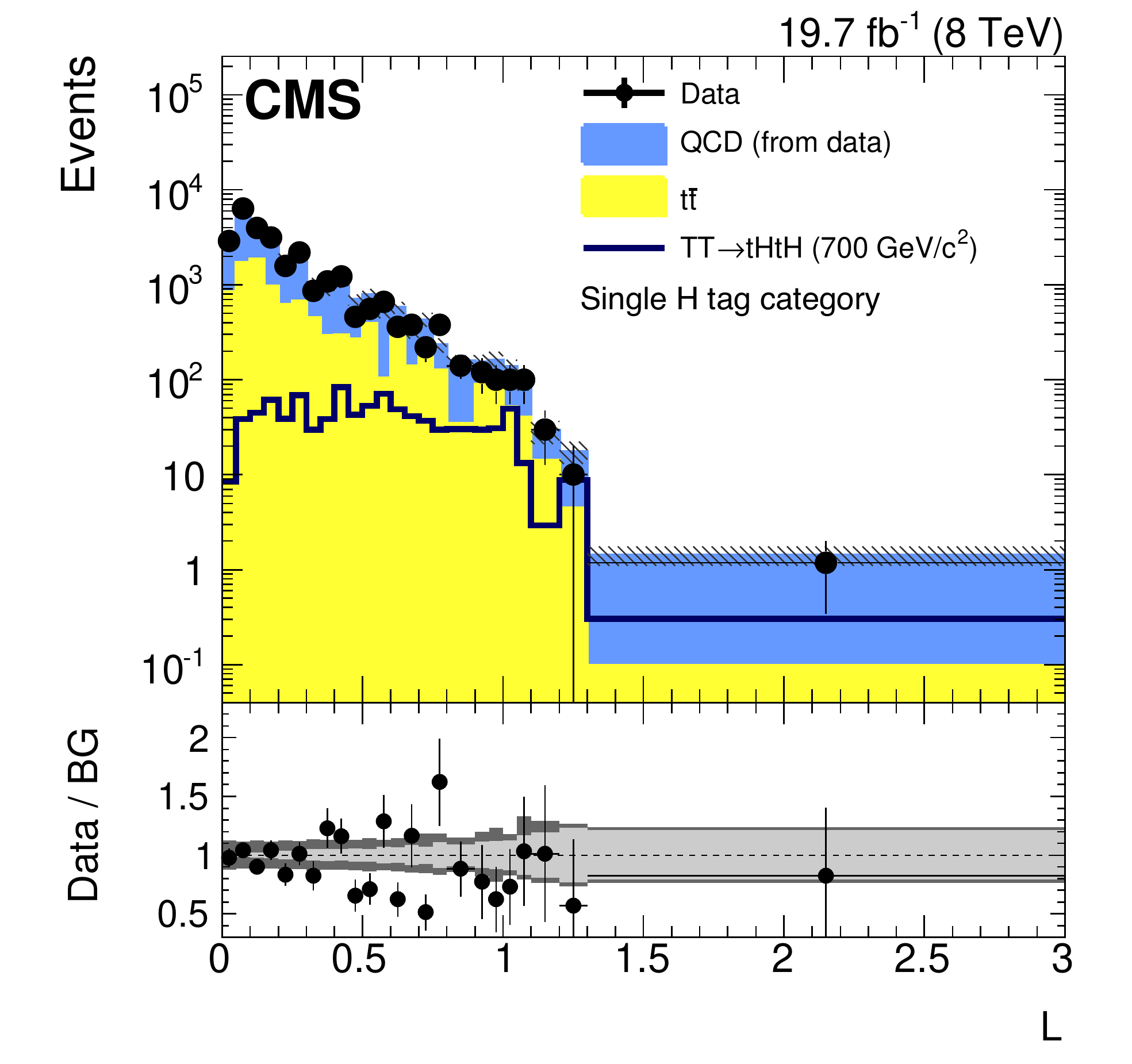}
 \includegraphics[width=0.32\textwidth]{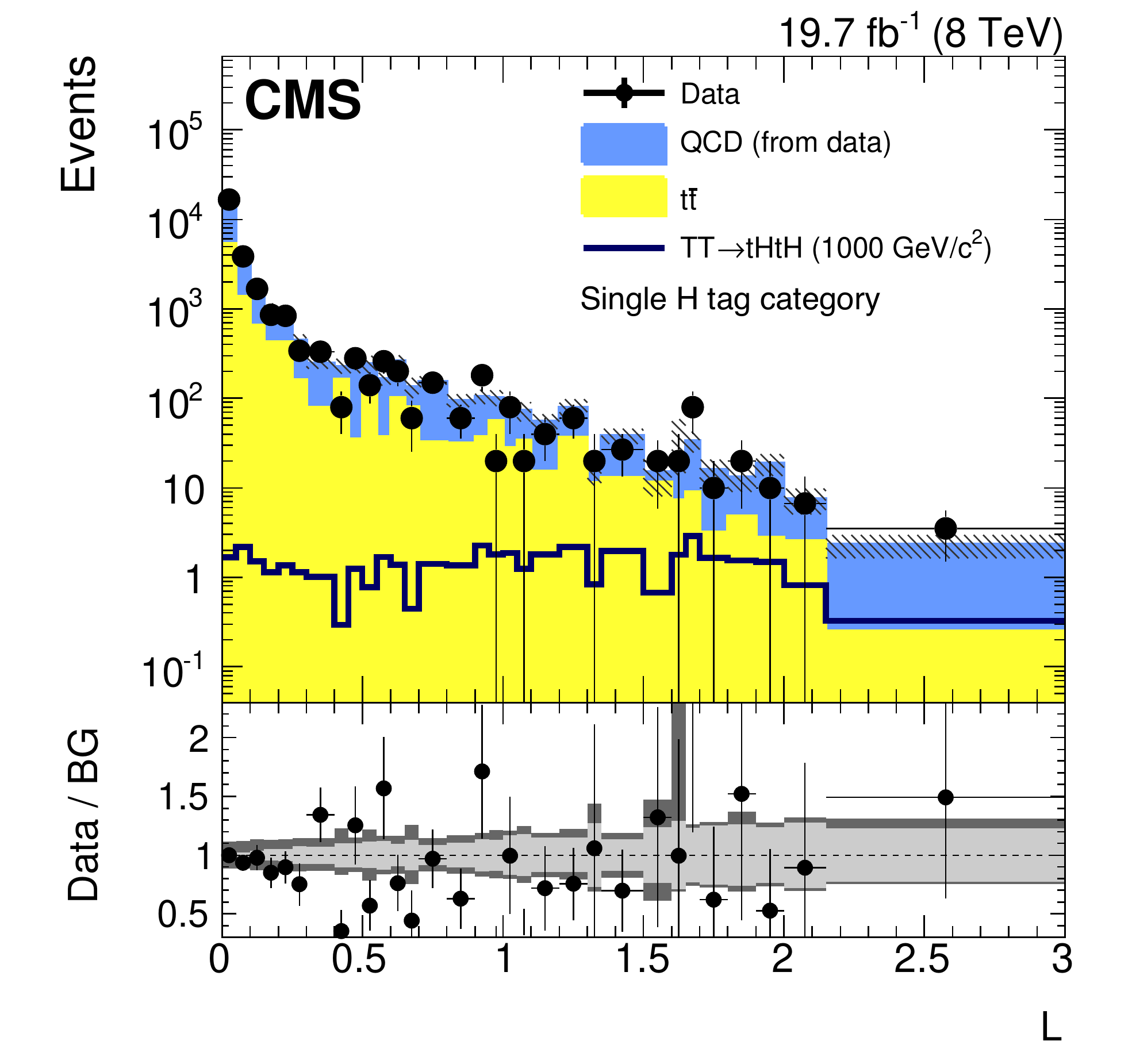}
 \includegraphics[width=0.32\textwidth]{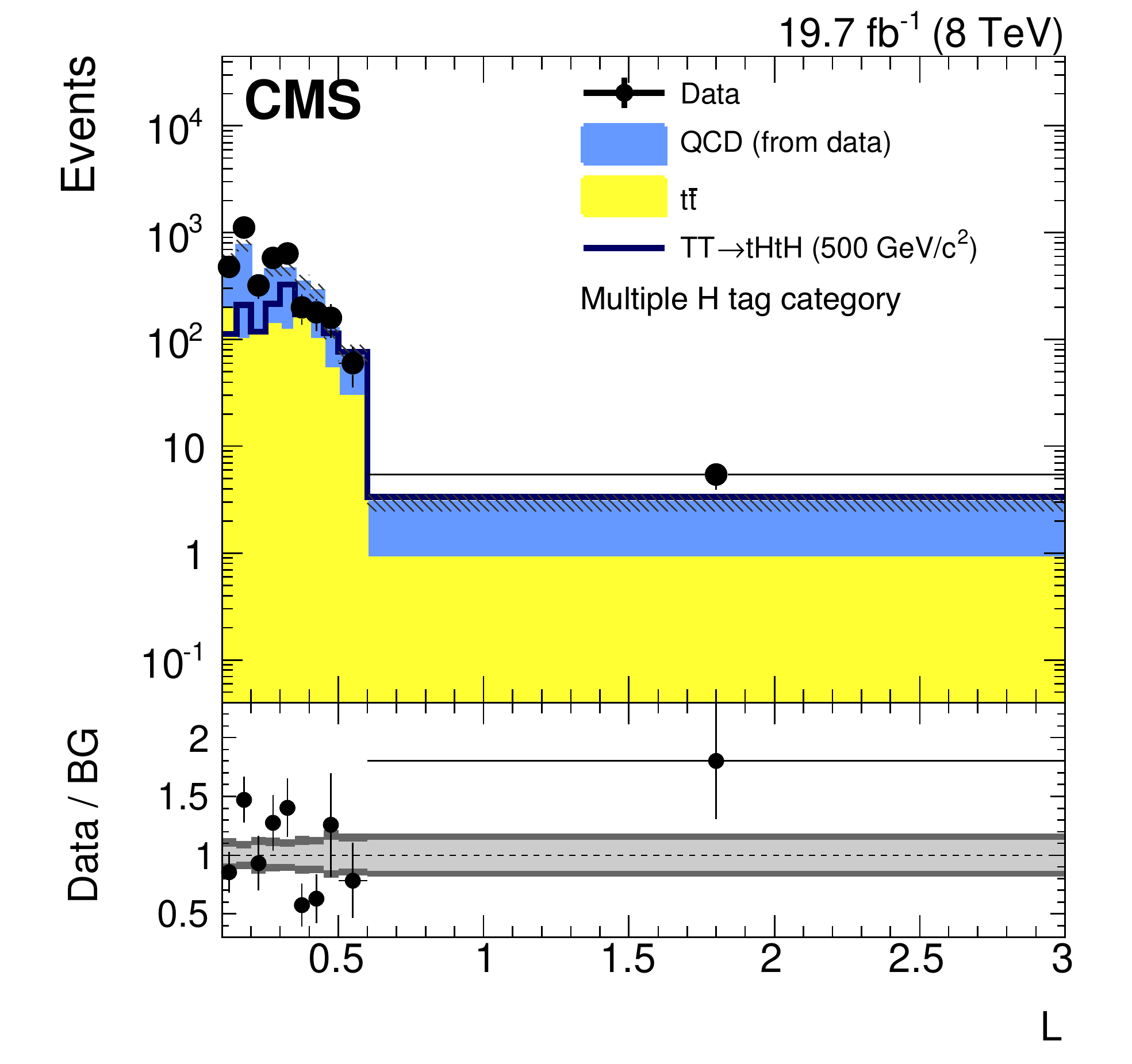}
 \includegraphics[width=0.32\textwidth]{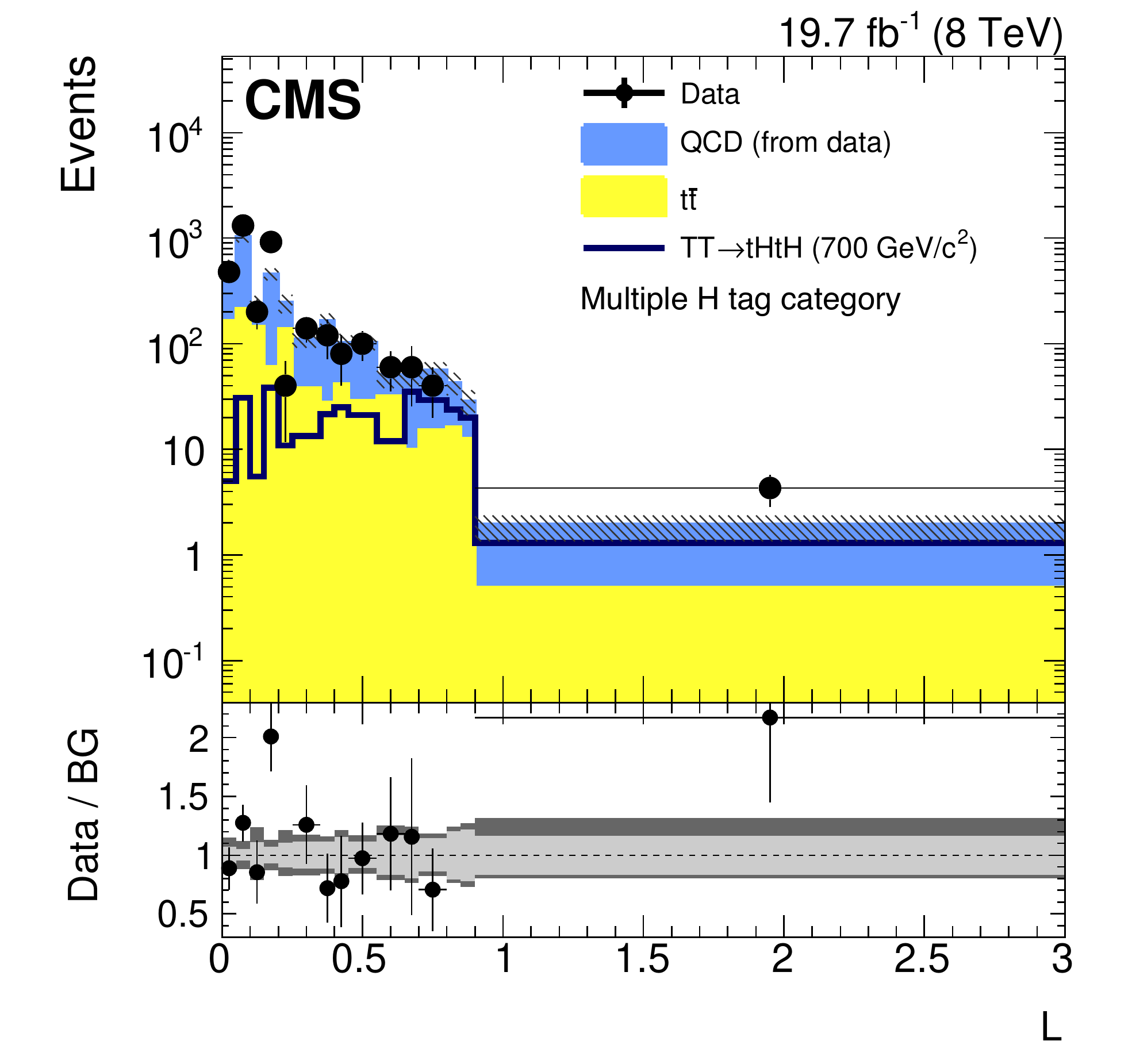}
 \includegraphics[width=0.32\textwidth]{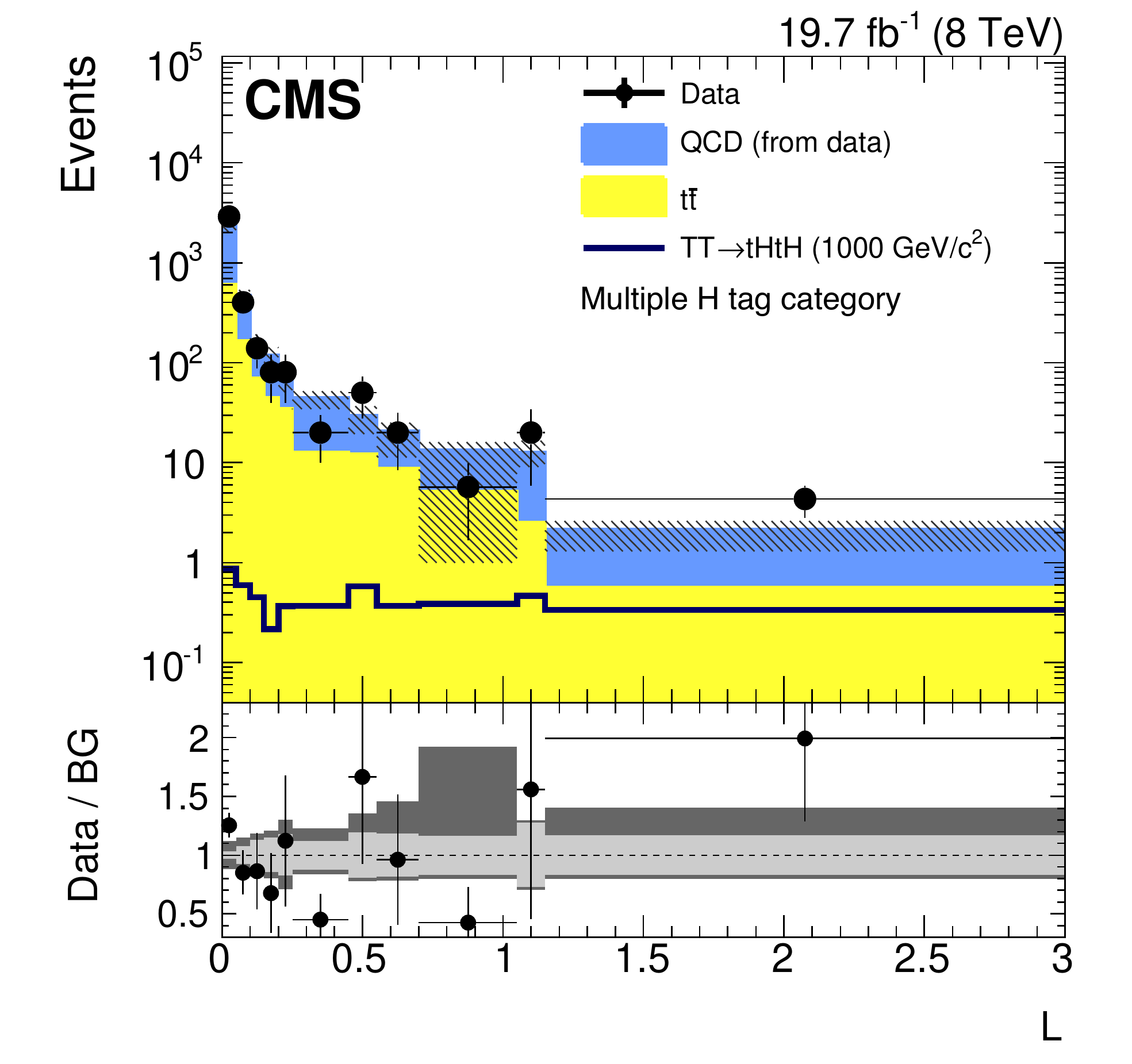}
}
        \caption{Discriminating variable $L$ constructed from both
          \HT and $m_{\bbbar}$ for the single (top) and the multiple (bottom) H tag categories.  The
          three signal hypotheses with 500, 700, and 1000\GeVcc  are
          shown on the left, middle, and right, respectively. The QCD multijet
          background  is derived from data. The $\ttbar$ background is derived from simulation. The hatched error bands
show the quadratic sum of all systematic and statistical uncertainties
in the background. In the ratio plot, the statistical uncertainty in
the background is depicted by the inner central band, while the
outer  band shows the quadratic sum of all systematic and
statistical uncertainties.}
    \label{figFinalResultLSingleMultiH}
\end{figure}

Based on the expected distributions for the background and signal models for \HT and $m_{\bbbar}$, a discriminating quantity $L$ is calculated for each event, where
\begin{linenomath}
\begin{equation}
L = \ln \left( 1+ \frac{ P_\text{sig}(\HT) }{ P_\text{back}(\HT) } \frac{ P_\text{sig}(m_{\bbbar}) }{ P_\text{back}(m_{\bbbar}) } \right).
\end{equation}
\end{linenomath}
The $P$ variables represent the probability densities for the signal
or background hypotheses. The $P_\text{back}$ values are obtained from
the sum of the simulated $\ttbar$ and QCD multijet
background distributions because other background contributions are
found to be negligible, as discussed in Section
\ref{sec:background}. For the signal hypothesis, the $P_\text{sig}$
values are obtained from simulated \HT and $m_{\bbbar}$
distributions for each signal mass point.  A binned likelihood method
is used where the values for the $P$ variables are taken from
histograms. The distribution of this variable is shown in
Fig.~\ref{figFinalResultLSingleMultiH} for data compared to the
background prediction and signal hypotheses, for both the single and
multiple H tag categories. As the signal model is included in the
discriminator, each signal mass hypothesis has its own definition of
$L$. The mass points 500, 700, and 1000\GeVcc are shown in these
figures. The spikes in these distributions are due to the
likelihood definition, that is obtained by taking values from  binned distributions.

\begin{figure}[Htb]
\begin{center}
                \includegraphics[width=.6\textwidth]{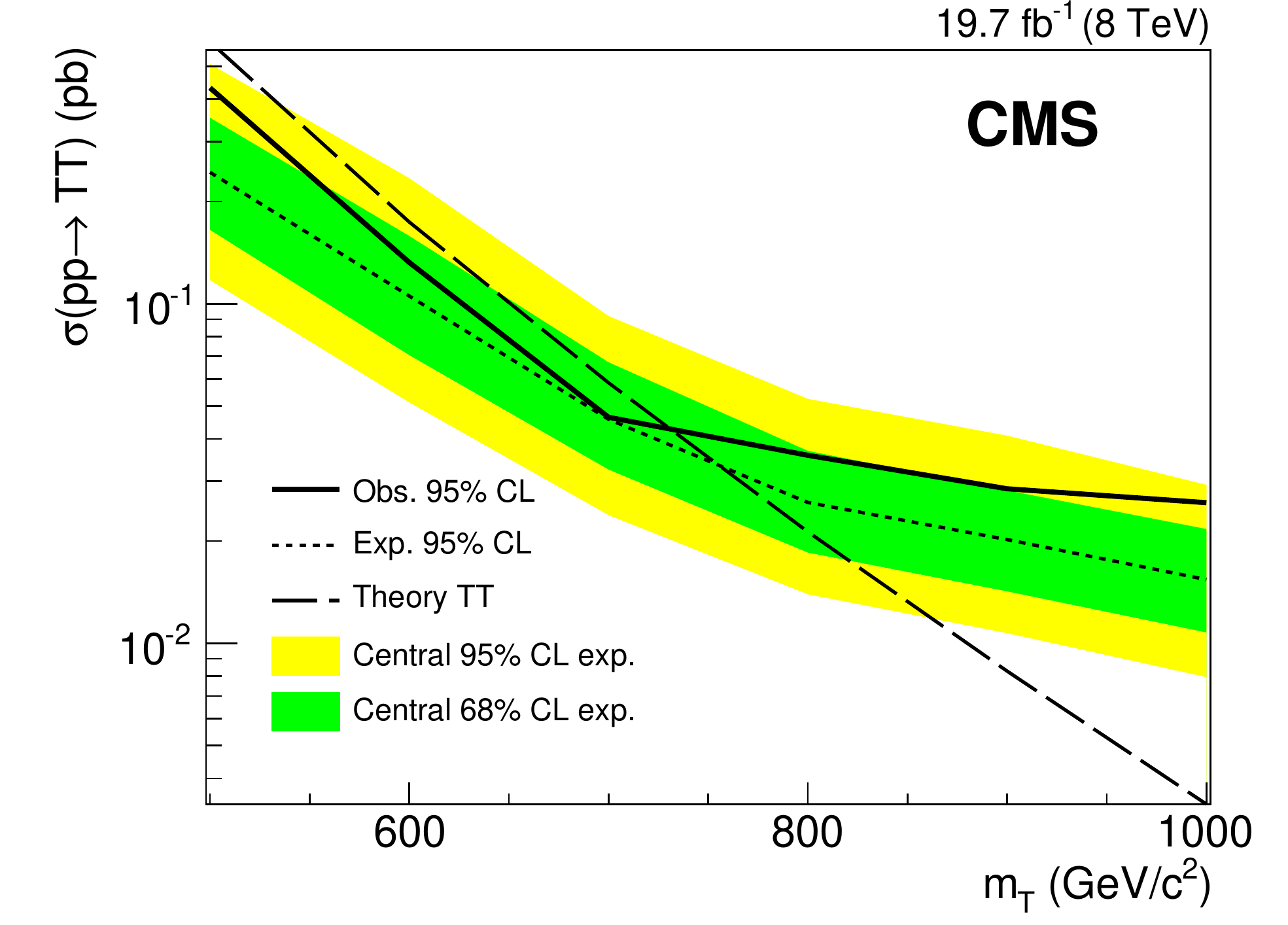}
\end{center}
        \caption{Observed (solid line) and expected (dotted line)
          Bayesian upper limits on the T quark production cross section determined from the variable $L$ for the combination
          of the single and multiple H tag categories, for the hypothesis of
 an exclusive branching fraction $\mathcal{B}(\T\to \PQt\PH) = 100\%$. The green (inner)
          and yellow (outer) bands show the   $1\,\sigma$ ($2\,\sigma$)
          uncertainty ranges, respectively. The
          dashed line shows the prediction of the theory as discussed
          in Section \ref{sec:samples}.}
    \label{fig:LimitsLCombined}
\end{figure}

No signal-like excess is observed in data.  Bayesian upper limits \cite{bayesian} on the T quark production cross section  are obtained with the Theta framework~\cite{theta_web}. The nuisance parameters are assigned to the sources of systematic uncertainties reported in Section \ref{sec:systematics}, which are taken into account as  global normalization
 uncertainties and as shape uncertainties where applicable. The shape
 uncertainties are taken into account by interpolating  between the
 nominal and ${\pm}1\,\sigma$ templates of the likelihood distributions. Figure \ref{fig:LimitsLCombined} shows the observed and expected limits on the T pair production cross section, for the hypothesis of
 an exclusive branching fraction $\mathcal{B}(\T\to \PQt\PH) = 100 \%$ using the
 combination of both the single and multiple H tag event
 categories. T quarks exclusively decaying into tH and with mass values below 745\GeVcc are excluded at 95\% confidence level (CL),
 with an expected exclusion limit of 773\GeVcc. Due to the lower background contamination, the multiple H tag event category provides the largest contribution to the achieved sensitivity.

In evaluating limits, the other decay modes of the T quark must be considered. For
mixed branching fractions  there are six distinct final states: tHtH, tHtZ, tHbW,
bWbW, bWtZ, tZtZ. Three of these final states contain at least one tH
decay. This means that the single H tag category of this analysis is sensitive also to non-exclusive
branching fractions. Furthermore, we also expect some sensitivity to tZ decays
because the mass of the Z boson differs from the mass of the Higgs
boson by only 35\GeVcc and because it decays into b quark pairs with a branching
fraction of 15.6\%.  A selection efficiency of 4.5\% is found for the
tHtZ final state, 3\% for tHbW, and 2\%  for
tZtZ for a T quark mass of 800\GeVcc. These efficiencies are
calculated in the same way as those for tHtH in Table
\ref{tab:signalEfficiency}.

A dedicated optimization is not performed for the non-exclusive decay
modes. Nevertheless, exclusion limits are calculated for all branching
fractions from a scan of all allowed values. Simulated signal
samples have been produced for each set of branching fractions used in
the scan.

\begin{figure}[Htb]
\includegraphics[width=.49\textwidth]{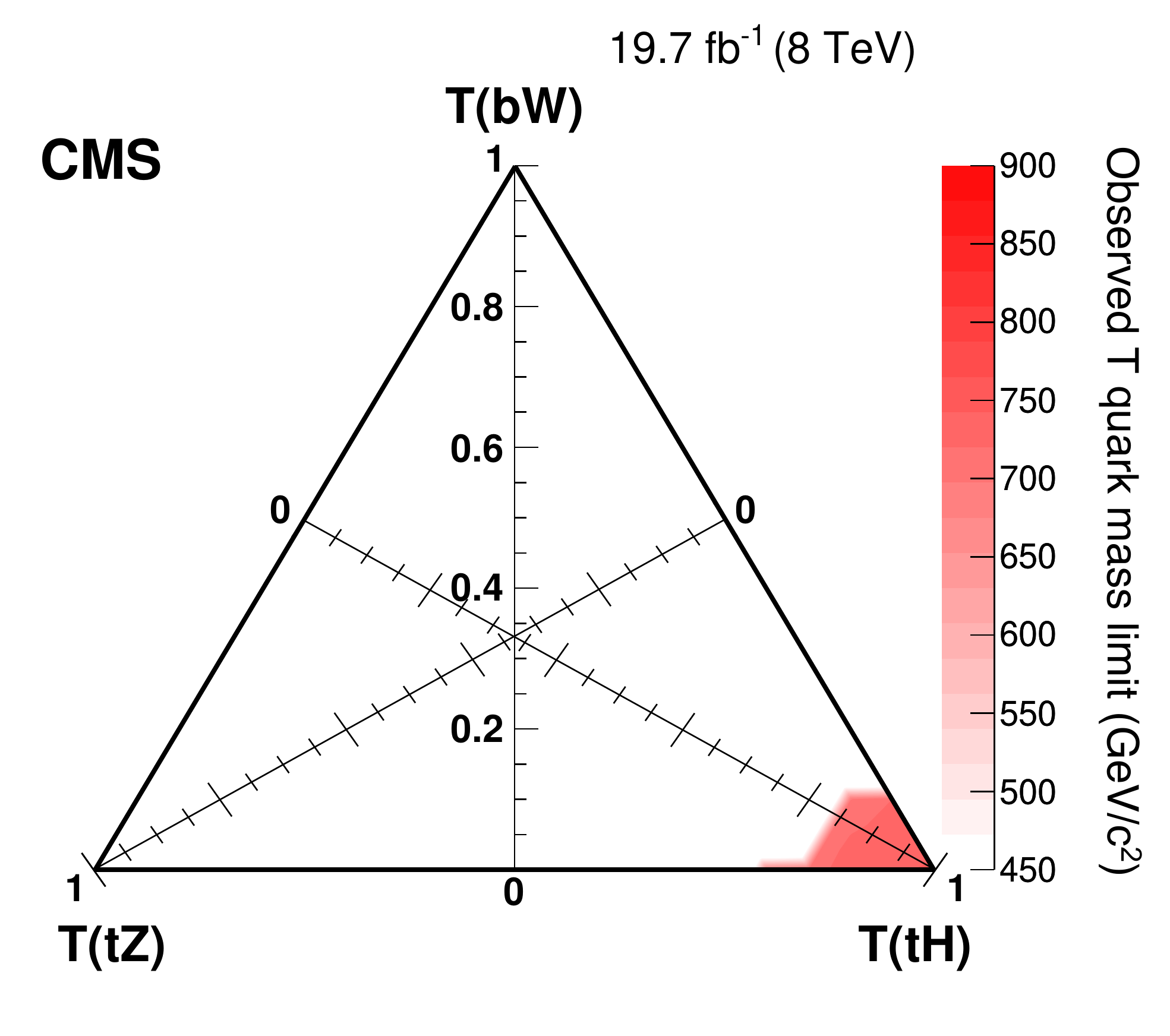}
\includegraphics[width=.49\textwidth]{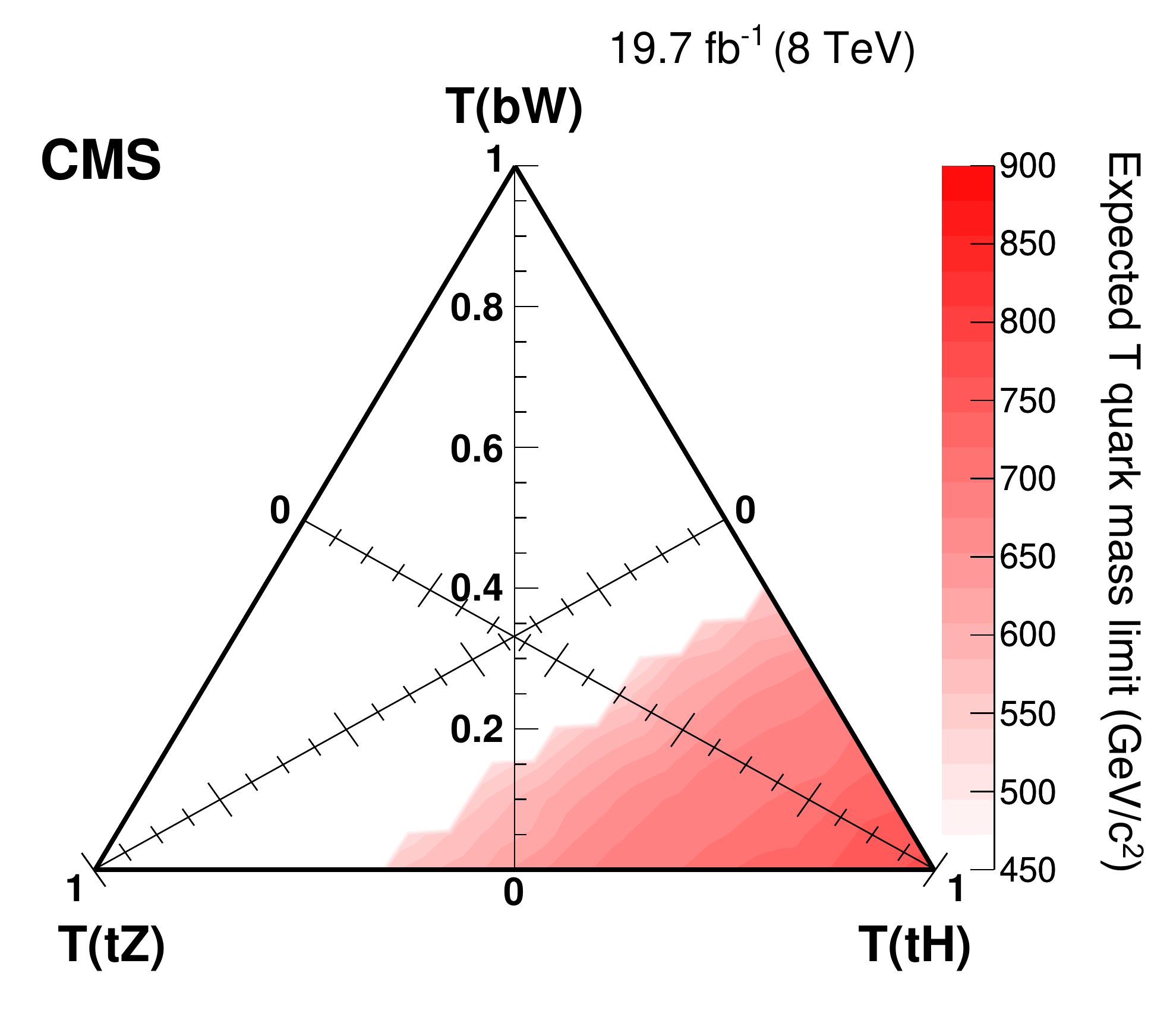}
        \caption{Branching fraction triangle with observed upper limits
          (left) and expected  limits (right) for the T quark
          mass. Every point in the triangle corresponds to a
          particular set of branching fraction values subject to the
          constraint that all three add up to one. The branching
          fraction for each mode decreases from one at the corner
          labelled with the specific decay mode to zero at the opposite side of the triangle.}
    \label{fig:triangleplots}
\end{figure}

\begin{figure}[Htb]
    \includegraphics[width=.33\textwidth]{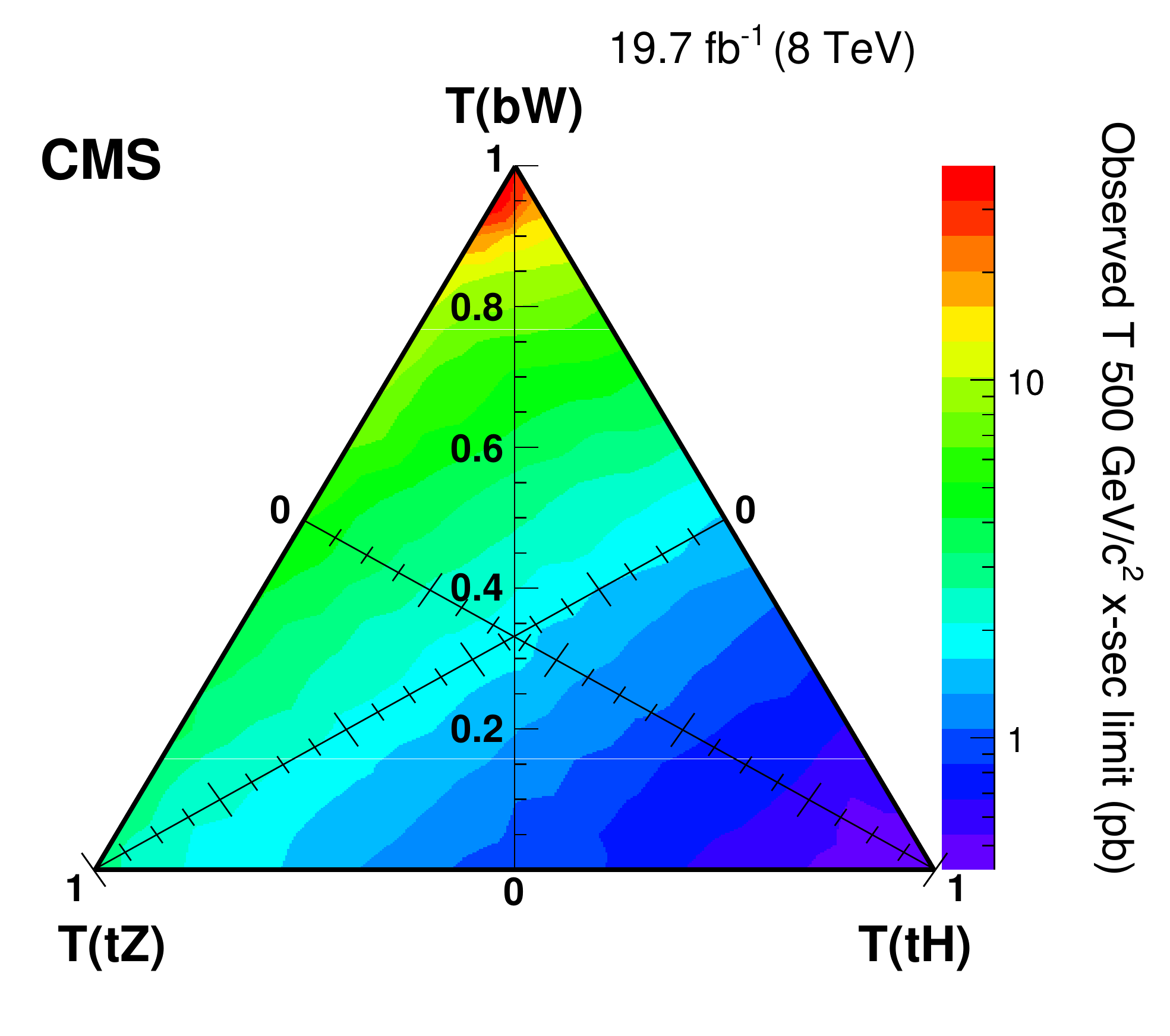}
  \includegraphics[width=.33\textwidth]{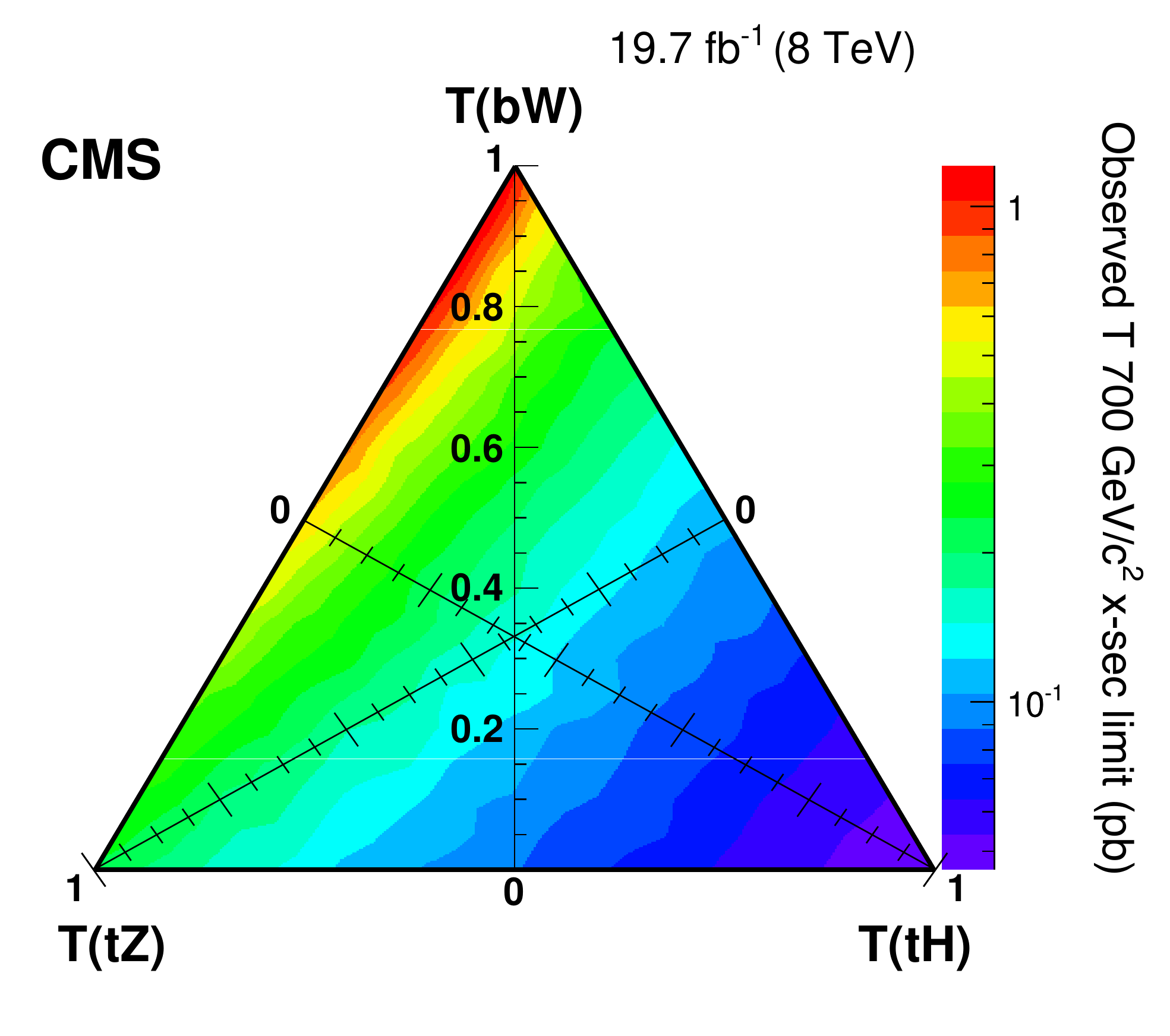}
  \includegraphics[width=.33\textwidth]{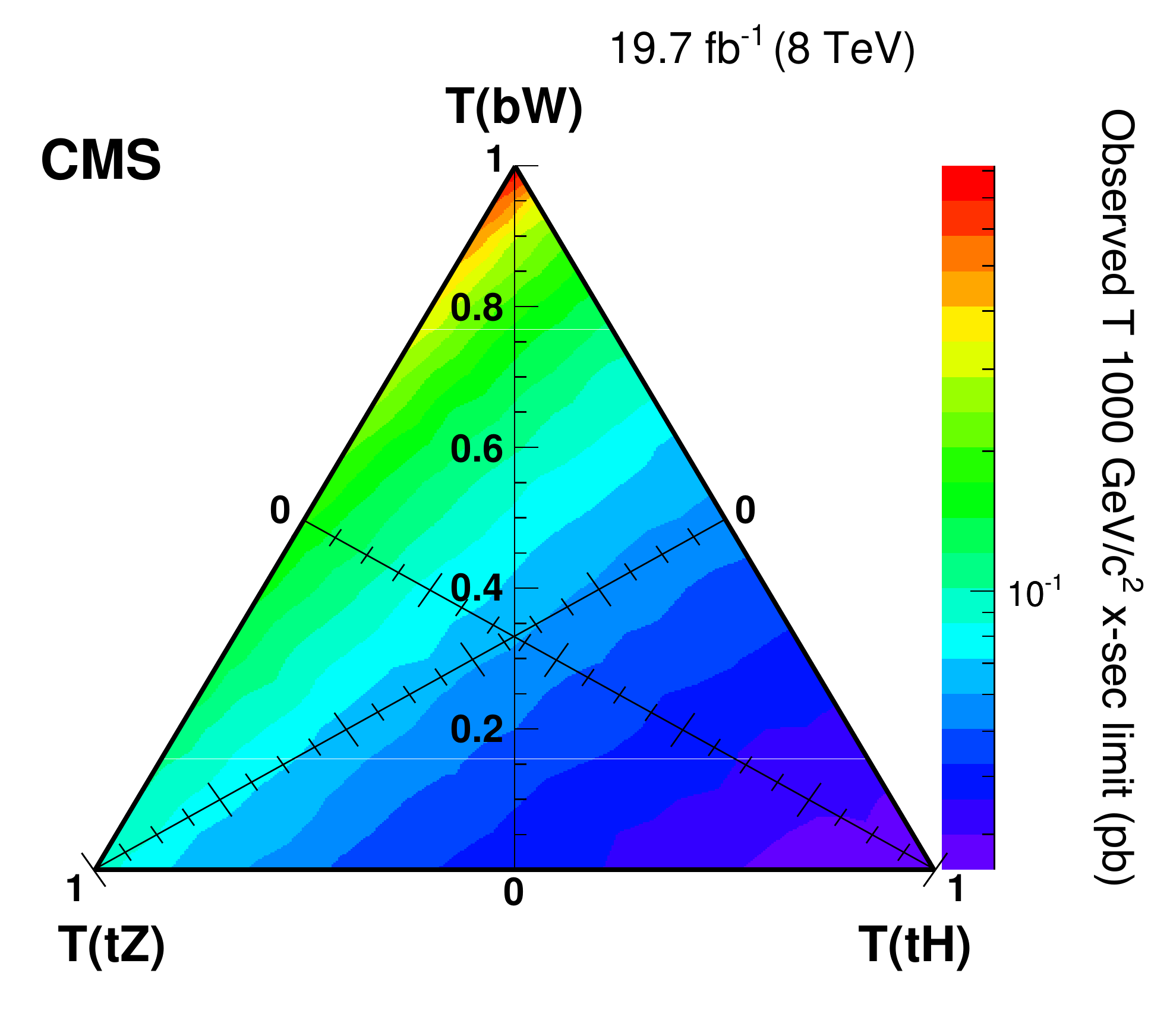}
  \includegraphics[width=.33\textwidth]{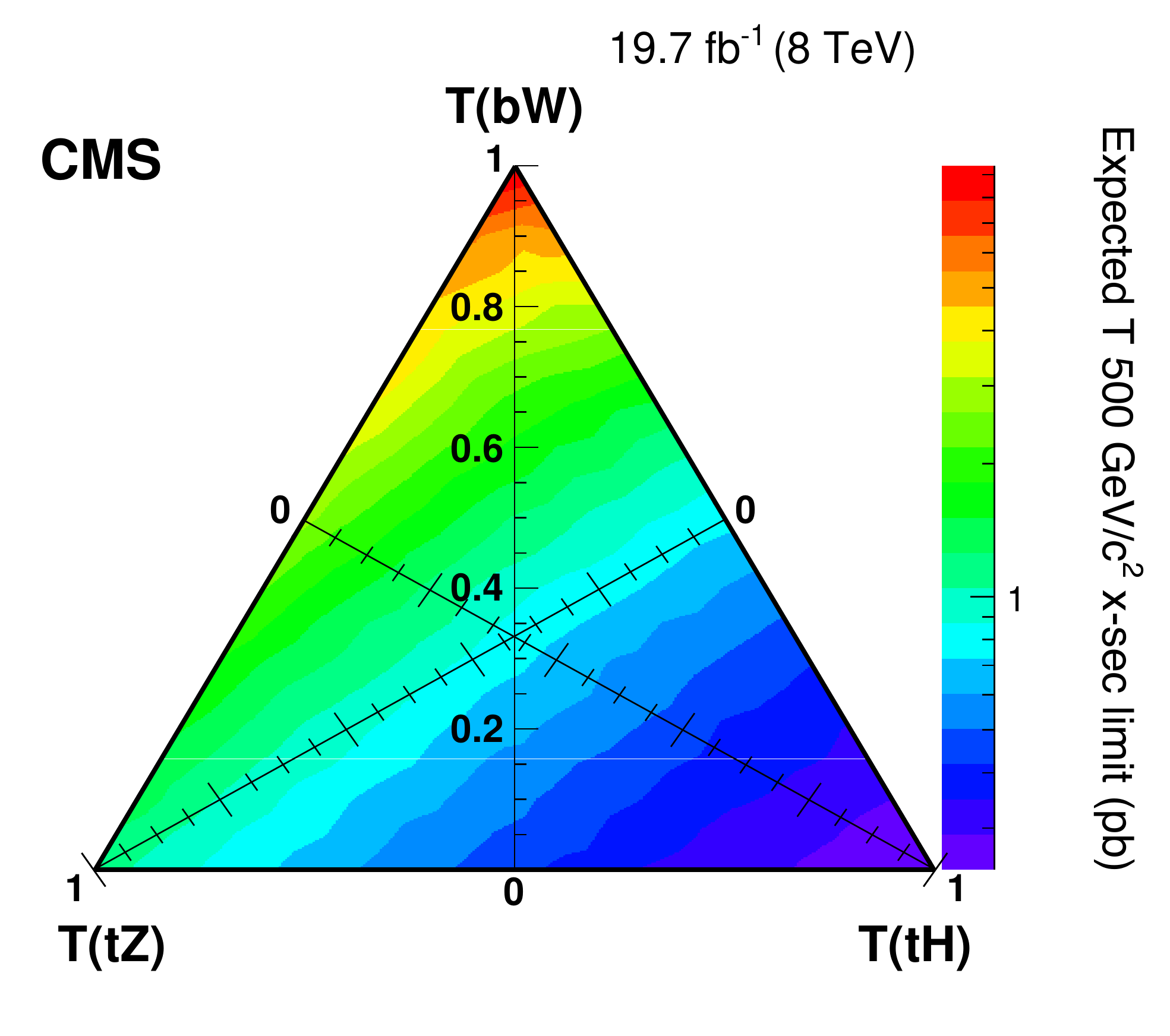}
  \includegraphics[width=.33\textwidth]{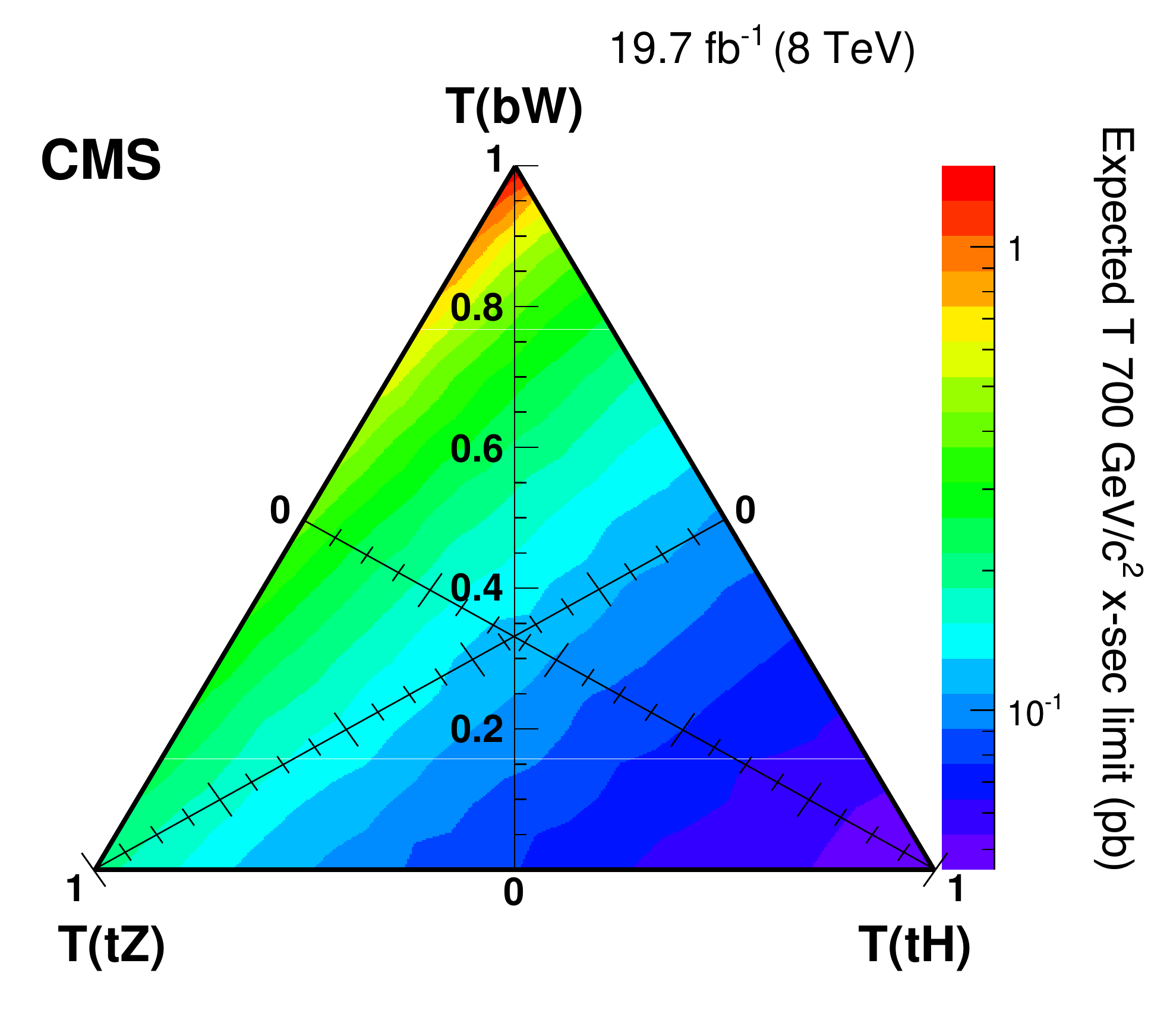}
  \includegraphics[width=.33\textwidth]{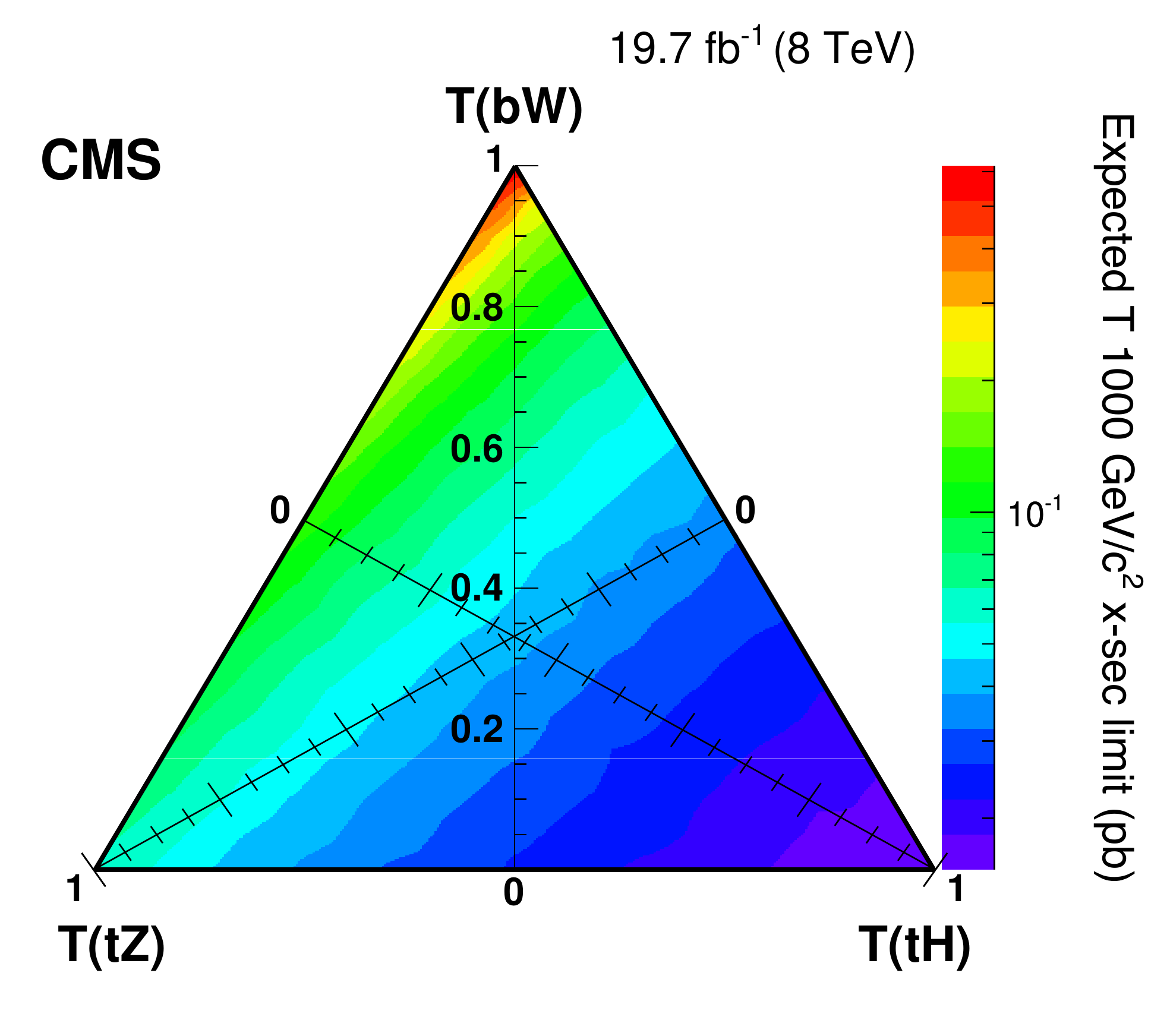}
        \caption{Branching fraction triangle with observed (top) and
          expected (bottom) limits on the T quark pair production
          cross section for three different T quark mass hypotheses: 500 (left), 700 (middle), and 1000\GeVcc
          (right). Every point in the triangle corresponds to a
          particular set of branching fraction values subject to the
          constraint that all three add up to one. The branching
          fraction for each mode decreases from one at the corner
          labelled with the specific decay mode to zero at the opposite side of the triangle.}
    \label{fig:triangleplotscrosssectionexpobs}
\end{figure}

Observed and expected lower  limits on the mass of the T quark for
different branching fractions are listed in Table~\ref{branchingsInScanTable} and shown in
Fig.~\ref{fig:triangleplots}.  Table~\ref{branchingsInScanTable} shows
only those branching fractions for which actual mass limits exist
(where the theory curve crosses the limit curve). A good sensitivity
is achieved for $\T\to\PQt\PH$ branching fractions down to  80\%. The observed and expected limits on the production cross section for
different branching fractions are given in Table
\ref{branchingsInScanTableMass} and shown in Fig.~\ref{fig:triangleplotscrosssectionexpobs}.

\begin{table}[Htb]
 \centering
\topcaption{Observed and expected lower limits on the mass of the T quark
  (in \GeVccns{}) for a range of T quark branching fraction hypotheses
  listed in the first three columns. Only combinations for which an   observed limit is found are reported. When the limit lies below the scanned
  mass region between 500 and 1000\GeVcc a value of $<500$ is indicated.}
\begin{tabular}{lllcccr}
\hline
bW &  tZ & tH  & observed limit& expected limit &
expected$\pm$1$\sigma$ & expected$\pm$2$\sigma$\\
\hline
0.0 & 0.2 & 0.8 & 698 & 732 & [596,795] & [$<$500,851] \\
0.0 & 0.15 & 0.85 & 715 & 734 & [633,798] & [$<$500,857] \\
0.0 & 0.1 & 0.9 & 725 & 751 & [639,806] & [$<$500,862] \\
0.0 & 0.05 & 0.95 & 739 & 763 & [655,827] & [538,873] \\
0.0 & 0.0 & 1.0 & 745 & 773 & [664,832] & [557,875] \\
0.05 & 0.1 & 0.85 & 716 & 732 & [619,798] & [$<$500,856] \\
0.05 & 0.05 & 0.9 & 724 & 749 & [633,812] & [503,858] \\
0.05 & 0.0 & 0.95 & 731 & 757 & [650,817] & [534,865] \\
0.1 & 0.05 & 0.85 & 708 & 730 & [595,795] & [$<$500,849] \\
0.1 & 0.0 & 0.9 & 720 & 737 & [599,799] & [$<$500,859] \\
\hline
\end{tabular}
\label{branchingsInScanTable}
\end{table}

\begin{table}[Htb]
 \centering \topcaption{Branching fractions  (first three columns) and
   the observed and expected upper  limits on the cross section  for
   different mass values of
  the T quark. The expected limits are quoted with their corresponding
  uncertainties while the observed limits are quoted without
  uncertainties.  The cross section
  limits are given in units of pb, while the T quark mass values are given
  in units of\GeVcc.}
\label{branchingsInScanTableMass}
\begin{tabular}{*{9}{c}}
\hline
 bW & tZ & tH &500 & 600 & 700 & 800 & 900 & 1000\\
\hline
&&&0.432&0.132&0.046&0.036&0.029&0.026\\
0&0&1& 0.244$^{+0.109}_{-0.079}$ & 0.105$^{+0.053}_{-0.035}$ & 0.046$^{+0.021}_{-0.014}$ & 0.026$^{+0.011}_{-0.008}$ & 0.020$^{+0.008}_{-0.006}$ & 0.015$^{+0.007}_{-0.004}$ \\
&&&0.576&0.157&0.059&0.046&0.036&0.029\\
0&0.2&0.8& 0.299$^{+0.124}_{-0.100}$ & 0.118$^{+0.063}_{-0.036}$ & 0.054$^{+0.025}_{-0.014}$ & 0.032$^{+0.014}_{-0.010}$ & 0.023$^{+0.011}_{-0.006}$ & 0.018$^{+0.008}_{-0.005}$ \\
&&&0.866&0.191&0.076&0.057&0.043&0.036\\
0&0.4&0.6& 0.389$^{+0.210}_{-0.122}$ & 0.143$^{+0.074}_{-0.043}$ & 0.067$^{+0.027}_{-0.019}$ & 0.041$^{+0.019}_{-0.012}$ & 0.030$^{+0.014}_{-0.009}$ & 0.023$^{+0.010}_{-0.007}$ \\
&&&0.656&0.155&0.061&0.049&0.038&0.033\\
0.2&0&0.8& 0.340$^{+0.174}_{-0.110}$ & 0.137$^{+0.062}_{-0.043}$ & 0.061$^{+0.027}_{-0.018}$ & 0.035$^{+0.015}_{-0.010}$ & 0.026$^{+0.011}_{-0.008}$ & 0.020$^{+0.009}_{-0.006}$ \\
&&&0.934&0.206&0.081&0.060&0.049&0.039\\
0.2&0.2&0.6& 0.459$^{+0.241}_{-0.150}$ & 0.165$^{+0.080}_{-0.052}$ & 0.076$^{+0.033}_{-0.022}$ & 0.045$^{+0.019}_{-0.014}$ & 0.033$^{+0.014}_{-0.010}$ & 0.025$^{+0.011}_{-0.007}$ \\
\hline
\end{tabular}
\end{table}

\clearpage
\section{Summary}
A search for heavy resonances decaying to top quarks and Higgs bosons
has been performed using proton-proton collisions recorded with the CMS detector at
$\sqrt{s}=8$\TeV, corresponding to an integrated luminosity of 19.7\fbinv.
The benchmark model considered is a heavy vector-like T quark that
decays into bW, tZ, and tH in all-hadronic final states. The analysis
makes use of jet substructure techniques including algorithms for the identification of boosted top quarks, boosted Higgs bosons, and subjet b tagging. Results are presented for exclusive T quark
decay modes as well as for non-exclusive branching fractions. If the heavy
T quark has a branching fraction  of 100\% for $\T\to\PQt\PH$, the observed
(expected) exclusion limit on the mass of the T quark is 745 (773)\GeVcc at  95\% confidence level.  This limit is similar to that obtained from leptonic final states \cite{tagkey2014149}. These results are the first to exploit the all-hadronic final state in the search for vector-like quarks and they facilitate the combination with other analyses to improve the mass reach.

\begin{acknowledgments}

We congratulate our colleagues in the CERN accelerator departments for the excellent performance of the LHC and thank the technical and administrative staffs at CERN and at other CMS institutes for their contributions to the success of the CMS effort. In addition, we gratefully acknowledge the computing centres and personnel of the Worldwide LHC Computing Grid for delivering so effectively the computing infrastructure essential to our analyses. Finally, we acknowledge the enduring support for the construction and operation of the LHC and the CMS detector provided by the following funding agencies: BMWFW and FWF (Austria); FNRS and FWO (Belgium); CNPq, CAPES, FAPERJ, and FAPESP (Brazil); MES (Bulgaria); CERN; CAS, MoST, and NSFC (China); COLCIENCIAS (Colombia); MSES and CSF (Croatia); RPF (Cyprus); MoER, ERC IUT and ERDF (Estonia); Academy of Finland, MEC, and HIP (Finland); CEA and CNRS/IN2P3 (France); BMBF, DFG, and HGF (Germany); GSRT (Greece); OTKA and NIH (Hungary); DAE and DST (India); IPM (Iran); SFI (Ireland); INFN (Italy); MSIP and NRF (Republic of Korea); LAS (Lithuania); MOE and UM (Malaysia); CINVESTAV, CONACYT, SEP, and UASLP-FAI (Mexico); MBIE (New Zealand); PAEC (Pakistan); MSHE and NSC (Poland); FCT (Portugal); JINR (Dubna); MON, RosAtom, RAS and RFBR (Russia); MESTD (Serbia); SEIDI and CPAN (Spain); Swiss Funding Agencies (Switzerland); MST (Taipei); ThEPCenter, IPST, STAR and NSTDA (Thailand); TUBITAK and TAEK (Turkey); NASU and SFFR (Ukraine); STFC (United Kingdom); DOE and NSF (USA).

Individuals have received support from the Marie-Curie programme and the European Research Council and EPLANET (European Union); the Leventis Foundation; the A. P. Sloan Foundation; the Alexander von Humboldt Foundation; the Belgian Federal Science Policy Office; the Fonds pour la Formation \`a la Recherche dans l'Industrie et dans l'Agriculture (FRIA-Belgium); the Agentschap voor Innovatie door Wetenschap en Technologie (IWT-Belgium); the Ministry of Education, Youth and Sports (MEYS) of the Czech Republic; the Council of Science and Industrial Research, India; the HOMING PLUS programme of the Foundation for Polish Science, cofinanced from European Union, Regional Development Fund; the Compagnia di San Paolo (Torino); the Consorzio per la Fisica (Trieste); MIUR project 20108T4XTM (Italy); the Thalis and Aristeia programmes cofinanced by EU-ESF and the Greek NSRF; and the National Priorities Research Program by Qatar National Research Fund.
\end{acknowledgments}

 \clearpage
\bibliography{auto_generated}

\cleardoublepage \appendix\section{The CMS Collaboration \label{app:collab}}\begin{sloppypar}\hyphenpenalty=5000\widowpenalty=500\clubpenalty=5000\textbf{Yerevan Physics Institute,  Yerevan,  Armenia}\\*[0pt]
V.~Khachatryan, A.M.~Sirunyan, A.~Tumasyan
\vskip\cmsinstskip
\textbf{Institut f\"{u}r Hochenergiephysik der OeAW,  Wien,  Austria}\\*[0pt]
W.~Adam, T.~Bergauer, M.~Dragicevic, J.~Er\"{o}, M.~Friedl, R.~Fr\"{u}hwirth\cmsAuthorMark{1}, V.M.~Ghete, C.~Hartl, N.~H\"{o}rmann, J.~Hrubec, M.~Jeitler\cmsAuthorMark{1}, W.~Kiesenhofer, V.~Kn\"{u}nz, M.~Krammer\cmsAuthorMark{1}, I.~Kr\"{a}tschmer, D.~Liko, I.~Mikulec, D.~Rabady\cmsAuthorMark{2}, B.~Rahbaran, H.~Rohringer, R.~Sch\"{o}fbeck, J.~Strauss, W.~Treberer-Treberspurg, W.~Waltenberger, C.-E.~Wulz\cmsAuthorMark{1}
% ...
\vskip\cmsinstskip
% ...
1:~~Also at Vienna University of Technology, Vienna, Austria\\
2:~~Also at CERN, European Organization for Nuclear Research, Geneva, Switzerland\\
% ...

\end{sloppypar}
\end{document}